\documentclass[graybox,envcountchap]{svmono}

\usepackage{amsmath}
\usepackage{amssymb}
\usepackage{nomencl,authblk} 

\usepackage{mathptmx}
\usepackage{helvet}
\usepackage{courier}
\usepackage{type1cm}         

\usepackage{makeidx}         
\usepackage{graphicx}        
\usepackage{multicol}        
\usepackage[bottom]{footmisc}

\makeindex             

\begin{document}

\author{\large{
		Shi-Ju Ran\footnote{\ \ Department of Physics, Capital Normal University, Beijing 100048, China} ,
		Emanuele Tirrito\footnote{\ \ ICFO - Institut de Ciencies Fotoniques, The Barcelona Institute of Science and Technology, Av. Carl Friedrich Gauss 3, 08860 Castelldefels (Barcelona), Spain} ,
		Cheng Peng\footnote{\ \ School of Physical Sciences, University of Chinese Academy of Sciences, P. O. Box 4588, Beijing 100049, China} , 
		Xi Chen$^{\ddagger}$, 
		Luca Tagliacozzo\footnote{\ \ Department of Physics and SUPA, University of Strathclyde, Glasgow G4 0NG, United Kingdom; Departament de F\'isica Qu\`antica i Astrof\'isica and Institut de Ci\`encies del Cosmos (ICCUB), Universitat deBarcelona, Mart\'ii Franqu\`es 1, 08028 Barcelona, Catalonia, Spain} , 
		Gang Su$^{\ddagger,}$\footnote{\ Kavli Institute for Theoretical Sciences and CAS Center for Excellence in Topological Quantum Computation} , 
		Maciej Lewenstein$^{\dag,}$\footnote{\ \ ICREA, Pg. Llu\'is Companys 23, 08010 Barcelona, Spain}}}

%

\title{Lecture Notes of Tensor Network Contractions}
\subtitle{}

\maketitle

\frontmatter

%
%

\preface
Tensor network (TN), a young mathematical tool of high vitality and great potential, has been undergoing extremely rapid developments in the last two decades, gaining tremendous success in condensed matter physics, atomic physics, quantum information science, statistical physics, and so on. In this lecture notes, we focus on the contraction algorithms of TN as well as some of the applications to the simulations of quantum many-body systems. Starting from basic concepts and definitions, we first explain the relations between TN and physical problems, including the TN representations of classical partition functions, quantum many-body states (by matrix product state, tree TN, and projected entangled pair state), time evolution simulations, etc. These problems, which are challenging to solve, can be transformed to TN contraction problems. We present then several paradigm algorithms based on the ideas of the numerical renormalization group and/or boundary states, including density matrix renormalization group, time-evolving block decimation, coarse-graining/corner tensor renormalization group, and several distinguished variational algorithms. Finally, we revisit the TN approaches from the perspective of multi-linear algebra (also known as tensor algebra or tensor decompositions) and quantum simulation. Despite the apparent differences in the ideas and strategies of different TN algorithms, we aim at revealing the underlying relations and resemblances in order to present a systematic picture to understand the TN contraction approaches. 

\vspace{\baselineskip}

%
%

\extrachap{Acknowledgements}
We are indebted to Mari-Carmen Ba\~nuls, Ignacio Cirac, Jan von Delft, Yichen Huang, Karl Jansen, Jos\'e Ignacio Latorre,  Michael Lubasch, Wei Li, Simone Montagero, Tomotoshi Nishino, Roman Or\'us, Didier Poilblanc, Guifre Vidal, Andreas Weichselbaum, Tao Xiang, and Xin Yan for helpful discussions and suggestions. SJR acknowledges Fundaci\'o Catalunya - La Pedrera $\cdot$ Ignacio Cirac Program Chair, and Beijing Natural Science Foundation (Grants No. 1192005). ET and ML acknowledge the Spanish Ministry MINECO (National Plan 15 Grant: FISICATEAMO No. FIS2016-79508-P, SEVERO OCHOA No. SEV-2015-0522, FPI), European Social Fund, Fundació Cellex, Generalitat de Catalunya (AGAUR Grant No. 2017 SGR 1341 and CERCA/Program), ERC AdG OSYRIS and NOQIA, EU FETPRO QUIC, and the National Science Centre, Poland-Symfonia Grant No. 2016/20/W/ST4/00314. LT was supported by the Spanish RYC-2016-20594 program from MINECO. SJR, CP, XC, and GS were supported by the NSFC (Grant No. 11834014). CP, XC, and GS were supported in part by the National Key R\&D Program of China (Grant No. 2018FYA0305800), the Strategic Priority Research Program of CAS (Grant No. XDB28000000), and Beijing Municipal Science and Technology Commission (Grant No. Z118100004218001).

\tableofcontents

\mainmatter
\chapter{Introduction}

\abstract{
	One characteristic that defines us, human beings, is the curiosity of the unknown. Since our birth, we have been trying to use any methods that human brains can comprehend to explore the nature: to mimic, to understand and to utilize in a controlled and repeatable way. One of the most ancient means lies in the nature herself, experiments, leading to tremendous achievements from the creation of fire to the scissors of genes. Then comes mathematics, a new world we made by numbers and symbols, where the nature is reproduced by laws and theorems in an extremely simple, beautiful and unprecedentedly accurate manner. With the explosive development of digital sciences, computer was created. It provided us the third way to investigate the nature, a digital world whose laws can be ruled by ourselves with codes and \textit{algorithms} to numerically mimic the real universe. In this chapter, we briefly review the history of tensor network algorithms and the related progresses made recently. The organization of our lecture notes is also presented.}

\section{Numeric renormalization group in one dimension}

Numerical simulation is one of the most important approaches in science, in particular for the complicated problems beyond the reach of analytical solutions. One distinguished example of the algorithms in physics as well as in chemistry is \textit{ab-initio} principle calculation, which is based on density function theory (DFT\index{DFT}) \cite{S10DFTRev,B12DFT,B14DFT}. It provides a reliable solution to simulate a wide range of materials that can be described by the mean-field theories and/or single-particle approximations. Monte Carlo method \cite{R14MCbook}, named after a city famous of gambling in Monaco, is another example that appeared in almost every corner of science. In contemporary physics, however, there are still many ``hard nuts to crack''. Specifically in quantum physics, numerical simulation faces un-tackled issues for the systems with strong correlations, which might lead to exciting and exotic phenomena like high-temperature superconductivity \cite{LNW06HighSCRev,KKNUZ15highTcRev} and fractional excitations \cite{L99FracRev}.

Tensor network (TN) \index{TN} methods in the context of many-body quantum systems have been developed recently. One could however identify some precursors of them in the seminal works of Krammers and Wannier \cite{kramers1941statistics,kramers1941statistics2}, Baxter  \cite{B68Prototype,NB86Baxterfomu}, Kelland \cite{kelland1976estimates}, Tsang \cite{tsang1979square}, Nighingale and Blote \cite{NB86Baxterfomu}, Derrida \cite{DE93MPS000,DEHP93MPS0}, as found by T. Nishino \cite{NO96CTMRG0, nishino1996numerical, nishino2001two,nishino2000self, nishino1998density, okunishi2000kramers, nishino1997numerical}. Here we start their history from the Wilson numerical renormalization group (NRG) \cite{W75NRGRev} \index{NRG}. The NRG aims at finding the ground state of a spin systems. The idea of the NRG is to start from a small system whose Hamiltonian  can be easily diagonalized. The system is then projected on few low-energy states of the Hamiltonian. A new system is then constructed by  adding several spins and a new low-energy effective Hamiltonian is obtained working only in the subspace spanned by the low-energy states of the previous step and the full Hilbert space of the new spins. In this way the low-energy effective Hamiltonian can be diagonalized again and its low energy states can be used to construct a new restricted Hilbert space. The procedure is then iterated. The original NRG has been improved for example by combining it with the expansion theory \cite{K90DMRGexp,XG92DMRGexpan,XG93DMRG}. As already shown in \cite{W75NRGRev} the NRG successfully tackles the Kondo problem in one dimension \cite{K64Kondo1D}, however, its accuracy is limited when applied to generic  strongly-correlated systems such as Heisenberg chains.

In the nineties, White and Noack  were able to related the poor NRG\index{NRG} accuracy with the fact that it fails to consider properly the boundary conditions \cite{WN92RG}. In 1992, White proposed the famous density matrix renormalization group (DMRG)\index{DMRG} that is as of today the most efficient and accurate algorithms for one-dimensional (1D) models \cite{W92DMRG,W93DMRG}. White used the largest eigenvectors of the reduced denisty matrix of a block as the states describing the relevant part of the low energy physics Hilbert space. The reduced density matrix is obtained by explicitly constructing the ground state of the system on a larger region.  In other words, the space of one block is renormalized by taking the rest of the system as an \textit{environment}. 

The simple idea of environment had revolutionary consequences in the RG-based algorithms. Important generalizations of DMRG\index{DMRG} were then developed, including the finite-temperature variants of matrix renormalization group \cite{BXG96TMRG, MC96FTDMRG, WX97TMRG, S97TMRG}, dynamic DMRG algorithms \cite{H95dDMRG, RPKSB97dDMRG, KW99dDMRG, J02dDMRG}, and corner transfer matrix renormalization group by Nishino and Okunishi \cite{NO96CTMRG0} \footnote{We recommend a web page built by Tomotoshi Nishino, \url{http://quattro.phys.sci.kobe-u.ac.jp/dmrg.html}, where one exhaustively can find the progresses related to DMRG.}.

About ten years later, TN\index{TN} was re-introduced in in simplest form of matrix-product states (MPS) \cite{FNW92TTN, FNW92MPS, DE93MPS000, DEHP93MPS0, klumper1991equivalence} in the the context of the theory of entanglement in quantum many-body systems; see, e.g., \cite{ON02DMRGent, V03EntState, VPC04DMRGQinfo, V04Qsimu} \footnote{For the general theory of entanglement and its role in the physics of quantum-many body systems, see for instance \cite{BD00EntRev, NC00EntBook, AFOV08EntRev, HHHH09EntRev}.}. In this context, the MPS encodes the coefficients of the wave functions in a product of matrices, and are thus defined as the contraction of a one dimensional TN. Each elementary tensor has three indexes: one physical index acting on the physical Hilbert space of the constituent, and two auxiliary indexes that will be contracted. The MPS\index{MPS} structure is chosen since it represents the states whose entanglement only scales with the boundary of a region rather than its volume, something called the ``area law'' of entanglement. Furthermore, an MPS gives only finite correlations, thus is well suited to represent the ground states of the gapped short-range Hamiltonians. The relation between these two facts was evinced in seminal contributions  \cite{H04CorFunDecExp-1, H04CorFunDecExp-2, H07EntArea, H15TNthesis, B73EntAreaLaw, S93Sarea, LRV04GSEnt, CC04EntAreaLaw, PEDC05EntAreaLaw, ECP10AreaLawRev} and led Verstraete and Cirac to prove that  MPS can provide  faithful representations of  the ground states of  1D gapped local Hamiltoinan  \cite{VC06MPSFaithfully}. 

These results  together with the previous works that identified the outcome of converged DMRG\index{DMRG} simulations with an MPS\index{MPS} description of the ground states \cite{OR95MPS} allowed to better understand the impressive performances of DMRG in terms of the correct scaling of entanglement of its underlying TN ansatz. The connection between DMRG and MPS stands in the fact that the projector onto the effective Hilbert space built along the DMRG iterations can be seen as an MPS. Thus, the MPS in DMRG can be understood as not only a 1D state ansatz, but also a TN representation of the RG flows (\cite{FNW92MPS,OR95MPS,RO97MPS,DMNS98MPS,M07DMRGsymme,S11DMRGRev}, as recently reviewed in  \cite{PVWC07MPSRev}). 

These results from the quantum information community fueled the search for better algorithms allowing to optimize variationally the MPS tensors in order to target specific states  \cite{VMC08MPSPEPSRev}. In this broader scenario, DMRG can be seen as an alternating-least-square optimization method. Alternative methods include the imaginary-time evolution from an initial state encoded as in an MPS base of the time-evolving block decimation (TEBD)\index{TEBD} \cite{V03TEBD,V04TEBD,V07iTEBD,OV08canonical}, and time-dependent variational principle of MPS \cite{HCOPVV11TDVP}. Note that these two schemes can be generalized to simulate also the short out-of-equilibrium evolution of a slightly entangled state. MPS have been used beyond ground states, for example in the context of finite-temperature and low-energy excitations based on MPS or its transfer matrix \cite{OR95MPS, BSZ03exciteMPS, CW09EntPurt, PHV12exciteMPS, HPWC+12MPSexcitations, ZVHMV18exciteMPS}.

MPS have further been used to characterize state violating the area law of entanglement, such as ground states of critical systems, and ground states of Hamiltonian with long-range interactions \cite{LRV04GSEnt,HLW94CFTRG, VLRK03CritEnt, TOIL08EntScaling, PMTM09EntScaling, PMJ2010entanglement, SHMTV15ScalingcMPS, RPLLS17Scaling2D, hauke2013spread, koffel2012entanglement}.

The relevance of MPS goes far beyond their use as a numerical ansatz. There have been numerous analytical studies that have led to MPS exact solutions such as the Affleck-Kennedy-Lieb-Tasaki (AKLT)\index{AKLT} state \cite{AKLT87AKLTState,AKLT88AKLTState}, as well as its higher-spin / higher-dimensional generalizations \cite{FNW92TTN,VPC04DMRGQinfo,NKZ97HLS3-2PT,NKZ00S2SLGS,VMC04EntLenDiv,KM08MPSVBS}. MPS has been crucial in understanding the classification of topological phases in 1D \cite{PT12SPT}. Here we will not talk about these important results, but we will focus on numerical applications even though the theory of MPS is still in full development and constantly new fields emerge such as the application of MPS to 1D quantum field theories \cite{VC10cMPS}.

\section{Tensor network states in two dimensions}

The simulations of two-dimensional (2D) systems, where analytical solutions are extremely rare and mean-field approximations often fail to capture the long-range fluctuations, are much more complicated and tricky. For numeric simulations, exact diagonalization can only access small systems; quantum Monte Carlo (QMC)\index{QMC} approaches are hindered by the notorious ``negative sign'' problem on frustrated spin models and fermionic models away from half-filling, causing an exponential increase of the computing time with the number of particles \cite{WSSLGS89SignQMC,TW05SignQMC}.

While very elegant and extremely powerful for 1D models, the 2D version of DMRG \cite{W96DMRG2D, WS98tjDMRG,XLS01DMRG2D,SW12DMRG2DRev}\index{DMRG} suffers several severe restrictions. The ground state obtained by DMRG is an MPS\index{MPS} that is essentially a 1D state representation, satisfying the 1D area law of entanglement entropy \cite{H07EntArea, H15TNthesis, S93Sarea, SWVC08MPSent}. However, due to the lack of alternative approaches, 2D DMRG is still one of the most important 2D algorithms, producing a large number of astonishing works including discovering the numeric evidence of quantum spin liquid \cite{M00QSLrev,B10QSLRev,SB17QSLRev} on kagom\'e lattice (see, e.g., \cite{JWS08DMRGKagome,YHW12DMRGkagome,JWB12TopoEntKagome,DMS12DMRGKagome,NSH12kagomemag,HZOP17kagomeDMRG}).

Besides directly using DMRG in 2D, another natural way is to extend the MPS representation, leading to the tensor product state \cite{NHOMAG01TPS0}, or projected entangled pair state (PEPS)\index{PEPS} \cite{VC04PEPS, VC06PEPSArxiv}. While an MPS is made up of tensors aligned in a 1D chain, a PEPS is formed by tensors located in a 2D lattice, forming a 2D TN. Thus, PEPS can be regarded as one type of 2D tensor network states (TNS)\index{TNS}. Note the work of Affleck \textit{et al} \cite{AKLT88PEPSorigin} can be considered as a prototype of PEPS.

The network structure of the PEPS allows us to construct 2D states that strictly fulfill the area law of entanglement entropy \cite{VWPC06PEPSfamous}. It indicates that PEPS can efficiently represents 2D gapped states, and even the critical and topological states, with only finite bond dimensions. Examples include resonating valence bond states \cite{VWPC06PEPSfamous,PSPC12PEPSRVB,SPCP12PEPSRVB,WPGWV13PEPSRVB,PCSC14PEPSRVB} originally proposed by Anderson \textit{et al} for super-conductivity \cite{A73RVB,A74RVB,A87RVB,BZA87RVBSC,ABZH87RVBSC}, string-net states \cite{GLSW09StringTPS,BAV09StringTPS,CZGCW10StringTPS} proposed by Wen \textit{et al} for gapped topological orders \cite{W89TopoOrder,W90TopoOrder,WN90TopoOrder,W95TopoOrder,LW05TopoOrder,LW05TopoOrderRev,W05TopoOrderRev}, and so on.

The network structure makes PEPS so powerful that it can encode difficult computational problems including non-deterministic polynomial (NP)\index{NP} hard ones \cite{VWPC06PEPSfamous,SWVC07PEPSAreaLaw,GL11TNNPhardarxiv}. What is even more important for physics is that PEPS provides an efficient representation as a variational ansatz for calculating ground states of 2D models. However, obeying the area law costs something else: the computational complexity rises \cite{VWPC06PEPSfamous,SWVC07PEPSAreaLaw,HWS13PEPSAreaLaw}. For instance, after having determined the ground state (either by construction or variation), one usually wants to extract the physical information by computing, e.g., energies, order parameters, or entanglement. For an MPS, most of the tasks are matrix manipulations and products which can be easily done by classical computers. For PEPS, one needs to contract a TN \index{TN} stretching in a 2D plain, unfortunately, most of which cannot be neither done exactly or nor even efficiently. The reason for this complexity is what brings the physical advantage to PEPS: the network structure. Thus, algorithms to compute the TN contractions need to be developed.

Other than dealing with the PEPS, TN provides a general way to different problems where the \textit{cost functions} is written as the contraction of a TN. A cost function is usually a scalar function, whose maximal or minimal point gives the solution of the targeted optimization problem. For example, the cost function of the ground-state simulation can be the energy (e.g., \cite{SV09TNQMC,VHCV16GrdPEPS}); for finite-temperature simulations, it can be the partition function or free energy (e.g., \cite{CCD12FTPEPS, RXLS13NCD}); for the dimension reduction problems, it can be the truncation error or the distance before and after the reduction (e.g., \cite{V04TEBD,LN07TRG,RLXZS12ODTNS}); for the supervised machine learning problems, it can be the accuracy (e.g., \cite{SS16TNML}). TN can then be generally considered as a specific mathematical structure of the parameters in the cost functions.

Before reaching the TN\index{TN} algorithms, there are a few more things worth mentioning. MPS\index{MPS} and PEPS\index{PEPS} are not the only TN representations in one or higher dimensions. As a generalization of PEPS, projected entangled simplex state was proposed, where certain redundancy from the local entanglement is integrated to reach a better efficiency \cite{XCYKNX14PESS,LXCLX+17kagome}. Except for a chain or 2D lattice, TN can be defined with some other geometries, such as trees or fractals. Tree TNS\index{TNS} is one example with non-trivial properties and applications \cite{FNW92TTN, NKZ97HLS3-2PT, F97TreeDMRG, LCP00TreeDMRG, MRS02DendrimersDMRG, SDV06TTN, NFGSS08TreeMPS, TEV09TTN, MVLN10TTN,LDX12TTN, NC13TTN, PVK13TTN, GSRF+14TTN, MVSNL15TTN}. Another example is multi-scale entanglement renormalization ansatz (MERA)\index{MERA} proposed by Vidal \cite{V07EntRenor,V08MERA,CDR08MERA,EV09EntRenor,AV08EntRenor,EV09EntRenorAlgor,CV09EntRenor,EV10EntRenorBoson,EV10EntRenorFermi}, which is a powerful tool especially for studying critical systems \cite{PEV09MERAcrit, MRGF09MERAcrit, ECV10EntRenorOpera, SGCSF10EntRenorCrit, EV13EntRenorCrit, BOBD15EntRenorCrit} and AdS/CFT theories (\cite{EV11AdS/CFT, S12AdS/CFT, B13AdS/CFT, Q13AdS/CFTArxiv,MNSTW15AdS/CFT,BCCCHP+15AdS/CFT,CLMS15AdS/CFTArxiv}, see \cite{N15AdS/CFTBook} for a general introduction of CFT\index{CFT}). TN has also been applied to compute exotic properties of the physical models on fractal lattices \cite{GGN16fractal,WRLZZS16Fractal}.

The second thing concerns the fact that some TN's can indeed be contracted exactly. Tree TN is one example, since there is no loop of a tree graph. This might be the reason that a tree TNS\index{TNS} can only have a finite correlation length \cite{NFGSS08TreeMPS}, thus cannot efficiently access criticality in two dimensions. MERA modifies the tree in a brilliant way, so that the criticality can be accessed without giving up the exactly contractible structure \cite{EV09EntRenorAlgor}. Some other exactly contractible examples have also been found, where exact contractibility is not due to the geometry, but due to some algebraic properties of the local tensors \cite{KRV09exactTRG, DBJC12TopoTNS}.

Thirdly, TN can represent operators, usually dubbed as TN operators. Generally speaking, a TN state can be considered as a linear mapping from the physical Hilbert space to a scalar given by the contraction of tensors. A TN operator is regarded as a mapping from the \textit{bra} to the \textit{ket} Hilbert space. Many algorithms explicitly employ the TN operator form, including the matrix product operator (MPO)\index{MPO} for representing 1D many-body operators and mixed states, and for simulating 1D systems in and out of equilibrium \cite{VGC04MPDO, ZV04MPO, PMCV10MPO, LRGZXY+11LTRG, BCL14MPOopen, MFS15MPOopen, CCB15MPOsteady, BKTMWH17FT1D, GIK17FT1D, HV17MPO, CPSV17MPO}, tensor product operator (also called projected entangled pair operators) in for higher-systems \cite{CCD12FTPEPS, RXLS13NCD, RLXZS12ODTNS, FND10PEPO, O12PEPO, CD15TPO, CD15FTPEPS, CDO16TPO, CRD16TPO,DSCBZ17FT2D, CDO17TPOQMC, KREO18PEPOthermal, CDC19TPO}, and multiscale entangled renormalization ansatz \cite{MIH13MERA, MNRT14cMERAft, GSW16FTmera}.

\section{Tensor renormalization group and tensor network algorithms}

Since most of TN's cannot be contracted exactly (with \#P-complete computational complexity \cite{GL11TNNPhardarxiv}), efficient algorithms are strongly desired. In 2007, Levin and Nave generalized the NRG\index{NRG} idea to TN and proposed tensor renormalization group (TRG)\index{TRG} approach \cite{LN07TRG}. TRG consists of two main steps in each RG iteration: contraction and truncation. In the contraction step, the TN is deformed by singular value decomposition (SVD)\index{SVD} of matrix in such a way that certain adjacent tensors can be contracted without changing the geometry of the TN graph. This procedure reduces the number of tensors $N$ to $N/\nu$, with $\nu$ an integer that depends on the way of contracting. After reaching the fixed point, one tensor represents in fact the contraction of infinite number of original tensors, which can be seen as the approximation of the whole TN.

After each contraction, the dimensions of local tensors increase exponentially, and then truncations are needed. To truncate in an optimized way, one should consider the ``environment'', a concept which appears in DMRG\index{DMRG} and is crucially important in TRG-based schemes to determine how optimal the truncations are. In the truncation step of Levin's TRG, one only keeps the basis corresponding to the $\chi$-largest singular values from the SVD\index{SVD} in the contraction step, with $\chi$ called dimension cut-off. In other words, the environment of the truncation here is the tensor that is decomposed by SVD. Such a local environment only permits local optimizations of the truncations, which hinders the accuracy of Levin's TRG on the systems with long-range fluctuations. Nevertheless, TRG is still one of the most important and computationally-cheap approaches for both classical (e.g., Ising and Potts models) and quantum (e.g., Heisenberg models) simulations in two and higher dimensions \cite{KRV09exactTRG, JWX08SimpleUpdate,GLW08TERG,GW09TERG, CY09TRGapp, ZXCWCX10TNR, HL10MultiEnt, LGZS10honeycomb, GH10TRG, GHB10TRG, WKS11TRG, CQCWN+11PottsTRG, XCQZYX12HOSRG, S12TRGboson, GL13TRG3D, QCXCY+14PottsTRG, WXCBX14TRG3D, RH15TRGdimer, ZXXI16finiteTRG}. It is worth mentioning that for 3D classical models, the accuracy of the TRG algorithms have surpassed other methods \cite{XCQZYX12HOSRG,WXCBX14TRG3D}, such as QMC\index{QMC}. Following the contraction-and-truncation idea, the further developments of the TN contraction algorithms concern mainly two aspects: more reasonable ways of contracting, and more optimized ways of truncating.

While Levin's TRG ``coarse-grains'' a TN in an exponential way (the number of tensors decreases exponentially with the renormalization steps), Vidal's TEBD\index{TEBD} scheme \cite{V03TEBD,V04TEBD,V07iTEBD,OV08canonical} implements the TN contraction with the help of MPS in a linearized way \cite{LRGZXY+11LTRG}. Then, instead of using the singular values of local tensors, one uses the entanglement of the MPS to find the optimal truncation, meaning the environment is a (non-local) MPS\index{MPS}, leading to a better precision than Levin's TRG. In this case, the MPS at the fixed point is the dominant eigenstate of the transfer matrix of the TN. Another group of TRG algorithms, called corner transfer matrix renormalization group (CTMRG)\index{CTMRG} \cite{OV09CTMRG}, are based on the corner transfer matrix idea originally proposed by Baxter in 1978 \cite{B78CTM}, and developed by Nishina and Okunishi in 1996 \cite{NO96CTMRG0}. In CTMRG, the contraction reduces the number of tensors in a polynomial way and the environment can be considered as a finite MPS defined on the boundary. CTMRG has a compatible accuracy compared with TEBD\index{TEBD}.

With a certain way of contracting, there is still high flexibility of choosing the environment, i.e. the reference to optimize the truncations. For example, Levin's TRG\index{TRG} and its variants \cite{LN07TRG, JWX08SimpleUpdate, GLW08TERG, GW09TERG,ZXCWCX10TNR, XCQZYX12HOSRG}, the truncations are optimized by local environments. The second renormalization group proposed by Xie \textit{et al} \cite{XCQZYX12HOSRG, XJCWX09SRG} employs TRG to consider the whole TN\index{TN} as the environments.

Besides the contractions of TN's, the concept of environment becomes more important for the TNS update algorithms, where the central task is to optimize the tensors for minimizing the cost function. According to the environment, the TNS\index{TNS} update algorithms are categorized as the simple \cite{RXLS13NCD, RLXZS12ODTNS, JWX08SimpleUpdate, XCQZYX12HOSRG, LCB14PEPScontract, JO18graphPEPS}, cluster \cite{ RXLS13NCD, LCB14PEPScontract, WV11PEPSclusterArxiv, RPPSL17AOP3D}, and full update \cite{XCQZYX12HOSRG, OV09CTMRG, XJCWX09SRG, JOVVC08PEPS, PWV11tPEPS, O12CTMRG, LCB14fPEPS, PBTCO15FastFullUpdate, C16vPEPS}. The simple update uses local environment, hence has the highest efficiency but limited accuracy. The full update considers the whole TN as the environment, thus has a high accuracy. Though with a better treatment of the environment, one drawback of the full update schemes is the expensive computational cost, which strongly limits the dimensions of the tensors one can keep. The cluster update is a compromise between simple and full update, where one considers a reasonable subsystem as the environment for a balance between the efficiency and precision.

It is worth mentioning that TN\index{TN} encoding schemes are found to bear close relations to the techniques in multi-linear algebra (MLA)\index{MLA} (also know as tensor decompositions or tensor algebra; see a review \cite{KB09MLA}). MLA was originally targeted on developing high-order generalization of the linear algebra (e.g., the higher-order version of singular value or eigenvalue decomposition \cite{LMV00Rank1,DDV00HOSVD,LV04MLA,L06MLA}), and now has been successfully used in a large number of fields, including data mining (e.g., \cite{ACKY05MLADM,LZYCLB+05MLADM,SZLLC05MLADM,ACY06MLADM,SPY06MLADM}), image processing (e.g., \cite{KBK05MLAIP,KB06MLAIP,BHK07MLAIP,DZZHT17MLAimage}), machine learning (e.g., \cite{SDLFHP+17TDML}), and so on. The interesting connections between the fields of TN and MLA (for example tensor-train decomposition \cite{O11TTD} and matrix product state representation) open new paradigm for the interdisciplinary researches that cover a huge range in sciences.

\section{Organization of lecture notes}

Our lectures are organized as following. In Chapter 2, we will introduce the basic concepts and definitions of tensor and TN states/operators, as well as their graphic representations. Several frequently used architectures of TN states will be introduced, including matrix product state, tree TN state, and PEPS\index{PEPS}. Then the general form of TN, the gauge degrees of freedom, and the relations to quantum entanglement will be discussed. Three special types of TN's that can be exactly contracted will be exemplified in the end of this chapter.

In Chapter 3, the contraction algorithms for 2D TN's will be reviewed. We will start with several physical problems that can be transformed to the 2D TN contractions, including the statistics of classical models, observation of TN states, and the ground-state/finite-temperature simulations of 1D quantum models. Three paradigm algorithms namely TRG\index{TRG}, TEBD\index{TEBD}, and CTMRG\index{CTMRG}, will be presented. These algorithms will be further discussed from the aspect of the exactly contractible TN's.

In Chapter 4, we will concentrate on the algorithms of PEPS\index{PEPS} for simulating the ground states of 2D quantum lattice models. Two general schemes will be explained, which are the variational approaches and the imaginary-time evolution. According to the choice of environment for updating the tensors, we will explain the simple, cluster, and full update algorithms. Particularly in the full update, the contraction algorithms of 2D TN's presented in Chapter 3 will play a key role to compute the non-local environments.

In Chapter 5, a special topic about the underlying relations between the TN methods and the MLA\index{MLA} will be given. We will start from the canonicalization of MPS\index{MPS} in one dimension, and then generalize to the super-orthogonalization of PEPS in higher dimensions. The super-orthogonalization that gives the optimal approximation of a tree PEPS\index{PEPS} in fact extends the Tucker decomposition from single tensor to tree TN\index{TN}. Then the relation between the contraction of tree TN's and the rank-1 decomposition will be discussed, which further leads to the ``zero-loop'' approximation of the PEPS on the regular lattice. Finally, we will revisit the infinite DMRG (iDMRG)\index{iDMRG}, infinite TEBD (iTEBD)\index{iTEBD}, and infinite CTMRG\index{CTMRG} in a unified picture indicated by the tensor ring decomposition, which is a higher-rank extension of the rank-1 decomposition.

In Chapter 6, we will revisit the TN simulations of quantum lattice models from the ideas explained in Sec. 5. Such a perspective, dubbed as quantum entanglement simulation (QES)\index{QES}, shows an unified picture for simulating one- and higher-dimensional quantum models at both zero \cite{RPPSL17AOP3D, R16AOP} and finite \cite{RXPS+18FTQES} temperatures. The QES implies an efficient way of investigating infinite-size many-body systems by simulating few-body models with classical computers or artificial quantum platforms. In Sec. 7, a brief summary is given.

As TN makes a fundamental language and efficient tool to a huge range of subjects, which has been advancing in an extremely fast speed, we cannot cover all the related progresses in this review. We will concentrate on the algorithms for TN contractions and the closely related applications. The topics that are not discussed or are only briefly mentioned in this review include: the hybridization of TN with other methods such as density functional theories and \textit{ab-initio} calculations in quantum chemistry \cite{WM99DMRGabini, MFOLP01DMRGDFT, MOMR08DMRGQchem, MR10DMRGQchem, CS11DMRGQchem, W14DMRGQchemy, SA15MPSchem, KVLE16fTNS, RLJC17DMRGinitio, YSLM18DMRGChem}, the dynamic mean-field theory \cite{GJMNN03DMRGDMFT, NGJ04DMRGDMFT, GHR04DMFTwithDMRG, K06DMRGDMFT, WMPS14MPSDMFT, GTVHE14MPSimpu, WMIS14MPSDMFT, WGMM+15MPSDMFT, GAET+15MPSDMFT, BZTAE17TNDMFT}, and the expansion/perturbation theories \cite{GTVHE14MPSimpu, HWHSV11ChMPS, WJMS15Chebyshev, HKM15ChMPS, CLCL17TNexpand, TRFML17DMRGPT, VMSVV19TNpert}; the TN algorithms that are less related to the contraction problems such as time-dependent variational principle \cite{HCOPVV11TDVP, HLOVV16TDVPuni}, the variational TN state methods \cite{HPWC+12MPSexcitations, C16vPEPS, MHOV13vMPS, HOV13MPStan, VMVH15PEPStan, ZVFVH18vuMPS, ZMV18uMPS, VHV19uMPS}, and so on; the TN methods for interacting fermions \cite{YWTSC15PEPSchiral, EV10EntRenorFermi, KVLE16fTNS, BPE09fTN, COBV10fPEPS, CJV10fPEPS, PV10fPEPS, KSVC10fPEPS, MBRTV10fTNS, CWVT11fPEPS, MR11TNSchem, G13GTN, CD14fPEPS, SK14GrassTN, ZBWC15fPEPS, WBE17fTN, BWHV17fPEPS}, quantum field theories \cite{TCL14TNgaugeF, RPDZM14GaugeF, HVSC15TNgauge, XA15TNgauge, PDRZM16TNgauge, BMHVV17TNgauge, ZO17TNgauge}, topological states and exotic phenomena in many-body systems (e.g., \cite{JWS08DMRGKagome, YHW12DMRGkagome, DMS12DMRGKagome, HZOP17kagomeDMRG, PSPC12PEPSRVB, SPCP12PEPSRVB, WPGWV13PEPSRVB, PCSC14PEPSRVB,  GLSW09StringTPS,BAV09StringTPS,PCBA+10AnyonTN, KB10AnyonER, WTSC13PEPSchiral,  DR15TNchiral, YWTSC15PEPSchiral, PCS15PEPSchiral, MOP16TNsymme, HW16SPTTN, GRSDM17QHETNarxiv, LXCLX+17kagomeQSL, PRLCS17octakagome, LRLYZS14Husimi, YLPVV+14boundary, WHTCS14cPEPS, RLGWDS15StarPEPS, WBMH15MPSSPT, JR17TNanyon}), the open/dissipative systems \cite{VGC04MPDO, BCL14MPOopen, MFS15MPOopen, CCB15MPOsteady, GIK17FT1D, PZ09MPSsteady, SC16MPSopen, WJSKC+16PTNS, KWO17MPSopen, JMC18MPSopen}, quantum information and quantum computation \cite{VPC04DMRGQinfo, J06TNcuicirt, GE07QCPEPS, AL10QCTN, GESP07MPSQC, MS08TNQcomp, GMF08TNchannel, FM12QCTN, JBCJ13TNccuit, FP14TNcorrect, DESB+18TNQsimu}, machine learning \cite{SS16TNML, B13MERAML, BPT15MPSML, NPOV15TTNN, LRWP+17MLTN, CCXWX17TNML, HM17TNML, HWFWZ17MPSML, LYCS17TNML, GO17TNML, GJLP18MPOML, CLOPZ+17TNML1rev, CLOPZ+17TNML2rev, GJLP18MPOML, GPC18gTNML, S18MERAML, CBKW18SVMTT, CWXZ19generateTTNML}, and other classical computational problems \cite{EHHS11TNMLA, C14TNbigdata, BMT15TNSAT, BSU16TNMLA, YKCMR18TNCC, KCMR18TNCC}; the TN theories/algorithms with nontrivial statistics and symmetries \cite{GLSW09StringTPS,BAV09StringTPS,CZGCW10StringTPS,ZBWC15fPEPS, HVSC15TNgauge, PCBA+10AnyonTN, MOP16TNsymme, PSGWC10symme, W12Qspace, SCP10PEPSsymme, SPV10TNsymme, SPV11symmeU1, O11advTN, BCOT11assyme, SV12SU2symme, TCL14TNsymme, O14TNadvRev, RDS15PEPSZ2, JR15TNsymme, LH16TNsymme, ZB16PEPSgauge}; several latest improvements of the TN algorithms for higher efficiency and accuracy \cite{PWV11tPEPS, PBTCO15FastFullUpdate, BHVC09folding, MCB12folding, HM15folding, YGW17LoopTN, XLHX+19TNcon}.

\chapter{Tensor Network: Basic Definitions and Properties}

\abstract{This chapter is to introduce some basic definitions and concepts of TN\index{TN}. We will show that the TN can be used to represent quantum many-body states, where we explain MPS\index{MPS} in 1D and PEPS\index{PEPS} in 2D systems, as well as the generalizations to thermal states and operators. The quantum entanglement properties of the TN states including the area law of entanglement entropy will also be discussed. Finally, we will present several special TN's that can be exactly contracted, and demonstrate the difficulty of contracting TN's in general cases.}

\section{Scalar, vector, matrix, and tensor}

\begin{figure}[tbp]
	\centering
	\includegraphics[angle=0,width=0.75\linewidth]{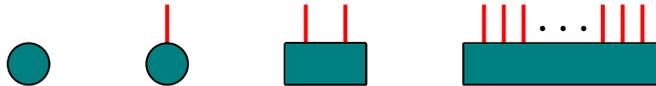}
	\caption{(Color online) From left to right, the graphic representations of a scalar, vector, matrix and tensor.}
	\label{fig-1Tensor}
\end{figure}

Generally speaking, a tensor is defined as a series of numbers labeled by $N$ indexes, with $N$ called the \textit{order} of the tensor \footnote{Note that in some references, $N$ is called the tensor \textit{rank}. Here, the word \textit{rank} is used in another meaning, which will be explained later.}. In this context, a scalar, which is one number and labeled by zero index, is a 0th-order tensor. Many physical quantities are scalars, including energy, free energy, magnetization and so on. Graphically, we use a dot to represent a scalar (Fig. \ref{fig-1Tensor}).

A $D$-component vector consists of $D$ numbers labeled by one index, and thus is a 1st-order tensor. For example, one can write the state vector of a spin-$1/2$ in a chosen basis (say the eigenstates of the spin operator $\hat{S}^{[z]}$) as
\begin{eqnarray}
|\psi\rangle  =C_1|0 \rangle +C_2|1 \rangle
=\sum_{s=0,1}C_s|s\rangle,
\end{eqnarray}
with the coefficients $C$ a two-component vector. Here, we use $|0\rangle$ and $|1\rangle$ to represent spin up and down states. Graphically, we use a dot with one open bond to represent a vector (Fig.\ref{fig-1Tensor}).

A matrix is in fact a 2nd-order tensor. Considering two spins as an example, the state vector can be written under a irreducible representation as a four-dimension vector. Instead, under the local basis of each spin, we write it as
\begin{eqnarray}
|\psi\rangle =C_{00}|0\rangle|0\rangle +C_{01}|0\rangle|1\rangle
+C_{10}|1\rangle|0\rangle+C_{11}|1\rangle|1\rangle =\sum_{ss'=0}^{1}C_{ss'}|s\rangle|s'\rangle,
\end{eqnarray}
with $C_{ss'}$ a matrix with two indexes. Here, one can see that the difference between a ($D \times D$) matrix and a $D^2$-component vector in our context is just the way of labeling the tensor elements. Transferring among vector, matrix and tensor like this will be frequently used later. Graphically, we use a dot with two bonds to represent a matrix and its two indexes (Fig.\ref{fig-1Tensor}).

It is then natural to define an $N$-th order tensor. Considering, e.g., $N$ spins, the $2^N$ coefficients can be written as a $N$-th order tensor $C$ \footnote{If there is no confuse, we use the symbol without all its indexes to refer to a tensor for conciseness, e.g., use $C$ to represent $C_{s_1...s_N}$.}, satisfying
\begin{eqnarray}
|\psi\rangle =\sum_{s_1 \cdots s_N=0}^{1}C_{s_1...s_N}|s_1\rangle...|s_N\rangle.
\end{eqnarray}
Similarly, such a tensor can be \textit{reshaped} into a $2^N$-component vector. Graphically, an $N$-th order tensor is represented by a dot connected with $N$ open bonds (Fig.\ref{fig-1Tensor}).

In above, we use states of spin-$1/2$ as examples, where each index can take two values. For a spin-$S$ state, each index can take $d=2S+1$ values, with $d$ called the \textit{bond dimension}. Besides quantum states, operators can also be written as tensors. A spin-$1/2$ operator $\hat{S}^{\alpha}$ ($\alpha=x,y,z$) is a ($2\times 2$) matrix by fixing the basis, where we have $ S^{\alpha}_{s_1's_2's_1s_2} =  \langle s_1's_2' |\hat{S}^{\alpha} |s_1s_2\rangle$. In the same way, an $N$-spin operator can be written as a $2N$-th order tensor, with $N$ \textit{bra} and $N$ \textit{ket} indexes \footnote{Note that here, we do not distinguish \textit{bra} and \textit{ket} indexes deliberately in a tensor, if not necessary}.

We would like to stress some conventions about the ``indexes'' of a tensor (including matrix) and those of an operator. A tensor is just a group of numbers, where their indexes are defined as the labels labeling the elements. Here, we always put all indexes as the lower symbols, and the upper ``indexes'' of a tensor (if exist) are just a part of the symbol to distinguish different tensors. For an operator which is defined in a Hilbert space, it is represented by a hatted letter, and there will be no ``true'' indexes, meaning that both upper and lower ``indexes'' are just parts of the symbol to distinguish different operators.

\section{Tensor network and tensor network states}

\subsection{A simple example of two spins and Schmidt decomposition}
\label{sec.2SVD}
\begin{figure}[tbp]
	\centering
	\includegraphics[angle=0,width=0.4\linewidth]{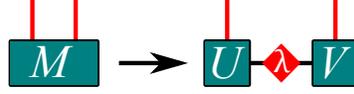}
	\caption{(Color online) The graphic representation of the Schmidt decomposition (singular value decomposition of a matrix). The positive-defined diagonal matrix $\lambda$, which gives the entanglement spectrum (Schmidt numbers), is defined on a virtual bond (dumb index) generated by the decomposition.}
	\label{fig-1SVD}
\end{figure}

After introducing tensor (and it diagram representation), now we are going to talk about TN, which is defined as the contraction of many tensors. Let us start with the simplest situation, two spins, and consider to study the quantum entanglement properties for instance. Quantum entanglement, mostly simplified as entanglement, can be defined by the \textit{Schmidt decomposition} \cite{S1907SD,EK95SD,P95SD} of the state (Fig. \ref{fig-1SVD}) as
\begin{eqnarray}
| \psi \rangle = \sum_{ss'=0}^{1} C_{ss'} |s \rangle |s' \rangle = \sum_{ss'=0}^{1} \sum_{a=1}^{\chi} U_{sa} \lambda_{a a'} V^{\ast}_{a s'} |s \rangle |s' \rangle,
\label{eq-1Schmidt}
\end{eqnarray}
where $U$ and $V$ are unitary matrices, $\lambda$ is a positive-defined diagonal matrix in descending order \footnote{Sometime, $\lambda$ is treated directly as a $\chi$-component vector.}, and $\chi$ is called the \textit{Schmidt rank}. $\lambda$ is called the \textit{Schmidt coefficients} since in the new basis after the decomposition, the state is written in a summation of $\chi$ product states as $| \psi \rangle = \sum_{a} \lambda_a | u \rangle_{a}  | v \rangle_{a}$, with the new basis $| u \rangle_{a} = \sum_{s} U_{s a}  |s \rangle$ and $| v \rangle_{a} = \sum_{s'} V^{\ast}_{s a}  |s' \rangle$.

Graphically, we have a small TN\index{TN}, where we use green squares to represent the unitary matrices $U$ and $V$, and a red diamond to represent the diagonal matrix $\lambda$. There are two bonds in the graph shared by two objects, standing for the summations (contractions) of the two indexes in Eq. (\ref{eq-1Schmidt}), $a$ and $a'$. Unlike $s$ (or $s'$), The space of the index $a$ (or $a'$) is not from any physical Hilbert space. To distinguish these two kinds, we call the indexes like $s$ the \textit{physical indexes} and those like $a$ the \textit{geometrical} or \textit{virtual indexes}. Meanwhile, since each physical index is only connected to one tensor, it is also called an \textit{open bond}.

Some simple observations can be made from the Schmidt decomposition. Generally speaking, the index $a$ (also $a'$ since $\lambda$ is diagonal) contracted in a TN carry the quantum entanglement \cite{NC10EntQIS}. In quantum information sciences, entanglement is regarded as a quantum version of correlation \cite{NC10EntQIS}, which is crucially important to understand the physical implications of TN. One usually uses the \textit{entanglement entropy} to measure the strength of the entanglement, which is defined as $S = - 2 \sum_{a=1}^{\chi} \lambda_{a}^2 \ln \lambda_{a}$. Since the state should be normalized, we have $\sum_{a=1}^{\chi}\lambda_{a}^2 = 1$. For $\dim(a) = 1$, obviously $| \psi \rangle = \lambda_1 | u \rangle_{1}  | v \rangle_{1}$ is a product state with zero entanglement $S=0$ between the two spins. For $\dim(a) = \chi$, the entanglement entropy $S \leq \ln \chi$, where $S$ takes its maximum if and only if $\lambda_1 = \cdots = \lambda_{\chi}$. In other words, the dimension of a geometrical index determines the upper bound of the entanglement.

Instead of Schmidt decomposition, it is more convenient to use another language to present later the algorithms: \textit{singular value decomposition} (SVD)\index{SVD}, a matrix decomposition in linear algebra. The Schmidt decomposition of a state is the SVD of the coefficient matrix $C$, where $\lambda$ is called the \textit{singular value spectrum} and its dimension $\chi$ is called the \textit{rank} of the matrix. In linear algebra, SVD gives the optimal lower-rank approximations of a matrix, which is more useful to the TN algorithms. Specifically speaking, with a given matrix $C$ of rank-$\chi$, the task is to find a rank-$\tilde{\chi}$ matrix $C'$ ($\tilde{\chi} \leq \chi$) that minimizes the norm
\begin{eqnarray}
\mathcal{D} = |M-M'| = \sqrt{\sum_{ss'} (M_{ss'} - M'_{ss'})^2}.
\end{eqnarray}
The optimal solution is given by the SVD as
\begin{eqnarray}
M'_{ss'} = \sum_{a=0}^{\chi'-1} U_{sa} \lambda_{a a} V^{\ast}_{s' a}.
\end{eqnarray}
In other words, $M'$ is the optimal rank-$\chi'$ approximation of $M$, and the error is given by
\begin{eqnarray}
\varepsilon = \sqrt{\sum_{a=\chi'}^{\chi-1} \lambda_{a}^2},
\label{eq-2trunerr}
\end{eqnarray}
which will be called the \textit{truncation error} in the TN\index{TN} algorithms.

\subsection{Matrix product state}

\begin{figure}[tbp]
	\centering
	\includegraphics[angle=0,width=0.8\linewidth]{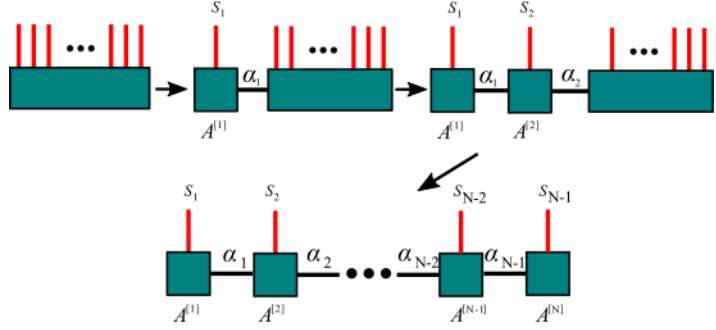}
	\caption{(Color online) An impractical way to obtain an MPS from a many-body wave-function is to repetitively use the SVD\index{SVD}.}
	\label{fig-1MPSSVD}
\end{figure}

\begin{figure}[tbp]
	\centering
	\includegraphics[angle=0,width=0.9\linewidth]{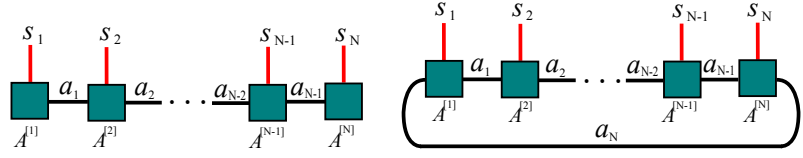}
	\caption{(Color online) The graphic representations of the matrix product states with open (left) and periodic (right) boundary conditions.}
	\label{fig-1MPSfinite}
\end{figure}

Now we take a $N$-spin state as an example to explain the MPS\index{MPS}, a simple but powerful 1D TN state. In an MPS, the coefficients are written as a TN given by the contraction of $N$ tensors. Schollw\"ock in his review \cite{S11DMRGRev} provides a straightforward way to obtain such a TN\index{TN} is by repetitively using SVD\index{SVD} or QR\index{QR} decomposition (Fig. \ref{fig-1MPSSVD}). First, we group the first $N-1$ indexes together as one large index, and write the coefficients as a $2^{N-1} \times 2$ matrix. Then implement SVD or any other decomposition (for example QR decomposition) as the contraction of $C^{[N-1]}$ and $A{[N]}$
\begin{eqnarray}
C_{s_1 \cdots s_{N-1}s_N} = \sum_{a_{N-1}} C^{[N-1]}_{s_1 \cdots s_{N-1},a_{N-1}} A^{[N]}_{s_{N}, a_{N-1} }.
\end{eqnarray}
Note that as a convention in this paper, we always put the physical indexes in front of geometrical indexes and use a comma to separate them. For the tensor $C^{[N-1]}$, one can do the similar thing by grouping the first $N-2$ indexes and decompose again as
\begin{eqnarray}
C_{s_1 \cdots s_{N-1} a_{N-1}} = \sum_{a_{N-2}} C^{[N-2]}_{s_1 \cdots s_{N-2},a_{N-2}} A^{[N-1]}_{s_{N-1}, a_{N-2}  a_{N-1}}.
\end{eqnarray}
Then the total coefficients becomes the contraction of three tensors as
\begin{eqnarray}
C_{s_1 \cdots s_{N-1}s_N} = \sum_{a_{N-2} a_{N-1}} C^{[N-2]}_{s_1 \cdots s_{N-2},a_{N-2}} A^{[N-1]}_{s_{N-1}, a_{N-2} a_{N-1}} A^{[N]}_{s_{N}, a_{N-1}}.
\end{eqnarray}
Repeat decomposing in the above way until each tensor only contains one physical index, we have the MPS\index{MPS} representation of the state as
\begin{eqnarray}
C_{s_1 \cdots s_{N-1}s_N} = \sum_{a_{1} \cdots a_{N-1}} A^{[1]}_{s_1, a_{1}} A^{[2]}_{s_2, a_{1} a_{2}} \cdots  A^{[N-1]}_{s_{N-1}, a_{N-2} a_{N-1}} A^{[N]}_{s_{N}, a_{N-1}}.
\label{eq-1MPS}
\end{eqnarray}
One can see that an MPS is a TN\index{TN} formed by the contraction of $N$ tensors. Graphically, MPS is represented by a 1D graph with $N$ open bonds. In fact, an MPS given by Eq. (\ref{eq-1MPS}) has open boundary condition, and can be generalized to periodic boundary condition (Fig. \ref{fig-1MPSfinite}) as
\begin{eqnarray}
C_{s_1 \cdots s_{N-1}s_N} = \sum_{a_{1} \cdots a_{N}} A^{[1]}_{s_1, a_{N} a_{1}} A^{[2]}_{s_2, a_{1} a_{2}} \cdots  A^{[N-1]}_{s_{N-1}, a_{N-2} a_{N-1}} A^{[N]}_{s_{N}, a_{N-1} a_{N}},
\label{eq-1MPS_PBC}
\end{eqnarray}
where all tensors are 3rd-order. Moreover, one can introduce translational invariance to the MPS, i.e. $A^{[n]} = A$ for $n=1,2,\cdots,N$. We use $\chi$, dubbed as \textit{bond dimension} of the MPS, to represent the dimension of each geometrical index.

MPS is an efficient representation of a many-body quantum state. For a $N$-spin state, the number of the coefficients is $2^N$ which increases exponentially with $N$. For an MPS given by Eq. (\ref{eq-1MPS_PBC}), it is easy to count that the total number of the elements of all tensors is $Nd\chi^2$ which increases only linearly with $N$. The above way of obtaining MPS with decompositions is also known as tensor train decomposition (TTD)\index{TTD} in MLA\index{MLA}, and MPS is also called tensor-train form \cite{O11TTD}. The main aim of TTD is investigating the algorithms to obtain the optimal tensor train form of a given tensor, so that the number of parameter can be reduced with well-controlled errors.

In physics, the above procedure shows that any states can be written in an MPS\index{MPS}, as long as we do not limit the dimensions of the geometrical indexes. However, it is extremely impractical and inefficient, since in principle, the dimensions of the geometrical indexes $\{a\}$ increase exponentially with $N$. In the following sections, we will directly applying the mathematic form of the MPS without considering the above procedure. 

Now we introduce a simplified notation of MPS that has been widely used in the community of physics. In fact with fixed physical indexes, the contractions of geometrical indexes are just the inner products of matrices (this is how its name comes from). In this sense, we write a quantum state given by Eq. (\ref{eq-1MPS}) as
\begin{eqnarray}
|\psi \rangle =tTr A^{[1]} A^{[2]} \cdots A^{[N]} |s_1s_2 \cdots s_N \rangle = tTr \prod_{n=1}^{N} A^{[n]} |s_n \rangle.
\label{eq-1MPSsimple}
\end{eqnarray}
$tTr$ stands for summing over all shared indexes. The advantage of Eq. (\ref{eq-1MPSsimple}) is to give a general formula for an MPS of either finite or infinite size, with either periodic or open boundary condition.

\begin{figure}[tbp]
	\centering
	\includegraphics[angle=0,width=0.45\linewidth]{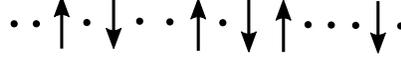}
	\caption{(Color online) One possible configuration of the sparse anti-ferromagnetic ordered state. A dot represents the $S=0$ state. Without looking at all the $S=0$ states, the spins are arranged in the anti-ferromagnetic way.}
	\label{fig-1SparseAF}
\end{figure}

\subsection{Afflect-Kennedy-Lieb-Tasaki state}

MPS\index{MPS} is not just a mathematic form. It can represent nontrivial physical states. One important example can be found with AKLT\index{AKLT} model proposed in 1987, a generalization of spin-$1$ Heisenberg model \cite{AKLT87AKLTState}. For 1D systems, Mermin-Wagner theorem forbids any spontaneously breaking of continuous symmetries at finite temperature with sufficiently short-range interactions. For the ground state of AKLT model called AKLT state, it possesses the \textit{sparse anti-ferromagnetic order} (Fig. \ref{fig-1SparseAF}), which provides a non-zero excitation gap under the framework of Mermin-Wagner theorem. Moreover, AKTL state provides us a precious exactly-solvable example to understand edge states and (symmetry-protected) topological orders.

AKLT state can be exactly written in an MPS with $\chi=2$ (see \cite{PVWC07MPSRev} for example). Without losing generality, we assume periodic boundary condition. Let us begin with the AKLT Hamiltonian that can be given by spin-1 operators as
\begin{eqnarray}
\hat{H}=\sum_n\left[\frac{1}{2} \hat{S}_n\cdot \hat{S}_{n+1}+\frac{1}{6} (\hat{S}_n\cdot \hat{S}_{n+1})^2+\frac{1}{3}\right].
\label{eq-1AKLTH}
\end{eqnarray}
By introducing the non-negative-defined projector $\hat{P}_2(\hat{S}_n+\hat{S}_{n+1})$ that projects the neighboring spins to the subspace of $S=2$, Eq. (\ref{eq-1AKLTH}) can be rewritten in the summation of projectors as
\begin{eqnarray}
\hat{H}=\sum_n \hat{P}_2(\hat{S}_n+\hat{S}_{n+1}).
\end{eqnarray}
Thus, the AKLT Hamiltonian is non-negative-defined, and its ground state lies in its kernel space, satisfying $\hat{H}|\psi_{AKLT}\rangle = 0$ with a zero energy.

\begin{figure}[tbp]
	\centering
	\includegraphics[angle=0,width=0.75\linewidth]{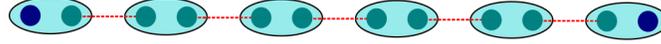}
	\caption{(Color online) An intuitive graphic representation of the AKLT\index{AKLT} state. The big circles representing $S=1$ spins, and the small ones are effective $S=\frac{1}{2}$ spins. Each pair of spin-$1/2$ connecting by a red bond forms a singlet state. The two  ``free'' spin-$1/2$ on the boundary give the edge state.}
	\label{fig-1AKLT}
\end{figure}

Now we construct a wave-function which has a zero energy. As shown in Fig. \ref{fig-1AKLT}, we put on each site a projector that maps two (effective) spins-$1/2$ to a \textit{triplet}, i.e. the physical spin-1, where the transformation of the basis obeys
\begin{eqnarray}
|+\rangle&=&|00\rangle \\
|\tilde{0}\rangle&=&\frac{1}{\sqrt{2}}(|01\rangle+|10\rangle),\\
|-\rangle&=&|11\rangle.
\end{eqnarray}
The corresponding projector is determined by the Clebsch-Gordan coefficients \cite{S63CG}, and is a ($3 \times 4$) matrix. Here, we rewrite it as a ($3 \times 2 \times 2$) tensor, whose three components (regarding to the first index) are the ascending, $z$-component and descending Pauli matrices of spin-$1/2$ \footnote{Here, one has some degrees of freedom to choose different projectors, which is only up to a gauge transformation. But once one projector is fixed, the other is also fixed.},
\begin{eqnarray}
\sigma^+ = \left[ {\begin{array}{*{30}c}
	0 \hspace{0.4cm} 1  \\
	0 \hspace{0.4cm} 0  \\
	\end{array}} \right],  \hspace{0.6cm}
\sigma^{z} = \left[ {\begin{array}{*{30}c}
	1 \hspace{0.4cm} 0 \\
	0 \hspace{0.08cm} -1  \\
	\end{array}} \right],  \hspace{0.6cm}
\sigma^- = \left[ {\begin{array}{*{30}c}
	0 \hspace{0.4cm} 0  \\
	1 \hspace{0.4cm} 0  \\
	\end{array}} \right].  \hspace{0.6cm}
\label{eq-1sigma}
\end{eqnarray}
In the language of MPS\index{MPS}, we have the tensor $A$ satisfying
\begin{eqnarray}
A_{0,a a'} = \sigma^+_{aa'}, \hspace{0.4cm} A_{1,a a'} = \sigma^z_{aa'}, \hspace{0.4cm} A_{2,a a'} = \sigma^-_{aa'}.
\label{eq-1AKLT-A}
\end{eqnarray}

Then we put another projector to map two spin-$1/2$ to a singlet, i.e. a spin-0 with
\begin{eqnarray}
|\bar{0}\rangle = \frac{1}{\sqrt{2}}(|01\rangle-|10\rangle)
\end{eqnarray}
The projector is in fact a ($2 \times 2$) identity with the choice of Eq. (\ref{eq-1sigma}),
\begin{eqnarray}
I = \left[ {\begin{array}{*{30}c}
	1 \hspace{0.4cm} 0  \\
	0 \hspace{0.4cm} 1  \\
	\end{array}} \right].
\label{eq-1Identity22}
\end{eqnarray}
Now, the MPS of the AKLT state with periodic boundary condition (up to a normalization factor) is obtained by Eq. (\ref{eq-1MPS_PBC}), with every tensor $A$ given by Eq. (\ref{eq-1AKLT-A}). For such an MPS, every projector operator $\hat{P}_2(\hat{S}_n+\hat{S}_{n+1})$ in the AKLT Hamiltonian is always acted on a singlet, then we have $\hat{H}|\psi_{AKLT}\rangle = 0$.

\subsection{Tree tensor network state (TTNS)\index{TTNS} and projected entangled pair state (PEPS)\index{PEPS}}
\label{sec.peps}

\begin{figure}[tbp]
	\centering
	\includegraphics[angle=0,width=\linewidth]{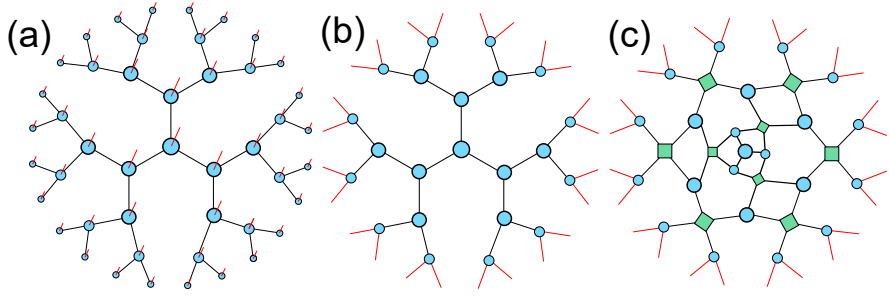}
	\caption{(Color online) The illustration of (a) and (b) two different TTNS's and (c) MERA\index{MERA}.}
	\label{fig-1TTNS}
\end{figure}

TTNS is a generalization of the MPS\index{MPS} that can code more general entanglement states. Unlike an MPS where the tensors are aligned in a 1D array, a TTNS is given by a tree graph. Figs. \ref{fig-1TTNS} (a) and (b) show two examples of TTNS with the coordination number $z=3$. The red bonds are the physical indexes and the black bonds are the geometrical indexes connecting two adjacent tensors. The physical ones may locate on each tensor or put on the boundary of the tree. A tree is a graph that has no loops, which leads to many simple mathematical properties that parallel to those of an MPS. For example, the partition function of a TTNS can be efficiently exactly computed. A similar but more power TN state called MERA\index{MERA} also has such a property [Figs. \ref{fig-1TTNS} (c)]. We will get back to this in Sec. \ref{Sec-exactTN}. Note an MPS can be treated as a tree with $z = 2$.



\begin{figure}[tbp]
	\centering
	\includegraphics[angle=0,width=0.8\linewidth]{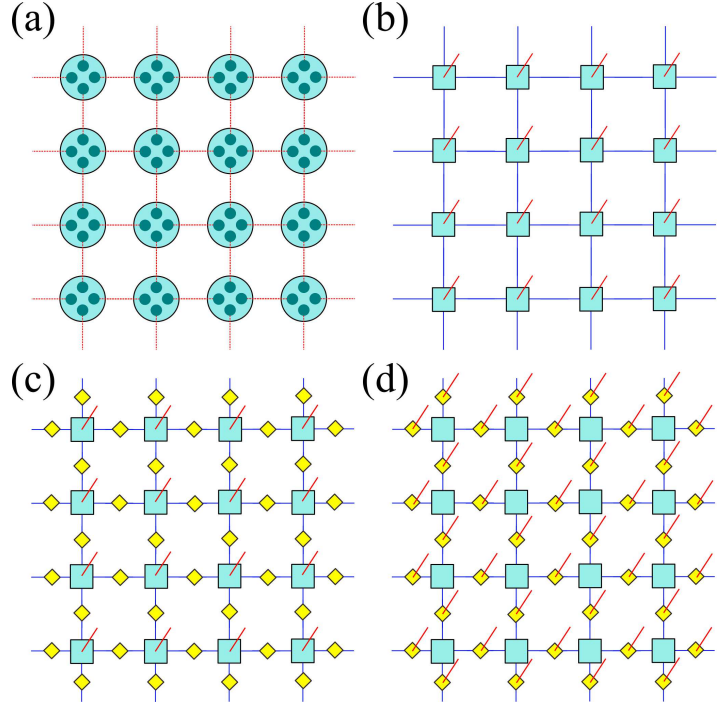}
	\caption{(Color online) (a) An intuitive picture of the projected entangled pair state. The physical spins (big circles) are projected to the virtual ones (small circles), which form the maximally entangled states (red bonds). (b)-(d) Three kinds of frequently used PEPS's.}
	\label{fig-1PEPSgraph}
\end{figure}

An important generalization to the TN's of loopy structures is known as projected entangled pair state (PEPS), proposed by Verstraete and Cirac \cite{VC04PEPS, VC06PEPSArxiv}. The tensors of a PEPS\index{PEPS} are located in, instead of a 1D chain or a tree graph, a d-dimensional lattice, thus graphically forming a $d$-dimensional TN\index{TN}. An intuitive picture of PEPS is given in Fig. \ref{fig-1PEPSgraph}, i.e., the tensors can be understood as projectors that map the physical spins into virtual ones. The virtual spins form the maximally entangled state in a way determined by the geometry of the TN. Note that such an intuitive picture was firstly proposed with PEPS \cite{VC04PEPS}, but it also applies to TTNS.

Similar to MPS\index{MPS}, a TTNS or PEPS can be formally written as
\begin{eqnarray}
|\Psi\rangle = tTr \prod_{n} P^{[n]} |s_n\rangle,
\label{eq-2PEPSsimple}
\end{eqnarray}
where $tTr$ means to sum over all geometrical indexes. Usually, we do not write the formula of a TTNS or PEPS, but give the graph instead to clearly show the contraction relations.

Such a generalization makes a lot of senses in physics. One key factor regards the \textit{area law of entanglement entropy} \cite{B73EntAreaLaw,S93Sarea,LRV04GSEnt,CC04EntAreaLaw,PEDC05EntAreaLaw,ECP10AreaLawRev} which we will talk about later in this chapter. In the following as two straightforward examples, we show that PEPS can indeed represents non-trivial physical states including nearest-neighbor \textit{resonating valence bond} (RVB)\index{RVB} and \textit{Z$_2$ spin liquid} states. Note these two types of states on trees can be similarly defined by the corresponding TTNS.

\subsection{PEPS can represent non-trivial many-body states: examples}

\begin{figure}[tbp]
	\centering
	\includegraphics[angle=0,width=1\linewidth]{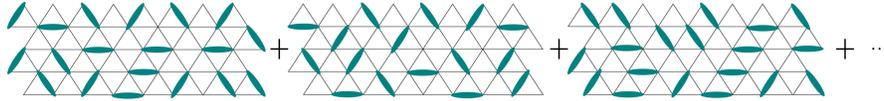}
	\caption{(Color online) The nearest-neighbor RVB\index{RVB} state is the superposition of all possible configurations of nearest-neighbor singlets.}
	\label{fig-1NNRVB}
\end{figure}

RVB state was firstly proposed by Anderson to explain the possible disordered ground state of the Heisenberg model on triangular lattice \cite{A73RVB,A74RVB}. RVB state is defined as the superposition of macroscopic configurations where all spins are paired to form the singlet states (dimers). The strong fluctuations are expected to restore all symmetries and lead to a spin liquid state without any local orders. The distance between two spins in a dimer can be short range or long range. For nearest-neighbor RVB, the dimers are only be nearest neighbors (Fig. \ref{fig-1NNRVB}, also see \cite{B10QSLRev}). RVB state is supposed to relate to high-$T_c$ copper-oxide-based superconductor. By doping the singlet pairs, the insulating RVB state can translate to a charged superconductive state \cite{A87RVB,BZA87RVBSC,ABZH87RVBSC}.


For the nearest-neighbor situation, an RVB state (defined on an infinite square lattice, for example) can be exactly written in a PEPS of $\chi=3$. Without losing generality, we take the translational invariance, i.e., the TN\index{TN} is formed by infinite copies of several inequivalent tensors. Two different ways have been proposed to construct the nearest-neighbor RVB\index{RVB} PEPS\index{PEPS} \cite{VWPC06PEPSfamous, SPCP12PEPSRVB}. In addition, Wang \textit{et al} proposed a way to construct the PEPS with long-range dimers \cite{WPGWV13PEPSRVB}. In the following, we explain the way proposed by Verstraete \textit{et al} to construct the nearest-neighbor one \cite{VWPC06PEPSfamous}. There are two inequivalent tensors: the tensor defined on each site whose dimensions are ($2 \times 3 \times 3 \times 3 \times 3$) only has five nonzero elements,
\begin{eqnarray}
P_{0,0000}=1, \ \ P_{1,2111}=1, \ \ P_{1,1211}=1, \ \ P_{1,1121}=1, \ \ P_{1,1112}=1.
\end{eqnarray}
We use the language of strings to understand physical meaning of the projector: the spin-up state (with the physical index $s=0$) stands for the vacuum state and the spin-down ($s=1$) for the occupied state of a string. In this sense, the first element $P_{0,0000}$ means it is vacuum in the physical space, thus all the geometrical spaces are vacuum. For the rest four elements, the physical space is occupied by a string that is mapped to one of the geometrical space with the same amplitude, leaving the rest three to be vacuum. For example, $P_{1,1211}=1$ means one possibility, where the physical string is mapped to the second geometrical space while the rest three remain vacuum \footnote{Note that for a geometrical space, $0$ and $1$ are to distinguish the vacuum states with vacuum and occupied physical states, respectively.}. The rest elements are all zero, which means the corresponding configurations are forbidden.

The tensor $P$ only maps physical strings to geometrical spaces. Then a projector $B$ is put on each geometrical bond to form the singlets in the RVB picture. $B$ is a ($3 \times 3$) matrix with only three nonzero elements
\begin{eqnarray}
B_{00}=1, \ \ B_{12}=1, \ \ B_{21}=-1.
\end{eqnarray}
Similarly, the first one means a vacuum state on this geometrical bond, and the rest two simply give a singlet state ($|12\rangle - |21\rangle$).

Then the infinite PEPS (iPEPS)\index{iPEPS} of the nearest-neighbor RVB\index{RVB} is given by the contraction of infinite copies of $P$'s on the sites and $B$'s (Fig.\ref{fig-1PEPSgraph}) on the bonds as
\begin{eqnarray}
|\Psi\rangle = \sum_{\{s,a\}} \prod_{n \in sites} P_{s_n, a_n^1 a_n^2 a_n^3 a_n^4} \prod_{m \in bonds} B_{a^1_m a^2_m }  \prod_{j \in sites} |s_j\rangle.
\label{eq-2RVBPEPS}
\end{eqnarray}

After the contraction of all geometrical indexes, the state is the super-position of all possible configurations consisting of nearest-neighbor dimers. This iPEPS looks different from the one given in Eq. (\ref{eq-2PEPSsimple}) but they are essentially the same, because one can contract the $B$'s into $P$'s so that the PEPS is only formed by tensors defined on the sites.

Another example is the Z$_2$ spin liquid state, which is one of simplest string-net states \cite{GLSW09StringTPS,BAV09StringTPS,CZGCW10StringTPS}, firstly proposed by Levin and Wen to characterize gapped topological orders \cite{LW05TopoOrder}. Similarly with the picture of strings, the Z$_2$ state is the super-position of all configurations of string loops. Writing such a state with TN\index{TN}, the tensor on each vertex is ($2 \times 2 \times 2 \times 2$) satisfying
\begin{eqnarray}
P_{a_1 \cdots a_N}&=&
\left\{
\begin{array}{lll}
1, \ \ a_1+\cdots + a_N=even, \\
0, \ \ otherwise.
\end{array}
\right.
\end{eqnarray}
The tensor $P$ forces the \textit{fusion rules} of the strings: the number of the strings connecting to a vertex must be even, so that there are no loose ends and all strings have to form loops. It is also called in some literatures the \textit{ice rule} \cite{BG01SpinIce,BF1933IceRule} or \textit{Gauss' law} \cite{FDWSPB+09SpinIce}. In addition, the square TN formed solely by the tensor $P$ gives the famous \textit{eight-vertex model}, where the number ``eight'' corresponds to the eight non-zero elements (i.e. allowed sting configurations) on a vertex \cite{B71EightVertex}.

The tensors $B$ are defined on each bond to project the strings to spins, whose non-zero elements are
\begin{eqnarray}
B_{0,00}=1, \ \ B_{1,11}=1.
\end{eqnarray}
The tensor $B$ is a projector that maps the spin-up (spin-down) state to the occupied (vacuum) state of a string.

\subsection{Tensor network operators}
\label{sec-MPO}

The MPS\index{MPS} or PEPS\index{PEPS} can be readily generalized from representations of states to those of operators called MPO\index{MPO} \cite{VGC04MPDO,ZV04MPO, PMCV10MPO, LRGZXY+11LTRG, BKTMWH17FT1D, GIK17FT1D, HV17MPO, CPSV17MPO} or projected entangled pair operator (PEPO)\index{PEPO} \footnote{Generally, a representation of an operator with a TN can be called tensor product operator (TPO)\index{TPO}. MPO and PEPO are two examples.} \cite{CCD12FTPEPS,RXLS13NCD,RLXZS12ODTNS,FND10PEPO,O12PEPO,CD15TPO,CDO16TPO,CRD16TPO,DSCBZ17FT2D,CDO17TPOQMC}. Let us begin with MPO, which is also formed by the contraction of local tensors as
\begin{eqnarray}
\hat{O} = \sum_{\{s,a\}} \prod_n W_{s_n s_n', a_n a_{n+1}}^{[n]}|s_n\rangle \langle s_n'|.
\label{eq-2MPO}
\end{eqnarray}

\begin{figure}
	\begin{center}
		\includegraphics[width=0.6\textwidth]{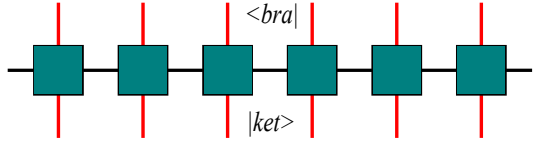}
	\end{center}
	\caption{(Color online) The graphic representation of a matrix product operator, where the upward and downward indexes represent the \textit{bra} and \textit{ket} space, respectively.}
	\label{fig-2MPO}
\end{figure}

\begin{figure}
	\begin{center}
		\includegraphics[width=0.5\textwidth]{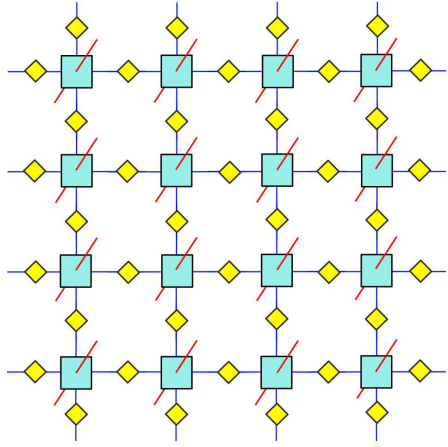}
	\end{center}
	\caption{(Color online) The graphic representation of a projected entangled pair operator, where the upward and downward indexes represent the \textit{bra} and \textit{ket} space, respectively.}
	\label{fig-2TPO}
\end{figure}

Different from MPS\index{MPS}, each tensor has two physical indexes, of which one is a \textit{bra} and the other is a \textit{ket} index (Fig. \ref{fig-2TPO}). An MPO may represent several non-trivial physical models, for example the Hamiltonian. Crosswhite and Bacon \cite{CB08automata} proposed a general way of constructing an MPO from called \textit{automata}. Now we show how to construct the MPO of an Hamiltonian using the properties of a \textit{triangular MPO}. Let us start from a general lower-triangular MPO satisfying $W^{[n]}_{::,00}=C^{[n]}$, $W^{[n]}_{::,01}=B^{[n]}$, and $W^{[n]}_{::,11}=A^{[n]}$ with $A^{[n]}$, $B^{[n]}$, and $C^{[n]}$ some $d \times d$ square matrices. We can write $W^{[n]}$ in a more explicit $2 \times 2$ block-wise form as
\begin{eqnarray}
W^{[n]}=
\begin{pmatrix}
C^{[n]} & 0 \\
B^{[n]} & A^{[n]}
\end{pmatrix}
\end{eqnarray}
If one puts such a $W^{[n]}$ in Eq. (\ref{eq-2MPO}), it will give the summation of all terms in the form of
\begin{eqnarray}
O &=& \sum_{n=1}^N A^{[1]} \otimes \cdots \otimes A^{[n-1]} \otimes B^{[n]} \otimes C^{[n+1]} \otimes \cdots \otimes C^{[N]} \nonumber \\
&=& \sum_{n=1}^N \prod_{\otimes i=1}^{n-1} A^{[i]} \otimes B^{[n]} \otimes \prod_{\otimes j=n+1}^{N} C^{[j]},
\label{eq-2UpTriMPO}
\end{eqnarray}
with $N$ the total number of tensors and $\prod_{\otimes}$ the tensor product \footnote{For $n=0$, $A^{[0]}$ (or $B^{[0]}$, $C^{[0]}$) does not exist but can be defined as a scalar 1, for simplicity of the formula.}. Such a property can be easily generalized to a $W$ formed by $D \times D$ blocks.

Imposing Eq. (\ref{eq-2UpTriMPO}), we can construct as an example the summation of one-site local terms, i.e., $\sum_n X^{[n]}$ \footnote{Note that $X^{[n_1]}$ and $X^{[n_2]}$ are not defined in a same space with $n_1 \neq n_2$, Thus, precisely speaking, $\sum$ here is the direct sum. We will not specify this when it causes no confuse}, with
\begin{eqnarray}
W^{[n]}=
\begin{pmatrix}
I & 0 \\
X^{[n]} & I
\end{pmatrix},
\end{eqnarray}
with $X^{[n]}$ a $d \times d$ matrix and $I$ the $d \times d$ identity.

If two-body terms are included, such as $\sum_m X^{[m]} + \sum_n Y^{[n]} Z^{[n+1]}$, we have
\begin{eqnarray}
W^{[n]}=
\begin{pmatrix}
I & 0 & 0 \\
Z^{[n]} & 0 & 0 \\
X^{[n]} & Y^{[n]} & I
\end{pmatrix}.
\end{eqnarray}
This can be obviously generalized to $L$-body terms. With open boundary conditions, the left and right tensors are
\begin{eqnarray}
W^{[1]}=
\begin{pmatrix}
I & 0 & 0
\end{pmatrix}, \ \ \\
W^{[N]}=
\begin{pmatrix}
0 \\ 0 \\ I
\end{pmatrix}.
\end{eqnarray}

Now we apply the above technique on a Hamiltonian of, e.g., the Ising model in a transverse field
\begin{eqnarray}
\hat{H} = \sum_n \hat{S}^{z}_n \hat{S}^{z}_{n+1} + h \sum_m \hat{S}^{x}_m.
\end{eqnarray}
Its MPO is given by
\begin{eqnarray}
W^{[n]}=
\begin{pmatrix}
I & 0 & 0 \\
\hat{S}^z & 0 & 0 \\
h \hat{S}^x & \hat{S}^z & I
\end{pmatrix}.
\end{eqnarray}



Such a way of constructing an MPO\index{MPO} is very useful. Another example is the Fourier transformation to the number operator of Hubbard model in momentum space $\hat{n}_k = \hat{b}_k^{\dagger} \hat{b}_{k}$. The Fourier transformation is written as
\begin{eqnarray}
\hat{n}_k = \sum_{m,n=1}^{N} e^{i(m-n)k} \hat{b}_m^{\dagger} \hat{b}_{n},
\end{eqnarray}
with $\hat{b}_n$ ($\hat{b}_n^{\dagger}$) the annihilation (creation) operator on the $n$-th site. The MPO representation of such a Fourier transformation is given by
\begin{eqnarray}
\hat{W}_n = \begin{pmatrix}
\hat{I} & 0 & 0 & 0 \\
\hat{b}^{\dagger} & e^{ik} \hat{I} & 0 & 0 \\
\hat{b} & 0 & e^{-ik} \hat{I} & 0 \\
\hat{b}^{\dagger} \hat{b} & e^{+ik} \hat{b}^{\dagger} & e^{-ik} \hat{b} & \hat{I}
\end{pmatrix}
\end{eqnarray}
with $\hat{I}$ the identical operator in the corresponding Hilbert space.

The MPO formulation also allows for a convenient and efficient representation of the Hamiltonians with longer range interactions \cite{CDV08MPOLR}. The geometrical bond dimensions will in principle increase with the interaction length. Surprisingly, a small dimension is needed to approximate the Hamiltonian with long-range interactions that decay polynomially \cite{FND10PEPO}.

Besides, MPO can be used to represent the time evolution operator $\hat{U}(\tau) = e^{-\tau \hat{H}}$ with \textit{Trotter-Suzuki decomposition}, where $\tau$ is a small positive number called \textit{Trotter-Suzuki step} \cite{V04TEBD,V07iTEBD}. Such an MPO is very useful in calculating real, imaginary, or even complex time evolutions, which we will present later in detail. An MPO can also give a mixed state.

Similarly, PEPS\index{PEPS} can also be generalized to projected entangled pair operator (PEPO\index{PEPO}, Fig. \ref{fig-2TPO}), which on a square lattice for instance can be written as
\begin{eqnarray}
\hat{O} = \sum_{\{s,a\}} \prod_n W_{s_n s_n', a_n^1 a_n^2 a_n^3 a_n^4}^{[n]} |s_n\rangle \langle s_n'|.
\label{eq-2PEPO}
\end{eqnarray}
Each tensor has two physical indexes (\textit{bra} and \textit{ket}) and four geometrical indexes. Each geometrical bond is shared by two adjacent tensors and will be contracted.

\subsection{Tensor network for quantum circuits}

\begin{figure}[t]
	\begin{center}
		\includegraphics[scale=0.55]{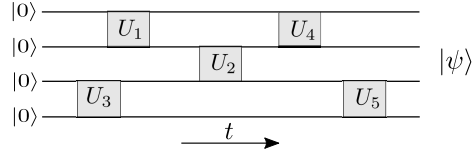}
	\end{center}
	\caption{The TN \index{TN} representation of a quantum circuit. Two-body unitaries act on a product state of a given number of constituents $|0\rangle\otimes\cdots\otimes|0\rangle$ and transform it into a target entangled state $|\psi\rangle$.\label{fig:quantum circuit}}
\end{figure}

A special case of TN\index{TN} are quantum circuits \cite{MS08TNQcomp}.
Quantum circuits encode computations made on qubits (or qudits in
general). Fig. \ref{fig:quantum circuit} demonstrates
the TN representation of a quantum circuit made by unitary gates that act on a product state of many constituents initialized as $\prod_{\otimes} |0\rangle$. 

\textbf{An example of quantum circuits}. In order
to make contact with TN, we will consider the specific
case of quantum circuits where all the gates act on at most two neighbors.
An example of such circuit is the Trotterized evolution of a system
described by a nearest-neighbor Hamiltonian $\hat{H}=\sum_{i,i+1} \hat{h}_{i,i+1}$,
where $i,i+1$ label the neighboring constituents of a one-dimensional
system. The evolution operator for a time $t$ is $\hat{U}(t)=exp(-i\hat{H}t)$, and
can be decomposed into a sequence of infinitesimal time evolution
steps \cite{V03TEBD} (more details will be given in Sec. \ref{sec.2.evolution})
\begin{equation}
\hat{U}(t)=lim_{N\to\text{\ensuremath{\infty}}} \exp(-i\frac{t}{N}\hat{H})^{N}.
\end{equation}
In the limit, we can decompose the evolution into a product of two-body evolution 
\begin{equation}
\hat{U}(t)=lim_{\tau \to 0}\prod_{i,i+1} \hat{U}(\tau)_{i,i+1}.
\end{equation}
where $\hat{U}_{i,i+1}(\tau)=\exp(-i\tau \hat{h}_{i,i+1})$ and $\tau = t/N$. This is obviously
a quantum circuit made by two-qubit gates with depth $N$. Conversely,
any quantum circuit naturally possesses an arrow of time; it transforms a product state into an entangled state after
a sequence of two-body gates. 

\begin{figure}[t]
	\includegraphics[scale=0.37]{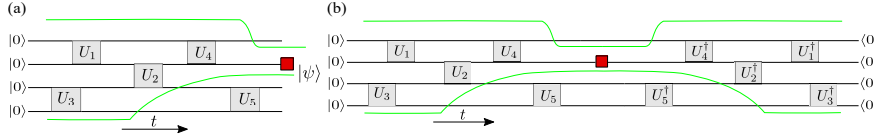}
	\caption{The past casual cone of the red site ($A$). The unitary gate $U_{5}$ does not affect the reduced density matrix of the red site. This is verified by computing explicitly $\rho_{A}$ by tracing over all the others constituents.  In the TN\index{TN} of $\rho_{A}$, $U_{5}$ is contracted with $U_{5}^{\dagger}$, which gives an identity. \label{fig:past-casual}}
\end{figure}

\textbf{Casual cone}. One interesting concept in a quantum circuit is that of the causal
cone illustrated in Fig. \ref{fig:past-casual}, which becomes explicit with the TN\index{TN} representations. Given a quantum circuit
that prepares (i.e., evolves the initial state to) the state $|\psi\rangle$, we can ask a question: which subset of the
gates affect the reduced density matrix of a certain sub-region $A$ of $|\psi\rangle$? This can be seen by constructing the reduced
density matrix of the sub-region $A$ $\rho_{A}=tr_{\bar{A}}|\psi\rangle\langle\psi|$ with $\bar{A}$ the rest part of the system besides $A$.

The TN of the reduced density matrix is formed by a set of unitaries that define the past causal cone of the region $A$ (see the area between the green lines in Fig. \ref{fig:past-casual}). The rest unitaries (for instance the $\hat{U}_5$ and its conjugate in the right sub-figure of Fig. \ref{fig:past-casual}) will be eliminated in the TN of the reduced density matrix. The contraction of
the causal cone can thus be rephrased in terms of the multiplication
of a set of transfer matrices, each performing the computation from
$t$ to $t-1$. The maximal width of these transfer matrices defines the width
of the causal cone, which can be used as a good measure of the complexity
of computing $\rho_{A}$ \cite{jozsarichard2003}.
The best computational strategy one can find to compute exactly $\rho_{A}$
will indeed always scale exponentially with the width of the cone \cite{MS08TNQcomp}.

\begin{figure}[t]
	\includegraphics[scale=0.55]{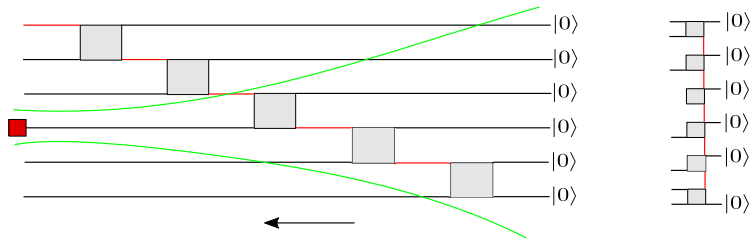}
	\caption{The MPS\index{MPS} as a quantum circuit. Time flows from right to left so that the lowest constituent is the first to interact with the auxiliary $D$-level system. Here we show the past causal cone of a single constituent. Similarly, the past causal cone of $A$ made by adjacent constituent has the same form starting from the upper boundary of $A$.
		\label{fig:MPS-as-quantum-circuit}}
\end{figure}

\begin{figure}[t]
	\begin{center}
		\includegraphics[scale=0.55]{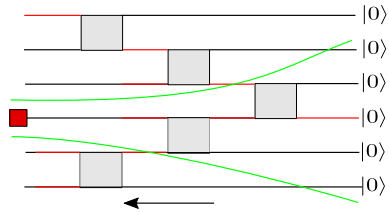}
	\end{center}
	\caption{Using the gauge degrees of freedom of an MPS, we can modify its past causal cone structure to make its region as small as possible, in such a way decreasing the computational complexity of the actual computation of specific $\rho_{A}$.
		A convenient choice is the center gauge used in
		iDMRG\index{iDMRG} \label{fig:center-gauge}}
\end{figure}

\textbf{Unitary tensor networks and quantum circuits.} The simplest TN\index{TN}, the MP\index{MPS} can be interpreted as a sequential quantum
circuit \cite{schon2005}. The idea is that one can think of the MPS
as a sequential interaction between each constituent (a $d$- level
system) an ancillary $D$-level system (the auxiliary qDit, red bonds). The first
constituent interacts (say the bottom one shown in Fig. \ref{fig:MPS-as-quantum-circuit})
and then sequentially all the constituents interact with the same
$D-$level system. With this choice, the past causal cone of a constituent
is made by all the MPS matrices below it. Interestingly in the MPS
case, the causal cone can be changed using the gauge transformations (see sec. \ref{sec.gauge}),
something very different to what happens in two dimensional TN's\index{TN}. This amounts to finding appropriate unitary transformations
acting on the auxiliary degrees of freedom that allow to re-order the interactions
between the $D-$level system and the constituents. In such a way, a desired constituent can be made to interact first, then followed
by the others. An example of the causal cone in the center gauge used
in iDMRG calculation \cite{M08iDMRGarxiv} is presented in Fig. \ref{fig:center-gauge}.
This idea allows to minimize the number of tensors in the
causal cone of a given region. However, the scaling of
the computational cost of the contraction is not affected by such a
temporal reordering of the TN, since in this case the width of
the cone is bounded by one unitary in any gauge. The gauge choice
just changes the number of computational steps required to construct
the desired $\rho_{A}$. In the case that $A$ includes non-consecutive
constituents, the width of the cone increases linearly with the number
of constituents, and the complexity of computing $\rho_{A}$ increases
exponentially with the number of constituents.

Again, the gauge degrees of freedom can be used to modify the structure of the past
causal cone of a certain spin. As an example, the iDMRG center gauge
is represented in Fig. \ref{fig:center-gauge}.

\begin{figure}[t]
	\begin{center}
		\includegraphics[scale=0.55]{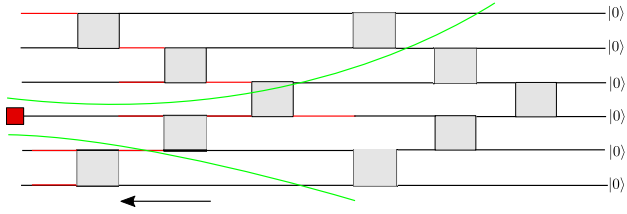}
	\end{center}
	\caption{The width of the causal cone increases as we increase the depth of the quantum circuit generating the MPS state.\label{fig:larger-causal-cone}}
\end{figure}

An example of a TN with a larger past causal cone can be obtained
by using more than one layers of interactions. Now the support of the causal
cone becomes larger since it includes transfer matrices acting on
two $D$-level systems (red bonds shown in Fig. \ref{fig:larger-causal-cone}). Notice that this TN has loops but it still exactly contractible since
the width of the causal cone is still finite.

\section{Tensor networks that can be contracted exactly}

\subsection{Definition of exactly contractible tensor network states}

The notion of the past causal cone can be used to classify TNS's\index{TNS} based on
the complexity of computing their contractions. It is important to remember that
the complexity strongly depends on the object that we want to compute, not just the TN. For example, the complexity of an MPS\index{MPS} for a $N$-qubit state scales only linearly with $N$. However, to compute the $n$-site reduced density matrix, the cost scales exponentially with $n$ since the matrix itself is an exponentially large object. Here we consider to compute scalar quantities, such as the observables of one- and two-site operators.

We define the a TNS\index{TNS} to be \emph{exactly contractible} when it is allowed to compute their contractions with a cost that is a polynomial
to the elementary tensor dimensions $D$. A more rigorous definition can be given in terms of their tree
width see e.g. \cite{MS08TNQcomp}. From the discussion of the previous
section, it is clear that such a TNS corresponds to a bounded causal
cone for the reduced density matrix of a local sub-region. In order to show this, we now focus on the cost of computing
the expectation value of local operators and their correlation functions on a few examples of TNS's\index{TNS}.

\begin{figure}[t]
	\begin{center}
		\includegraphics[scale=0.55]{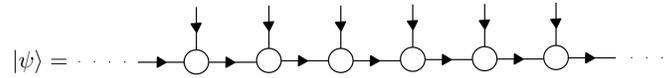}
	\end{center}
	\caption{The MPS wave-function representation in left canonical form.}
	\label{fig:wf_mps}
\end{figure}

The relevant objects are thus the reduced density matrix of a region
$A$ made of a few consecutive spins, and the reduced density matrix
of two disjoint blocks $A_{1}$ and $A_{2}$ of which each made of a few consecutive spins. Once we have the reduced density matrices of such regions, we can compute arbitrary expectation values of local operators by $\langle{\cal {\cal O}\rangle=}tr(\rho_{A}{\cal O})$ and $\langle{\cal {\cal O}}_{A_{1}}{\cal O}'_{A_{2}}\rangle=tr(\rho_{A_{1}\cup A_{2}}{\cal O}_{A_{1}}{\cal O}'_{A_{2}})$
with ${\cal O}_{A}$, ${\cal {\cal O}}_{A_{1}}$, ${\cal O}'_{A_{2}}$ arbitrary operators defined on the regions $A$, $A_{1}$, $A_{2}$.

\subsection{MPS\index{MPS} wave-functions}

\begin{figure}[t]
	\includegraphics[scale=0.55]{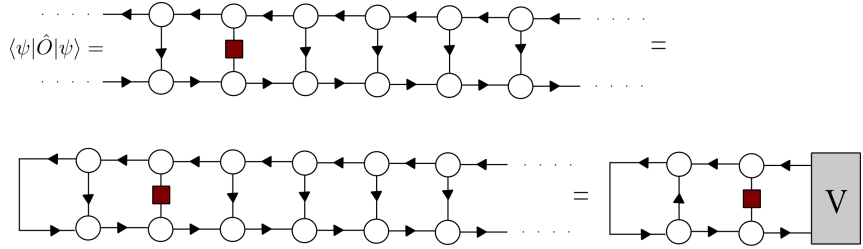}
	\caption{The expectation value of a single-site operator with an MPS\index{MPS} wave-function.
		\label{fig:obs_mps}}
\end{figure}

The simplest example of the computation of the expectation value of
a local operator is obtained by considering MPS wave-functions \cite{FNW92MPS,PVWC07MPSRev}.
Fig. \ref{fig:wf_mps} shows an MPS in the left-canonical form (see \ref{sec5.canon} for more details).
Rather than putting the arrows of time, here we put the direction
in which the tensors in the TN are isometric. In other words, an identity is obtained by contracting the inward bonds of a tensor in $|\psi \rangle$ with the outward bonds of its conjugate in $\langle \psi |$ (Fig. \ref{fig:obs_mps}). Note that $|\psi \rangle$ and $\langle \psi |$ have opposite arrows, by definition. These arrows are directly on the legs of the tensors. The arrows in $| \psi \rangle$  are in the opposite direction than the time, by comparing Fig. \ref{fig:MPS-as-quantum-circuit} with Fig. \ref{fig:obs_mps}. The two figures indeed represent the MPS in the same gauge. This means that the causal cone of an observable is on the right of that observable, as shown on the second line of Fig. \ref{fig:obs_mps}, where all the tensors on the left side are annihilated as a consequence of the isometric constraints. We immediately have that the causal cone
has at most the width of two. The contraction becomes a power of the transfer operator of the MPS $E=\sum_{i}A^{i}\otimes A^{i\dagger}$, where $A_{i}$ and $A^{i\dagger}$ represent the MPS tensors and its complex conjugate. The MPS transfer matrix $E$ only acts on two auxiliary degrees of freedom. Using the property that $E$ is a completely positive map and thus has a fixed point \cite{PVWC07MPSRev}, we can substitute the transfer operator by its largest eigenvector $v$, leading to the final TN\index{TN} diagram that encodes the expectation value of a local operator.

\begin{figure}[t]
	\includegraphics[scale=0.6]{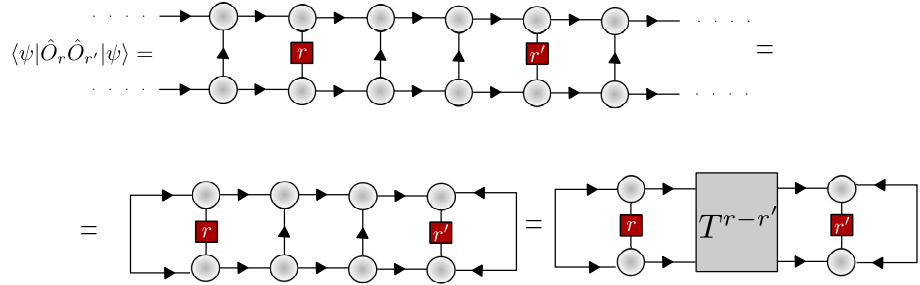}
	\caption{Two-point correlation function of an MPS wave-function. \label{fig:two-point-MPS}}
\end{figure}

In Fig. \ref{fig:two-point-MPS}, we show the TN\index{TN} representation of the expectation value
of the two-point correlation functions. Obviously, the past-causal cone width is bounded
by two auxiliary sites. Note that in the second line, the directions of the arrows on the right side are changed. This in general does not happen in more complicated
TN's as we will see in the next subsection. Before going
there, we would like to comment the properties of the two-point correlation
functions of MPS. From the calculation we have just performed, we see
that they are encoded in powers of the transfer matrix that evolve
the system in the real space. If that matrix can be diagonalized, we can immediately
see that the correlation functions naturally decay exponentially with the ratio of the first to the second eigenvalue. Related details can be found in Sec. \ref{sec5.extract}.

\subsection{Tree tensor network wave-functions}

An alternative kind of wave-functions are the TTNS's\index{TTNS} \cite{SDV06TTN,TEV09TTN,alba2010,alba2011,calabrese2013,ferris2012,gliozzi2009}. In a TTNS, one can add the physical bond on each of the tensor, and use it as a many-body state defined on a Caley-tree lattice \cite{SDV06TTN}. Here, we we will focus on the TTNS with physical bonds only on the outer leafs of the tree.

\begin{figure}[t]
	\begin{center} \label{fig:tree-tensor}
		\includegraphics[scale=0.4]{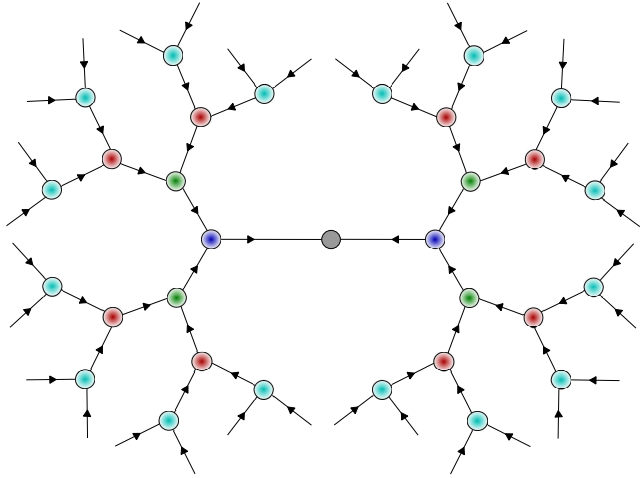}
	\end{center}
	\caption{A binary TTNS made of several layers of third-order tensors. Different layers are identified with different colors. The arrows flow in the opposite direction of the time while being interpreted as a quantum circuit.}
\end{figure}

The calculations with a TTNS normally correspond to the contraction of tree TN's.
A specific case
of a two-to-one TTNS is illustrated in Fig. \ref{fig:tree-tensor}, named binary Caley tree. This TN can be interpreted
as a quantum state of multiple spins with different boundary
conditions. It can also be considered as a hierarchical TN, in which
each layer corresponds to a different level of coarse-graining renormalization group (RG)\index{RG} transformation \cite{TEV09TTN}.
In the figure, different layers are colored differently. In the first layer,
each tensor groups two spins into one and so on. The tree TN can thus be interpreted
a specific RG transformation. Once more, the arrows on the tensors indicate
the isometric property of each individual tensor that; the directions are opposite as the time, if we interpret the tree TN as a quantum circuit. Note again that $|\psi \rangle$ and $\langle \psi |$ have opposite arrows, by definition.

\begin{figure}[t]
	\includegraphics[scale=0.32]{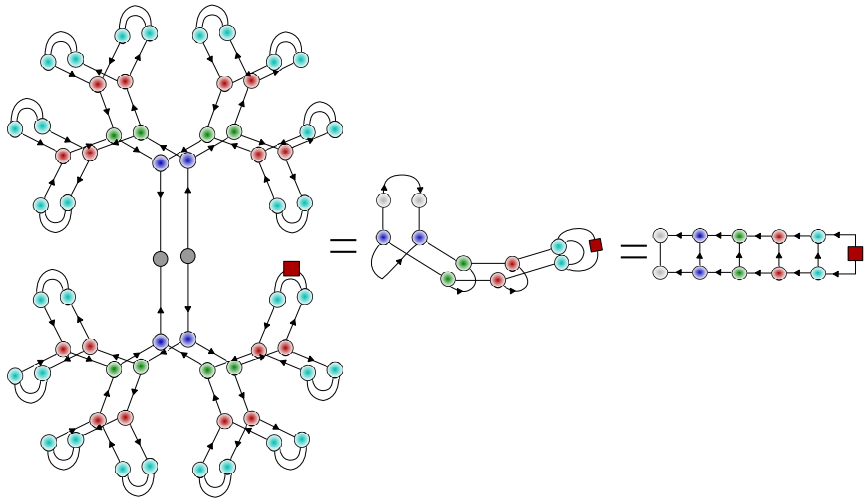}
	\caption{The expectation value of a local operator of a TTNS\index{TTNS}. We see that after applying the isometric properties of the tensors, the past causal cone of a single site has a bounded width. The calculation again boils down to a calculation of transfer matrices. This time the transfer matrices evolve between different layers of the tree.\label{fig:one-site-ttn}}
\end{figure}

The expectation value of a one-site operator is in fact a tree TN\index{TN} shown in Fig. \ref{fig:one-site-ttn}. We see that many of the tensors are completely
contracted with their Hermitian conjugates, which simply give identities. What are left is again a bounded causal cone. If we now build an infinite TTNS\index{TTNS} made by infinitely many layers, and
assume the scale invariance, the multiplication of infinitely many power
of the scale transfer matrix can be substituted with the corresponding
fixed point, leading to a very simple expression for the TN that encodes the expectation value of a single site operator.

\begin{figure}[t]
	\includegraphics[scale=0.26]{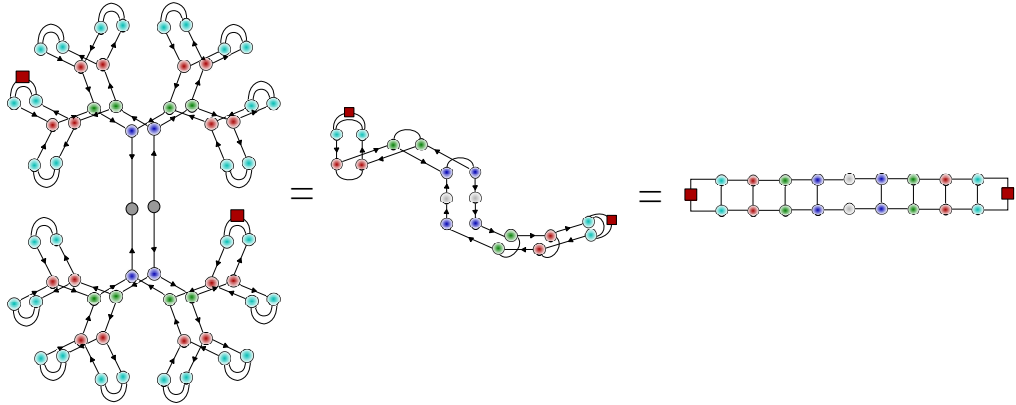}
	\caption{The computation of the correlation function of two operators separated by a given distance boils down  to the computation of a certain power of a transfer matrices. The computation of the casual cone can be simplified in a sequential way, as depicted in the last two sub-figures. \label{fig:two_poin_ttn}}
\end{figure}

Similarly, if we compute the correlation function of local operators
at a given distance, as shown in Fig. \ref{fig:two_poin_ttn}, we
can once more get rid of the tensors outside the casual cone.
Rigorously we see that the causal cone width now increases to four sites,
since it consists of two different two-site branches. However, if
we order the contraction as shown in the middle, we see that the
contractions boil down again to a two-site causal cone. Interestingly,
since the computation of two-point correlations at very large distance
involve the power of transfer matrices that translate in scale rather
than in space, one would expect that these matrices are
all the same (as a consequence of scale-invariance for example). Thus, we
would get polynomially decaying correlations \cite{silvi2010}.

\subsection{MERA\index{MERA} wave-functions}

\begin{figure}
	\begin{center}
		\includegraphics[scale=0.40]{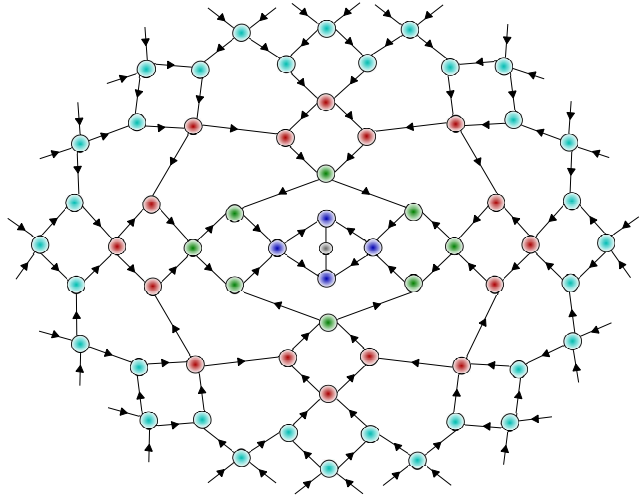}
	\end{center}
	\caption{The TN\index{TN} of MERA\index{MERA}. The MERA has a hierarchical structure consisting of several layers of disentanglers and isometries. The computational time
		flows from the center towards the edge radially, when considering MERA as a quantum circuit. The unitary and isometric tensors and the network
		geometry are chosen in order to guarantee that the width of the causal
		cone is bounded. \label{fig:MERA}}
\end{figure}

\begin{figure}[t]
	\includegraphics[height=5.5cm, width=12cm]{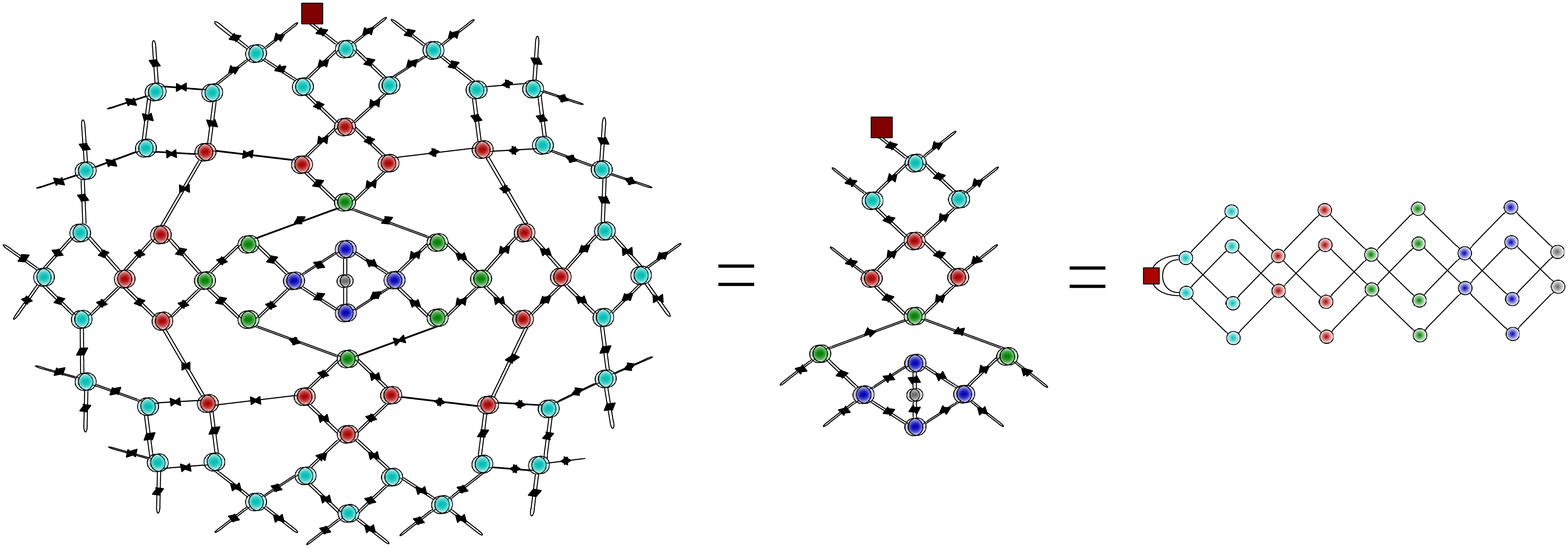}
	\caption{Past causal cone of a single-site operator for a MERA\index{MERA}. \label{fig:one_site_MERA}}
\end{figure}

Until now, we have discussed with the TN's\index{TN} that, even if they
can be embedded in a 2D space, they contain no loops. In the context
of network complexity theory, they are called mean-field networks \cite{bapst2013}.
However, there are also TN's with loops that are exactly contractible \cite{MS08TNQcomp}. A particular case is that
of a 1D MERA (and its generalizations) \cite{V07EntRenor,V08MERA,EV09EntRenor,EV09EntRenorAlgor,tagliacozzo2011}.
The MERA is again a TN that can be embedded in a 2D plane,
and that is full of loops as seen in Fig. \ref{fig:MERA}. This TN
has a very peculiar structure, again, inspired from RG\index{RG} transformation\cite{vidal2009}.
MERA can also be interpreted as a quantum circuit where the time
evolves radially along the network, once more opposite to the arrows
that indicate the direction along which the tensors are unitary. The
MERA is a layered TN, with where layer (in different colors in the figure) is
composed by the appropriate contraction of some third-order tensors
(isometries) and some forth-order tensors (disentangler). The concrete form of the network is not really important \cite{tagliacozzo2011}.
In this specific case we are plotting a two-to-one MERA that was
discussed in the original version of Ref. \cite{EV09EntRenorAlgor}. Interestingly, an
operator defined on at most two sites gives a bounded past-causal cone as shown in Fig. \ref{fig:one_site_MERA} and . 

As in the case of the TTNS\index{TTNS}, we can indeed perform the explicit calculation
of the past causal cone of a single site operators (Fig.\ref{fig:one_site_MERA}). There we show that
the TN\index{TN} contraction of the required expectation
value, and then simplify it by taking into account the contractions of the unitary
and isometric tensors outside the casual cone with a bounded width involving at most four auxiliary constituents.

\begin{figure}[t]
	\begin{center}
		\includegraphics[scale=0.37]{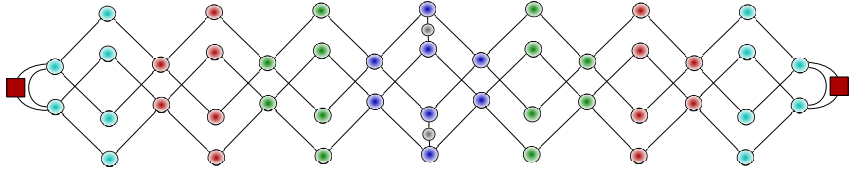}
	\end{center}
	\caption{Two-point correlation function in the MERA\index{MERA}.
		\label{fig:MERA_two_point}}
\end{figure}

The calculation of a two-point correlation function of local operators follows
a similar idea and leads to the contraction shown in Fig.\ref{fig:MERA_two_point}.
Once more, we see that the computation of the two-point correlation function
can be done exactly due to the bounded width of the corresponding casual cone. 

\subsection{Sequentially generated PEPS\index{PEPS} wave-functions}

The MERA\index{MERA} and TTNS\index{TTNS} can be generalized to two dimensional lattices \cite{TEV09TTN, EV09EntRenor}.
The generalization of MPS to 2D, on the other hand, gives rise to PEPS\index{PEPS}.
In general, it belongs to the 2D TN's\index{TN} that cannot be exactly contracted \cite{VWPC06PEPSfamous,nishio2004}. 

\begin{figure}[t]
	\begin{center}
		\includegraphics[scale=0.55]{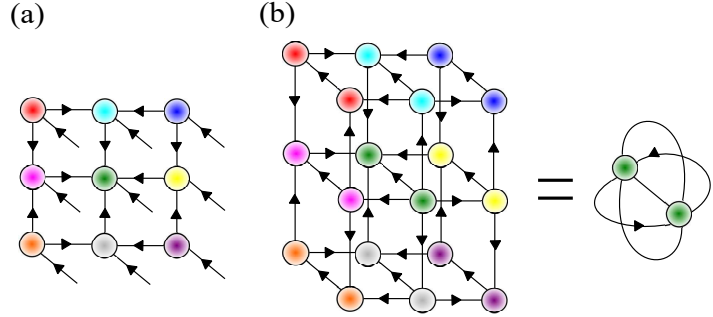}
	\end{center}
	\caption{(a) A sequentially generated PEPS. All tensors but the central
		one (green in the figure) are isometries, from the in-going bonds (marked with ingoing arrows) to the outgoing ones. The central tensor
		represents a normalized vector on the Hilbert space constructed by
		the physical Hilbert space and the four copies of auxiliary spaces,
		one for each of its legs. (b) The norm of such PEPS, after implementing the isometric constraints, boils down to the norm of its central
		tensor. \label{fig:sequentially-generated-peps}}
\end{figure}

\begin{figure}[t]
	\begin{center}
		\includegraphics[scale=0.55]{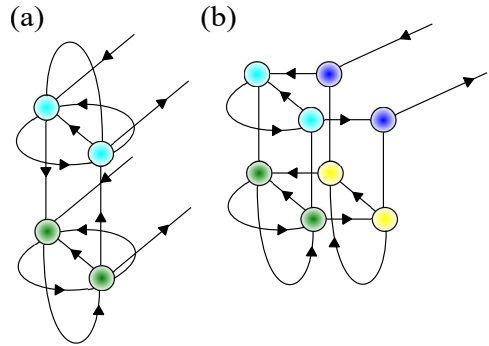}
	\end{center}
	\caption{(a) The reduced density matrices of a PEPS that is sequentially generated
		containing two consecutive spins (one of them is the central
		spin. (b) The reduced density matrix of a local
		region far from the central site is generally hard to
		compute, since it can give rise to an arbitrarily large causal cone.
		For the reduced density
		matrix of any of the corners with a $L\times L$ PEPS, which is the most consuming case, it leads to a causal cone with a width up to $L/2$. That means the
		computation is exponentially expensive with the size of the system.\label{fig:rdm_seq_peps}}
\end{figure}

However for a subclass
of PEPS, one can implement the contract exactly, which is called sequentially generated PEPS \cite{banuls2008}. Differently
from the MERA where the computation of the expectation value of any
sufficiently local operator leads to a bounded causal cone, sequentially
generated PEPS have a central site, and the local observables around the
central site can be computed easily. However, the local observables in other
regions of the TN give larger causal cones. For example, we represent in Fig.\ref{fig:sequentially-generated-peps} a sequentially
generated PEPS for a $3\times3$ lattice. The norm of the state is
computed in (b), where the TN boils down to the norm of
the central tensor. Some of the reduced density matrices of the system
are also easy to compute, in particular those of the central site and
its neighbors [Fig. \ref{fig:rdm_seq_peps} (a)].
Other reduced density matrices, such as those of spins close
to the corners, are much harder to compute. As illustrated in Fig. \ref{fig:rdm_seq_peps} (b),  the causal cone of a corner site in a $3\times\text{3}$ PEPS has a width $2$. In general for an $L\times L$ PEPS, the casual cone would
have a width $L/2$.

Differently from MPS, the causal cone of a PEPS cannot be transformed
by performing a gauge transformation. However, as firstly observed by
F. Cucchietti (private communication), one can try to approximate
a PEPS of a given causal cone with another one of
a different causal cone, by for example moving the center site. This is not an exact operation, and  the approximations
involved in such a transformation need to be addressed numerically.
The systematic study of the effect of these approximations have been
studied recently in \cite{zaletel2019, haghshenas2019}. In general,
we have to say that the contraction of a PEPS wave-function
can only be performed exactly with exponential resources. Therefore, efficient approximate contraction schemes are necessary to deal with PEPS.

\subsection{Exactly contractible tensor networks} \label{Sec-exactTN}

\begin{figure}
	\begin{center}
		\includegraphics[width=\textwidth]{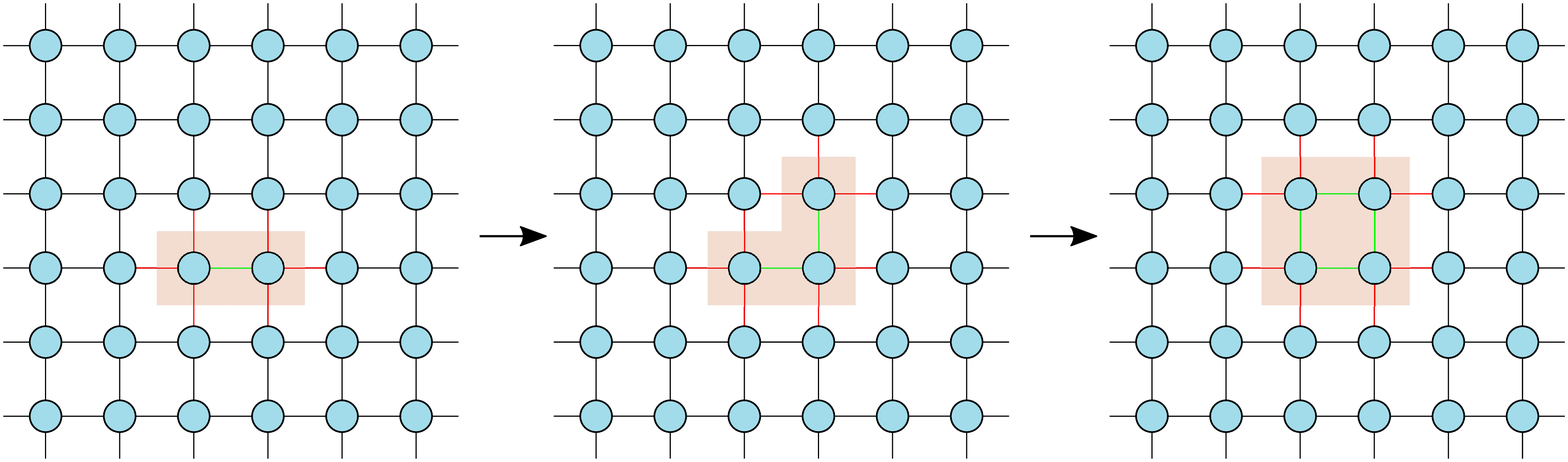}
	\end{center}
	\caption{(Color online)	If one starts with contracting an arbitrary bond, there will be a tensor with six bonds. As the contraction goes on, the number of bonds increases linearly with the boundary $\partial$ of the contracted area, thus the memory increases exponentially as $O(\chi^{\partial})$ with $\chi$ the bond dimension.}
	\label{fig-3TNcontract}
\end{figure}

We have considered above, from the perspective of quantum circuits, whether a TNS\index{TNS} can be contracted exactly by the width of the casual cones. Below, we re-consider this issue from the aspect of TN.

Normally, a TN cannot be contracted without approximation. Let us consider a square TN, as shown in Fig. \ref{fig-3TNcontract}. We start from contracting an arbitrary bond in the TN\index{TN} (yellow shadow). Consequently, we obtain a new tensor with six bonds that contains $\chi^6$ parameters ($\chi$ is the bond dimension). To proceed, the bonds adjacent to this tensor are probably a good choice to contract next. Then we will have to restore a new tensor with eight bonds. As the contraction goes on, the number of bonds increases linearly with the boundary $\partial$ of the contracted area, thus the memory increases exponentially as $O(\chi^{\partial})$. For this reason, it is impossible to exactly contract a TN, even if it only contains a small number of tensors. Thus, approximations are inevitable. This computational difficulty is closely related to the area law of entanglement entropy \cite{ECP10AreaLawRev} (also see Sec. \ref{sec.TNent}), or the width of the casual cone as in the case of PEPS. Below, we give three examples of TN's that can be exactly contracted.

\begin{figure}
	\begin{center}
		\includegraphics[width=0.95\textwidth]{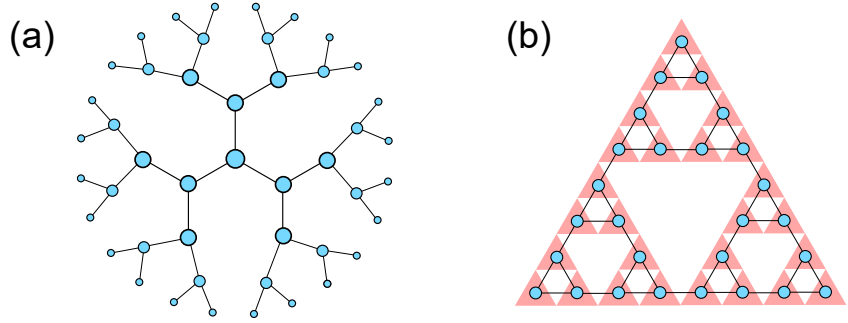}
	\end{center}
	\caption{(Color online)	Two kinds of TN's that can be exactly contracted: (a) tree and (b) fractal TN's. In (b), the shadow shows the Sierpi\'nski gasket, where the tensors are defined in the triangles.}
	\label{fig-3ExactTN}
\end{figure}

\textbf{\textit{Tensor networks on tree graphs}}. We here consider a scalar tree TN [Fig. \ref{fig-3ExactTN} (a)] with $N_L$ layers of third-order tensors. Some vectors are put on the out-most boundary. An example that a tree TN may represent is an observable of a TTNS. A tree TN is written as
\begin{eqnarray}
Z = \sum_{\{a\}} \prod_{n=1}^{N_L} \prod_{m=1}^{M_n} T^{[n,m]}_{a_{n,m,1},a_{n,m,2},a_{n,m,3}} \prod_k v^{[k]}_{a_k},
\end{eqnarray}
with $T^{[n,m]}$ the $m$-th tensor on the $n$-th layer, $M_n$ the number of tensors of the $n$-th layer, and $v^{[k]}$ the $k$-th vectors on the boundary.

Now we contract each of the tensor on the $N_L$-th layer with the corresponding two vectors on the boundary as
\begin{eqnarray}
v_{a_3}' = \sum_{a_1a_2} T^{[N_L m]}_{a_1a_2a_3} v^{[k_1]}_{a_1} v^{[k_2]}_{a_2}.
\end{eqnarray}
After the vectors are updated by the equation above, and the number of layers of the tree TN becomes $N_L-1$. The whole tree TN can be exactly contracted by repeating this procedure.

We can see from the above contraction that if the graph does not contain any loops, i.e. has a tree like structure, the dimensions of the obtained tensors during the contraction will not increase unboundedly. Therefore, the TN defined on it can be exactly contracted. This is again related to the area law of entanglement entropy that a loop-free TN satisfies: to separate a tree-like TN into two disconnecting parts, the number of bonds that needs to be cut is only one. Thus the upper bond of the entanglement entropy between these two parts is constant, determined by the dimension of the bond that is cut. This is also consistent with the analyses based on the maximal width of the casual cones.

\textbf{\textit{Tensor networks on fractals}}. Another example that can be exactly contracted is the TN defined on the fractal called Sierpi\'nski gasket [Fig. \ref{fig-3ExactTN} (b)] (see, e.g. \cite{GGN16fractal,WRLZZS16Fractal}). The TN can represent the partition function of the statistical model defined on the Sierpi\'nski gasket, such as Ising and Potts model. As explained in Sec. II, the tensor is given by the probability distribution of the three spins in a triangle.

Such a TN can be exactly contracted by iteratively contracting each three of the tensors located in a same triangle as
\begin{eqnarray}
T_{a_1a_2a_3}' = \sum_{b_1b_2b_3} T_{a_1b_1b_2} T_{a_2b_2b_3} T_{a_3b_3b_1}.
\end{eqnarray}
After each round of contractions, the dimension of the tensors and the geometry of the network keep unchanged, but the number of the tensors in the TN decreases from $N$ to $N/3$. It means we can exactly contract the whole TN by repeating the above process.

\textbf{\textit{Algebraically contractible tensor networks}}. The third example is called algebraically contractible TN's\index{TN} \cite{KRV09exactTRG, DBJC12TopoTNS}. The tensors that form the TN possess some special algebraic properties, so that even the bond dimensions increase after each contraction, the rank of the bonds is kept unchanged. It means one can introduce some projectors to lower the bond dimension without causing any errors.

The simplest algebraically contractible TN is the one formed by the \textit{super-diagonal tensor} $I$ defined as
\begin{eqnarray}
I_{a_1,\cdots, a_N}&=&
\left\{
\begin{array}{lll}
1, \ \ a_1 = \cdots = a_N, \\
0, \ \ \text{otherwise}.
\end{array}
\right.
\end{eqnarray}
$I$ is also called \textit{copy tensor}, since it forces all its indexes to take a same value.

For a square TN of an arbitrary size formed by the fourth-order $I$'s, obviously we have its contraction $Z=d$ with $d$ the bond dimension. The reason is that the contraction is the summation of only $d$ non-zero values (each equals to 1).

\begin{figure}[tbp]
	\centering
	\includegraphics[angle=0,width=0.55\linewidth]{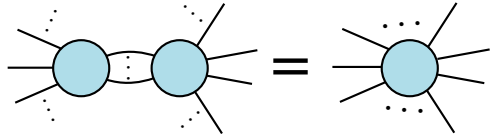}
	\caption{(Color online) The fusion rule of the copy tensor: the contraction of two copy tensors of $N_1$-th and $N_2$-th order gives a copy tensor of $(N_1+N_2-N)$-th order, with $N$ the number of the contracted bonds. }
	\label{fig-3DeltaFusion}
\end{figure}

To demonstrate its contraction, we will need one important property of the copy tensor (Fig. \ref{fig-3DeltaFusion}): if there are $n \geq 1$ bonds contracted between two copy tensors, the contraction gives a copy tensor,
\begin{eqnarray}
I_{a_1 \cdots b_1 \cdots} = \sum_{c_1 \cdots} I_{a_1\cdots c_1 \cdots} I_{b_1 \cdots c_1 \cdots}.
\end{eqnarray}
This property is called the \textit{fusion rule}, and can be understood in the opposite way: a copy tensor can be decomposed as the contraction of two copy tensors.

With the fusion rule, one will readily have the property for the dimension reduction: if there are $n \geq 1$ bonds contracted between two copy tensors, the contraction is identical after replacing the $n$ bonds with one bond,
\begin{eqnarray}
\sum_{c_1 \cdots c_n} I_{a_1\cdots c_1 \cdots c_n} I_{b_1 \cdots c_1 \cdots} = \sum_{c} I_{a_1\cdots c} I_{b_1 \cdots c}.
\end{eqnarray}
In other words, the dimension of the contracting bonds can be exactly reduced from $\chi^n$ to $\chi$. Applying this property to TN contraction, it means each time when the bond dimension increases after contracting several tensors into one tensor, the dimension can be exactly reduced to $\chi$, so that the contraction can continue until all bonds are contracted.

From the TN\index{TN} of the copy tensors, a class of exactly contractible TN can be defined, where the local tensor is the multiplication of the copy tensor by several unitary tensors. Taking the square TN as example, we have
\begin{eqnarray}
T_{a_1a_2a_3a_4} = \sum_{b_1b_2b_3b_4} X_{b_1} I_{b_1b_2b_3b_4} U_{a_1b_1} V_{a_2b_2} U^{\ast}_{a_3b_3} V^{\ast}_{a_4b_4},
\end{eqnarray}
with $U$ and $V$ two unitary matrices. $X$ is an arbitrary $d$-dimensional vector that can be understood as the ``weights'' (not necessarily to be positive to define the tensor). After putting the tensors in the TN, all unitary matrices vanish to identities. Then one can use the fusion rule of the copy tensor to exactly contract the TN, and the contraction gives $Z=\prod_b (X_{b})^{N_T}$ with $N_T$ the total number of tensors.

The unitary matrices are not trivial in physics. If we take $d=2$ and
\begin{eqnarray}
U=V=
\begin{bmatrix}
\sqrt{2}/2 & \sqrt{2}/2  \\
\sqrt{2}/2 & -\sqrt{2}/2 \\
\end{bmatrix},
\end{eqnarray}
the TN is in fact the inner product of the $Z_2$ topological state (see the definition of $Z_2$ PEPS\index{PEPS} in Sec. 2.2.3). If one cuts the system into two sub-regions, all the unitary matrix vanish into identities inside the bulk. However, those on the boundary will survive, which could lead to exotic properties such as topological orders, edge states and so on. Note that $Z_2$ state is only a special case. One can refer to a systematic picture given by X. G. Wen called the string-net states \cite{GLSW09StringTPS,BAV09StringTPS,CZGCW10StringTPS}.

\section{Some discussions}

\subsection{General form of tensor network}

One can see that a TN\index{TN} (state or operator) is defined as the contraction of certain tensors $\{T^{[n]}\}$ with a general form as
\begin{eqnarray}
\mathcal{T}_{\{s\}} = \sum_{\{a\}} \prod_{n} T^{[n]}_{s^n_1s^n_2 \cdots,a^n_1 a^n_2 \cdots}.
\label{eq-2TN}
\end{eqnarray}
The indexes $\{a\}$ are geometrical indexes, each of which is shared normally two tensors and will be contracted. The indexes $\{s\}$ are open bonds, each of which only belongs to one tensor. After contracting all the geometrical indexes, the TN\index{TN} represents a $\mathcal{N}$-th order tensor, with $\mathcal{N}$ the total number of the open indexes $\{s\}$.

Each tensor in the TN can possess different number of open or geometrical indexes. For an MPS\index{MPS}, each tensor has one open index (called physical bond) and two geometrical indexes; for PEPS\index{PEPS} on square lattice, it has one open and four geometrical indexes. For the generalizations of operators, the number of open indexes are two for each tensor. It also allows hierarchical structure of the TN, such as TTNS\index{TTNS} and MERA\index{MERA}.

One special kind of the TN's is the scalar TN with no open bonds, denoted as
\begin{eqnarray}
Z = \sum_{\{a\}} \prod_{n} T^{[n]}_{a^n_1 a^n_2 \cdots}.
\label{eq-2TNscalar}
\end{eqnarray}
It is very important because many physical problems can be transformed to computing the contractions of scalar TN's. A scalar TN can be obtained from the TN's that has open bonds, such as $Z = \sum_{\{s\}} \mathcal{T}_{\{s\}}$ or $Z = \sum_{\{s\}} \mathcal{T}_{\{s\}}^{\dagger} \mathcal{T}_{\{s\}}$, where $Z$ can be the cost function (e.g., energy or fidelity) to be maximized or minimized. The TN contraction algorithms mainly deal with the scalar TN's.

\subsection{Gauge degrees of freedom}
\label{sec.gauge}

For a given state, its TN\index{TN} representation is not unique. Let us take translational invariant MPS as an example. One may insert a (full-rank) matrix $U$ and its inverse $U^{-1}$ on each of the virtual bonds and then contracted them, respectively, into the two neighboring tensors. The tensors of new MPS\index{MPS} becomes $\tilde{A}^{[n]}_{s, a a'} = \sum_{bb'} U_{ab} A^{[n]}_{s, b b'} U^{-1}_{a' b'}$. In fact, we only put an identity $I = UU^{-1}$, thus do not implement any changes to the MPS. However, the tensors the form the MPS changes, meaning the TN representation changes. It is also the case when inserting an matrix and its inverse on any of the virtual bonds of a TN state, which changes the tensors without changing the state itself. Such degrees of freedom is known as the \textit{gauge degrees of freedom}, and the transformations are called \textit{gauge transformations}.

The gauge degrees of on the one hand may cause instability to TN simulations. Algorithms for finite and infinite PEPS were proposed to fix the gauge to reach higher stability \cite{LCB14fPEPS, PBTCO15FastFullUpdate,PMV15gaugePEPS}. On the other hand, one may use gauge transformation to transform a TN state to a special form, so that, for instance, one can implement truncations of local basis while minimizing the error non-locally \cite{OV08canonical, RLXZS12ODTNS} (we will go back to this issue later). Moreover, gauge transformation is closely related to other theoretical properties such as the global symmetry of TN states, which has been used to derive more compact TN representations \cite{SV13Gsymme}, and to classify many-body phases \cite{CGW11phase, SPC11PEPS} and to characterize non-conventional orders \cite{PWSC08symme, PTBO10EStopo}, just to name a few.

\subsection{Tensor network and quantum entanglement}
\label{sec.TNent}

\begin{figure}[tbp]
	\centering
	\includegraphics[angle=0,width=0.8\linewidth]{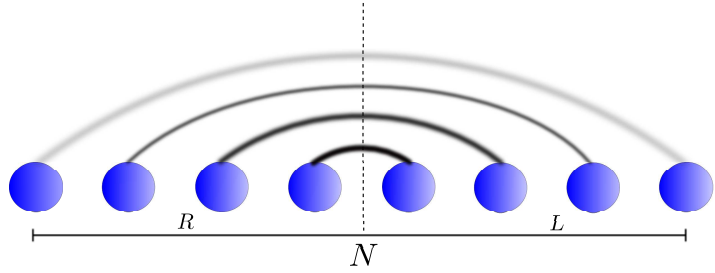}
	\caption{Bipartition of a 1D system into two half chains. Significant quantum correlations in gapped ground states occur only on short length scales.}
	\label{fig-Area_Law1}
\end{figure}

The numerical methods based on TN face great challenges, primarily that the dimension of the Hilbert space increases exponentially with the size. Such an ``\textit{exponential wall}'' has been treated in different ways by many numeric algorithms, including the DFT\index{DFT} methods \cite{K99DFT} and QMC\index{QMC} approaches \cite{TW05SignQMC}.

The power of TN\index{TN} has been understood in the sense of quantum entanglement: the entanglement structure of low-lying energy states can be efficiently encoded in TNS's\index{TNS}. It takes advantage of the fact that not all quantum states in the total Hilbert space of a many-body system are equally relevant to the low-energy or low-temperature physics. It has been found that the low-lying eigenstates of a gapped Hamiltonian with local interactions obey the area law of the entanglement entropy \cite{PKL99DMSpectra}.

More precisely speaking, for a certain subregion $\mathcal{R}$ of the system, its reduced density matrix is defined as $\hat{\rho}_{\mathcal{R}}= \rm Tr_{\mathcal{E}} (\hat{\rho})$, with $\mathcal{E}$ denotes the spatial complement of $\mathcal{R}$. The entanglement entropy is defined as
\begin{eqnarray}
S(\rho_{\mathcal{R}}) = - \rm Tr \lbrace \rho_{\mathcal{R}} \rm
log (\rho_{\mathcal{R}} ) \rbrace .
\end{eqnarray}
Then the area law of the entanglement entropy \cite{ECP10AreaLawRev,H15TNthesis} reads
\begin{eqnarray}
S(\rho_{\mathcal{R}})= O(\vert \partial \mathcal{R} \vert) ,
\end{eqnarray}
with $\vert \partial \mathcal{R}\vert$ the size of the boundary. In particular, for a $D$-dimensional system, one has
\begin{eqnarray}
S=O(l^{D-1}),
\label{eq-2arealaw}
\end{eqnarray}
with $l$ the length scale. This means that for 1D systems, $S= \rm \textit{const}$. The area law suggests that the low-lying eigenstates stay in a ``small corner'' of the full Hilbert space of the many-body system, and that they can be described by a much smaller number of parameters. We shall stress that the locality of the interactions is not sufficient to the area law. Vitagliano, \textit{el al} show that simple 1D spin models can exhibit volume law, where the entanglement entropy scales with the bulk \cite{VRL10arealaw, MS16arealaw}.

\begin{figure}[tbp]
	\centering
	\includegraphics[angle=0,width=0.8\linewidth]{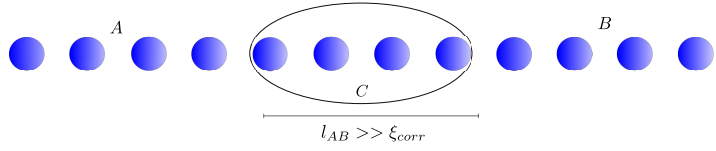}
	\caption{The argue the 1D area law, the chain is separated into three sub-systems denoted by $A$, $B$ and $C$. If the correlation length $\xi_{corr}$ is much larger than the size of $B$ (denoted by $l_{A C}$), the reduced density matrix by tracing $B$ approximately satisfies $\hat{\rho}_{AC} \simeq \hat{\rho}_{A} \otimes \hat{\rho}_{C}$.}
	\label{fig-Area_Law2}
\end{figure}

The area law of entanglement entropy is intimately connected to another fact that a non-critical quantum system exhibits a finite correlation length. The correlation functions between two blocks in a gapped system decay exponentially as a function of the distance of the blocks \cite{H04CorFunDecExp-1}, which is argued to lead to the area law. An intuitive picture can be seen in Fig. \ref{fig-Area_Law1}. Let us consider a 1D gapped quantum system whose ground state $| \psi_{ABC} \rangle$ possesses a correlation length $\xi_{corr}$. By dividing into three subregions $A$, $B$ and $C$, the reduced density operator $\hat{\rho}_{AC}$ is obtained when tracing out the block $B$, i.e. $\hat{\rho}_{AC} = \rm Tr_{B} | \psi_{ABC} \rangle \langle \psi_{ABC} |$ (see Fig. \ref{fig-Area_Law2}). In the limit of large distance between $A$ and $C$ blocks with $l_{AC} \gg \xi_{corr}$, one has the reduced density matrix satisfying
\begin{eqnarray}
\hat{\rho}_{AC} \simeq \hat{\rho}_{A} \otimes \hat{\rho}_{C},
\end{eqnarray}
up to some exponentially small corrections. Then $| \psi_{ABC} \rangle$ is a purification \footnote{Purification: Let $\rho$ be a density matrix acting an a Hilbert space $\mathcal{H}_A$ of finite dimension $n$. Then there exist a Hilbert space $\mathcal{H}_B$ and a pure state $\vert \psi \rangle \in \mathcal{H}_A \otimes \mathcal{H}_B$ such that the partial trace of $\vert \psi \rangle \langle \psi \vert$ with respect to $\mathcal{H}_B$: $\rho = Tr_B \vert \psi \rangle \langle \psi \vert$. We say that $\vert \psi \rangle$ is the purification of $\hat{\rho}$.} of a mixed state with the form $| \psi_{A B_l} \rangle \otimes | \psi_{B_r C} \rangle$ that has no correlations between $A$ and $C$; here $B_l$ and $B_r$ sit at the two ends of the block $B$, which together span the original block.

It is well known that all possible purifications of a mixed state are equivalent to each other up to a local unitary transformation on the virtual Hilbert space. This naturally implies that there exists a unitary operation $\hat{U}_B$ on the block $B$ that completely disentangles the left from the right part as
\begin{eqnarray}
\hat{I}_A \otimes \hat{U}_B \otimes \hat{I}_C \vert \psi_{ABC} \rangle \rightarrow \vert \psi_{A B_l} \rangle \otimes \vert \psi_{B_rC} \rangle .
\end{eqnarray}
$\hat{U}_B$ implies that there exists a tensor $B_{s,a a'}$ with $0\leq a, a', s \leq \chi-1$ and basis $\{| \psi^A \rangle\}$, $\{| \psi^B \rangle\}$, $\{| \psi^C \rangle\}$ defined on the Hilbert spaces belonging to $A$, $B$, $C$ such that
\begin{eqnarray}
\vert \psi_{ABC} \rangle \simeq \sum_{s a a'} B_{s,a a'} \vert \psi^A_{a} \rangle \vert \psi^B_{s} \rangle \vert \psi^C_{a'} \rangle .
\label{eq-3Btensor}
\end{eqnarray}

This argument directly leads to the MPS\index{MPS} description and gives a strong hint that the ground states of a gapped Hamiltonian is well represented by an MPS of finite bond dimensions, where $B$ in Eq. (\ref{eq-3Btensor}) is analog to the tensor in an MPS. Let us remark that every state of $N$ spins has an exact MPS representation if we allow $\chi$ grow exponentially with the number of spins \cite{VPC04DMRGQinfo}. The whole point of MPS is that a ground state can typically be represented by an MPS where the dimension $\chi$ is small and scales at most polynomially with the number of spins: this is the reason why MPS-based methods are more efficient than exact diagonalization.

For the 2D PEPS\index{PEPS}, it is more difficult to strictly justify the area law of entanglement entropy. However, we can make some sense of it from the following aspects. One is the fact that PEPS can exactly represent some non-trivial 2D states that satisfies the area law, such as the nearest-neighbor RVB and Z$_2$ spin liquid mentioned above. Another is to count the dimension of the geometrical bonds $\mathcal{D}$ between two subsystems, from which the entanglement entropy satisfies an upper bound as $S \leq \log \mathcal{D}$ \footnote{One can see this with simply a flat entanglement spectrum, $\lambda_n = 1/ \mathcal{D}$ for any $n$.}.

After dividing a PEPS\index{PEPS} into two subregions, one can see that the number of geometrical bonds $N_b$ increase linearly with the length scale, i.e. $N_b \sim l$. It means the dimension $\mathcal{D}$ satisfies $\mathcal{D} \sim \chi^{l}$, and the upper bound of the entanglement entropy fulfills the area law given by Eq. (\ref{eq-2arealaw}), which is
\begin{eqnarray}
S \leq O(l).
\end{eqnarray}
However, as we will see later, such a property of PEPS\index{PEPS} is exactly the reason that makes it computationally difficult.

\chapter{Two-dimensional tensor networks and contraction algorithms}
\label{sec3}

\abstract{In this section, we will first demonstrate in Sec. \ref{sec-phys2TN} that many important physical problems can be transformed to 2D TN's\index{TN}, and the central tasks become to compute the corresponding TN contractions. From Sec. \ref{sec.2.TRG} to \ref{transverse}, we will then present several paradigm contraction algorithms of 2D TN's including TRG\index{TRG}, TEBD\index{TEBD}, and CTMRG\index{CTMRG}. Relations to other distinguished algorithms and the exactly contractible TN's will also be discussed.}

\section{From physical problems to two-dimensional tensor networks}
\label{sec-phys2TN}

\subsection{Classical partition functions}
\label{Classical partition functions}

\begin{figure}[tbp]
	\centering
	\includegraphics[angle=0,width=1\linewidth]{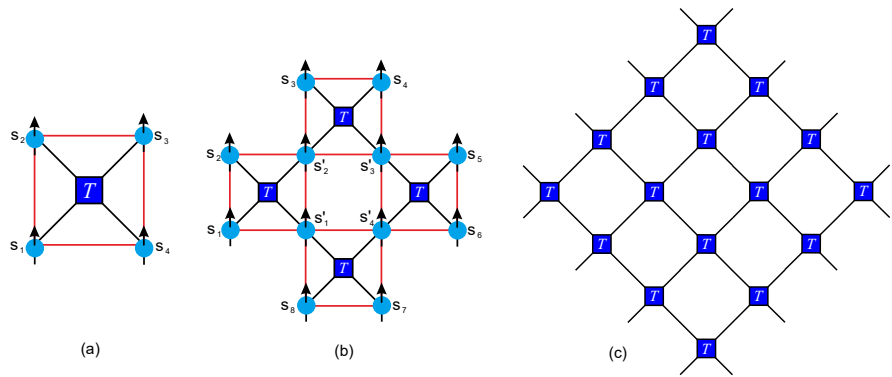}
	\caption{(Color online) (a) Four Ising spins (blue balls with arrows) sitting on a single square, and the red lines represent the interactions. The blue block is the tensor $T$ [Eq. (\ref{eq-3IsingT})], with the black lines denoting the indexes of $T$. (b) The graphic representation of the TN on a larger lattice with more than one squares. (c) The TN\index{TN} construction of the partition function on infinite square lattice.}
	\label{fig-3IsingSquare}
\end{figure}

Partition function, which is a function of the variables of a thermodynamic state such as temperature, volume, and etc., contains the statistical information of a thermodynamic equilibrium system. From its derivatives of different orders, we can calculate the energy, free energy, entropy, and so on. Levin and Nave pointed out in Ref. \cite{LN07TRG} that the partition functions of statistical lattice models (such as Ising and Potts models) with local interactions can be written in the form of TN. Without losing generality, we take square lattice as an example.

Let us start from the simplest case: the classical Ising model on a single square with only four sites. The four Ising spins denoted by $s_i$ ($i=1,2,3,4$) locate on the four corners of the square, as shown in Fig.\ref{fig-3IsingSquare} (a); each spin can be up or down, represented by $s_i=0$ and $1$, respectively. The classical Hamiltonian of such a system reads
\begin{eqnarray}
H_{s_1s_2s_3s_4}=J(s_1s_2 + s_2s_3 + s_3s_4 + s_4s_1) - h (s_1+s_2+s_3+s_4)
\end{eqnarray}
with $J$ the coupling constant and $h$ the magnetic field.

When the model reaches the equilibrium at temperature $\rm T$, the probability of each possible spin configuration is determined by the Maxwell-Boltzmann factor
\begin{eqnarray}
T_{s_1s_2s_3s_4}=e^{-\beta H_{s_1s_2s_3s_4}},
\label{eq-3IsingT}
\end{eqnarray}
with the inverse temperature $\beta=1/ \rm T$ \footnote{In this paper, we set Bolzmann constant $k_B = 1$ for convenience.}. Obviously, Eq. (\ref{eq-3IsingT}) is a forth-order tensor $T$, where each element gives the probability of the corresponding configuration.

The partition function is defined as the summation of the probability of all configurations. In the language of tensor, it is obtained by simply summing over all indexes as
\begin{eqnarray}
Z=\sum_{s_1s_2s_3s_4} T_{s_1s_2s_3s_4}.
\end{eqnarray}

Let us proceed a little bit further by considering four squares, whose partition function can be written in a TN\index{TN} with four tensors [Fig.\ref{fig-3IsingSquare}(b)] as
\begin{eqnarray}
Z = \sum_{\{ss'\}} T_{s_1s_2s_2's_1'} T_{s_2's_3s_4s_3'} T_{s_4's_3's_5s_6} T_{s_8s_1's_4's_7}.
\end{eqnarray}
Each of the indexes $\{s'\}$ inside the TN is shared by two tensors, representing the spin that appears in both of the squares. The partition function is obtained by summing over all indexes.

For the infinite square lattice, the probability of a certain spin configuration ($s_1, s_2, \cdots$) is given by the product of infinite number of tensor elements as
\begin{eqnarray}\label{eq1}
e^{-\beta H_{\{s\}}}=e^{-\beta H_{s_1s_2s_3s_4}} e^{-\beta H_{s_4s_5s_6s_7}} \cdots =T_{s_1s_2s_3s_4}T_{s_4s_5s_6s_7} \cdots
\end{eqnarray}
Then the partition function is given by the contraction of an infinite TN\index{TN} formed by the copies of $T$ [Eq. (\ref{eq-3IsingT})] as
\begin{eqnarray}\label{eq2}
Z=\sum_{\{s\}} \prod_{n} T_{s^n_1 s^n_2 s^n_3 s^n_4},
\label{eq-3IsingTN}
\end{eqnarray}
where two indexes satisfy $s^n_j = s^m_k$ if they refer to the same Ising spin. The graphic representation of Eq.\ref{eq-3IsingTN} is shown in Fig.\ref{fig-3IsingSquare}  (c). One can see that on square lattice, the TN still has the geometry of a square lattice. In fact, such a way will give a TN that has a geometry of the dual lattice of the system, and the dual of the square lattice is itself.

For the $Q$-state Potts model on square lattice, the partition function has the same TN representation as that of the Ising model, except that the elements of the tensor are given by the Boltzmann wight of the Potts model and the dimension of each index is $Q$.  Note that the Potts model with $q=2$ is equivalent to the Ising model.

Another example is the eight-vertex model proposed by Baxter in 1971 \cite{B71EightVertex}. It is one of the ``ice-type'' statistic lattice model, and can be considered as the classical correspondence of the Z$_2$ spin liquid state. The tensor that gives the TN of the partition function is also ($2 \times 2\times 2\times 2$), whose non-zero elements are
\begin{eqnarray}
T_{s_1,\cdots, s_N}&=&
\left\{
\begin{array}{lll}
1, \ \ s_1+\cdots + s_N=even, \\
0, \ \ \text{otherwise}.
\end{array}
\right.
\end{eqnarray}

We shall remark that there are more than one ways to define the TN of the partition function of a classical system. For example, when there only exist nearest-neighbor couplings, one can define a matrix $M_{ss'} = e^{-\beta H_{ss'}}$ on each bond and put on each site a \textit{super-digonal} tensor $I$ (or called copy tensor) defined as
\begin{eqnarray}
I_{s_1,\cdots, s_N}=
\begin{cases}
1, \ \ s_1 = \cdots = s_N;\\
0, \ \ \text{otherwise}.
\end{cases}
\end{eqnarray}
Then the TN of the partition function is the contraction of copies of $M$ and $I$, and possesses exactly the same geometry of the original lattice (instead of the dual one).

\subsection{Quantum observables}
\label{sec.QobTN}

With a TN\index{TN} state, the computations of quantum observables as $\langle \psi | \hat{O} | \psi \rangle$ and $\langle \psi | \psi \rangle$ is the contraction of a scalar TN, where $\hat{O}$ can be any operator. For a 1D MPS\index{MPS}, this can be easily calculated, since one only needs to deal with a 1D TN stripe. For 2D PEPS\index{PEPS}, such calculations become contractions of 2D TN's. Taking $\langle \psi | \psi \rangle$ as an example, the TN of such an inner product is the contraction of the copies of the local tensor [Fig. \ref{fig-3IsingSquare} (c)] defined as
\begin{eqnarray}
T_{a_1a_2a_3a_4} = \sum_{s} P_{s,a''_1a''_2a''_3a''_4}^{\ast} P_{s,a_1'a_2'a_3'a_4'},
\end{eqnarray}
with $P$ the tensor of the PEPS\index{PEPS} and $a_i = (a'_i, a''_i)$. There are no open indexes left and the TN\index{TN} gives the scalar $\langle \psi | \psi \rangle$. The TN for computing the observable $\langle \hat{O} \rangle$ is similar. The only difference is that we should substitute some small number of $T_{a_1a_2a_3a_4}$ in original TN of $\langle \psi |\psi \rangle$ with ``impurities'' at the sites where the operators locate. Taking one-body operator as an example, the ``impurity'' tensor on this site can be defined as
\begin{eqnarray}
\widetilde{T}_{a_1a_2a_3a_4}^{[i]} = \sum_{s,s'} P_{s,a''_1a''_2a''_3a''_4}^{\ast} \hat{O}_{s,s'}^{[i]} P_{s',a_1'a_2'a_3'a_4'},
\end{eqnarray}
In such a case, the single-site observables can be represented by the TN contraction of
\begin{eqnarray}
\frac{\langle \psi| \hat{O}^{[i]}|\psi \rangle}{\langle \psi| \psi \rangle} = \frac{\text{tTr }\ \widetilde{T}^{[i]} \prod_{n\neq i} T}{\text{tTr }\ \prod_{n=1}^{N} T},
\end{eqnarray}
For some non-local observables, e.g., the correlation function, the contraction of $\langle \psi| \hat{O}^{[i]} \hat{O}^{[j]} |\psi \rangle$ is nothing but adding another ``impurity'' by
\begin{eqnarray}
\langle \psi| \hat{O}^{[i]} \hat{O}^{[j]} |\psi \rangle = \text{tTr }\ \widetilde{T}^{[i]}\widetilde{T}^{[j]} \prod_{n\neq i,j}^{N} T,
\end{eqnarray}

\subsection{Ground-state and finite-temperature simulations}
\label{sec.2.evolution}

Ground-state simulations of 1D quantum models with short-range interactions can also be efficiently transferred to 2D TN contractions. When minimizing the energy
\begin{eqnarray}
E=\frac{\langle \psi | \hat{H} | \psi \rangle} {\langle \psi | \psi \rangle},
\end{eqnarray}
where we write $|\psi\rangle$ as an MPS\index{MPS}. Generally speaking, there are two ways to solve the minimization problem: (i) simply treat all the tensor elements as variational parameters; (ii) simulate the imaginary-time evolution
\begin{eqnarray}\label{eq3}
|\psi_{gs}\rangle = \lim_{\beta \rightarrow \infty} \frac{e^{-\beta \hat{H}}|\psi\rangle}{\parallel{e^{-\beta \hat{H}}|\psi\rangle}\parallel}.
\end{eqnarray}

The first way can be realized by, e.g., Monte Carlo methods where one could randomly change or choose the value of each tensor element to locate the minimal of energy. One can also use the Newton method and solve the partial-derivative equations $\partial E / \partial x_n = 0$ with $x_n$ standing for an arbitrary variational parameter. Anyway, it is inevitable to calculate $E$ (i.e., $\langle \psi | \hat{H} | \psi \rangle$ and $\langle \psi | \psi \rangle$) for most cases, which is to contraction the corresponding TN's as explained above.

We shall stress that without TN\index{TN}, the dimension of the ground state (i.e., the number of variational parameters) increases exponentially with the system size, which makes the ground-state simulations impossible for large systems.

The second way of computing the ground state with imaginary-time evolution is more or less like an ``annealing'' process. One starts from an arbitrarily chosen initial state and acts the imaginary-time evolution operator on it. The ``temperature'' is lowered a little for each step, until the state reaches a fixed point. Mathematically speaking, by using Trotter-Suzuki decomposition, such an evolution is written in a TN defined on ($D+1$)-dimensional lattice, with $D$ the dimension of the real space of the model.

Here, we take a 1D chain as an example. We assume that the Hamiltonian only contains at most nearest-neighbor couplings, which reads
\begin{eqnarray}\label{eq4}
\hat{H}=\sum_{n} \hat{h}_{n,n+1},
\end{eqnarray}
with $\hat{h}_{n,n+1}$ containing the on-site and two-body interactions of the $n$-th and $n+1$-th sites. It is useful to divide $\hat{H}$ into two groups, $\hat{H}=\hat{H}^e+\hat{H}^{o}$ as
\begin{eqnarray}\label{eq5}
\begin{split}
\hat{H}^e \equiv\sum_{even\ n}\hat{h}_{n,n+1}, \ \ \
\hat{H}^{o} \equiv\sum_{odd\ n}\hat{h}_{n,n+1}.
\end{split}
\end{eqnarray}

By doing so, each two terms in $\hat{H}^e$ or $\hat{H}^{o}$ commute with each other. Then the evolution operator $\hat{U}(\tau)$ for infinitesimal imaginary time $\tau \to 0$ can be written as
\begin{eqnarray}
\hat{U}(\tau) = e^{-\tau \hat{H}} = e^{-\tau \hat{H}^e} e^{-\tau \hat{H}^o} + O(\tau^2) [\hat{H}^e, \hat{H}^o]
\label{eq6}
\end{eqnarray}
If $\tau$ is small enough, the high-order terms are negligible, and the evolution operator becomes
\begin{eqnarray}
\hat{U}(\tau) \simeq \prod_{n} \hat{U}(\tau)_{n,n+1},
\label{eq-3trotter}
\end{eqnarray}
with the two-site evolution operator $\hat{U}(\tau)_{n,n+1} = e^{-\tau \hat{H}_{n,n+1}}$.

The above procedure is known as the first-order Trotter-Suzuki decomposition \cite{T59TrotterDecomp, SI87Trotter, IS88Trotter}. Note that higher-order decomposition can also be adopted. For example, one may use the second order Trotter-Suzuki decomposition that is written as
\begin{eqnarray}\label{eq7}
e^{-\tau \hat{H}} \simeq e^{-\frac{\tau}{2} \hat{H}^e} e^{-\tau \hat{H}^o} e^{-\frac{\tau}{2} \hat{H}^e}.
\end{eqnarray}

With Eq. (\ref{eq-3trotter}), the time evolution can be transferred to a TN, where the local tensor is actually the coefficients of $\hat{U}(\tau)_{n,n+1}$, satisfying
\begin{eqnarray}
T_{s_n s_{n+1} s'_n s'_{n+1}} = \langle s'_n s'_{n+1} | \hat{U}(\tau)_{n,n+1} | s_n s_{n+1}\rangle.
\label{eq-3evolveT}
\end{eqnarray}
Such a TN is defined in a plain of two dimensions that corresponds to the spatial and (real or imaginary) time, respectively. The initial state is located at the bottom of the TN\index{TN} ($\beta = 0$) and its evolution is to do the TN contraction which can efficient solved by TN algorithms (presented later).

In addition, one can readily see that the evolution of a 2D state leads to the contraction of a 3D TN. Such a TN scheme provides a straightforward picture to understand the equivalence between a ($d+1$)-dimensional classical and a $d$-dimensional quantum theory. Similarly, the finite-temperature simulations of a quantum system can be transferred to TN contractions with Trotter-Suzuki decomposition. For the density operator $\hat{\rho}(\beta) = e^{-\beta \hat{H}}$, the TN is formed by the same tensor given by Eq. (\ref{eq-3evolveT}).


\section{Tensor renormalization group}
\label{sec.2.TRG}

\begin{figure}[tbp]
	\centering
	\includegraphics[angle=0,width=\linewidth]{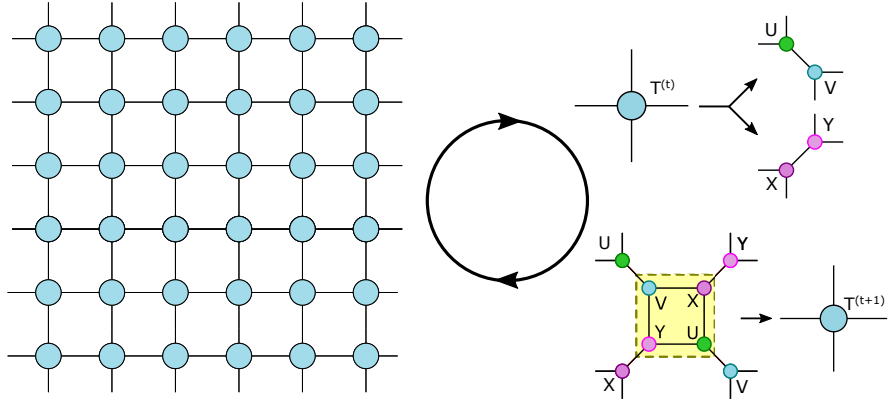}
	\caption{(Color online) For an infinite square TN with translational invariance, the renormalization in the TRG algorithm is realized by two local operations of the local tensor. After each iteration, the bond dimensions of the tensor and the geometry of the network keep unchanged.}
	\label{fig-3TRG}
\end{figure}

In 2007, Levin and Nave proposed TRG\index{TRG} approach \cite{LN07TRG} to contract the TN\index{TN} of 2D classical lattice models. In 2008, Gu \textit{et al} further developed TRG to handle 2D quantum topological phases \cite{GLW08TERG}. TRG can be considered as a coarse-graining contraction algorithm. To introduce the TRG algorithm, let us consider a square TN formed by infinite number of copies of a forth-order tensor $T_{a_1a_2a_3a_4}$ (see the left side of Fig. \ref{fig-3TRG}).

\textbf{\textit{Contraction and truncation}}. The idea of TRG is to iteratively ``coarse-grain'' the TN without changing the bond dimensions, the geometry of the network, and the translational invariance. Such a process is realized by two local operations in each iteration. Let us denote the tensor in the $t$-th iteration as $T^{(t)}$ (we take $T^{(0)}=T$). For obtaining $T^{(t+1)}$, the first step is to decompose $T^{(t)}$ by SVD\index{SVD} in two different ways [Fig. \ref{fig-3TRG}] as
\begin{eqnarray}
T^{(t)}_{a_1a_2a_3a_4} = \sum_{b} U_{a_1a_2b}V_{a_3a_4b},\\
T^{(t)}_{a_1a_2a_3a_4} = \sum_{b} X_{a_4a_1b}Y_{a_2a_3b}.
\label{eq-3TRGdecomp}
\end{eqnarray}
Note that the singular value spectrum can be handled by multiplying it with the tensor(s), and the dimension of the new index satisfies $dim(b)=\chi^2$ with $\chi$ the dimension of each bond of $T^{(t)}$.

The purpose of the first step is to deform the TN, so that in the second step, a new tensor $T^{(t+1)}$ can be obtained by contracting the four tensors that form a square [Fig. \ref{fig-3TRG}] as
\begin{eqnarray}
T^{(t+1)}_{b_1b_2b_3b_4} \leftarrow \sum_{a_1a_2a_3a_4} V_{a_1a_2b_1} Y_{a_2a_3b_2} U_{a_3a_4b_3} X_{a_4a_1b_4}.
\label{eq-3TRGnewT}
\end{eqnarray}
We use an arrow instead of the equal sign, because one may need to divide the tensor by a proper number to keep the value of the elements from being divergent. The arrows will be used in the same way below.

These two steps define the contraction strategy of TRG. By the first step, the number of tensors in the TN (i.e., the size of the TN\index{TN}) increases from $N$ to $2N$, and by the second step, it decreases from $2N$ to $N/2$. Thus, after $t$ times of each iterations, the number of tensors decreases to the $\frac{1}{2^t}$ of its original number. For this reason, TRG\index{TRG} is an \textit{exponential contraction algorithm}.

\textbf{\textit{Error and environment}}. The dimension of the tensor at the $t$-th iteration becomes $\chi^{2^t}$, if no truncations are implemented. that means that truncations of the bond dimensions are necessary. In its original proposal, the dimension is truncated by only keeping the singular vectors of the $\chi$-largest singular values in Eq. (\ref{eq-3TRGdecomp}). Then the new tensor $T^{(t+1)}$ obtained by Eq. (\ref{eq-3TRGnewT}) has exactly the same dimension as $T^{(t)}$.

Each truncation will absolutely introduce some error, which is called the \textit{truncation error}. Consistent with Eq. (\ref{eq-2trunerr}), the truncation error is quantified by the discarded singular values $\lambda$ as
\begin{eqnarray}
\varepsilon = \frac{\sqrt{\sum_{b=\chi}^{\chi^2-1} \lambda_{b}^2}}{\sqrt{\sum_{b=0}^{\chi^2-1} \lambda_{b}^2}}.
\label{eq-3TRGerr}
\end{eqnarray}
According to the linear algebra, $\varepsilon$ in fact gives the error of the SVD\index{SVD} given in Eq. (\ref{eq-3TRGdecomp}), meaning that such a truncation minimizes the error of reducing the rank of $T^{(t)}$, which reads
\begin{eqnarray}
\varepsilon = |T^{(t)}_{a_1a_2a_3a_4} - \sum_{b=0}^{\chi-1} U_{a_1a_2b} V_{a_3a_4b}|
\label{eq-3TRGerrT}
\end{eqnarray}
One may repeat the contraction-and-truncation process until $T^{(t)}$ converges. It usually only takes $\sim 10$ steps, after which one in fact contract a TN\index{TN} of $2^t$ tensors to a single tensor.

The truncation is optimized according to the SVD\index{SVD} of $T^{(t)}$. Thus, $T^{(t)}$ is called the \textit{environment}. In general, the tensor(s) that determines the truncations is called the environment. It is a key factor to the accuracy and efficiency of the algorithm. For those that use local environments, like TRG, the efficiency is relatively high since the truncations are easy to compute. But, the accuracy is bounded since the truncations are only optimized according to some local information (like in TRG the local partitioning $T^{(t)}$).

One may choose other tensors or even the whole TN\index{TN} as the environment. In 2009, Xie \textit{et al} proposed the second renormalization group (SRG)\index{SRG} algorithm \cite{XJCWX09SRG}. The idea is in each truncation step of TRG\index{TRG}, they define the global environment that is a forth-order tensor $\mathcal{E}_{a_1^{\tilde{n}} a_2^{\tilde{n}} a_3^{\tilde{n}} a_4^{\tilde{n}}} = \sum_{\{a\}} \prod_{n \neq \tilde{n}} T^{(n, t)}_{a_1^na_2^na_3^na_4^n}$ with $T^{(n, t)}$ the $n$-th tensor in the $t$-th step and $\tilde{n}$ the tensor to be truncated. $\mathcal{E}$ is the contraction of the whole TN after getting rid of $T^{(\tilde{n}, t)}$, and is computed by TRG. Then the truncation is obtained not by the SVD\index{SVD} of $T^{(\tilde{n}, t)}$, but by the SVD of $\mathcal{E}$. The word ``second'' in the name of the algorithm comes from the fact that in each step of the original TRG, they use a second TRG to calculate the environment. SRG is obviously more consuming, but bears much higher accuracy than TRG. The balance between accuracy and efficiency, which can be controlled by the choice of environment, is one main factor top consider while developing or choosing the TN algorithms.


\section{Corner transfer-matrix renormalization group}
\label{sec.CTMRG}

\begin{figure}[tbp]
	\centering
	\includegraphics[angle=0,width=0.7\linewidth]{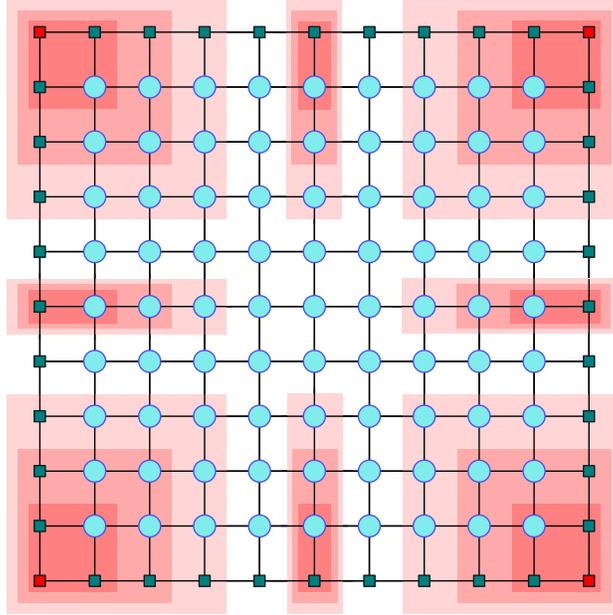}
	\caption{(Color online) Overview of the CTMRG\index{CTMRG} contraction scheme. The tensors in the TN are contracted to the variational tensors defined on the edges and corners.}
	\label{fig-3CTMRGcontract}
\end{figure}

\begin{figure}[tbp]
	\centering
	\includegraphics[angle=0,width=\linewidth]{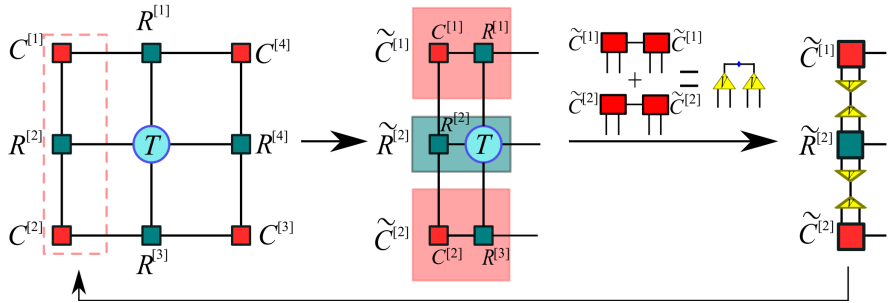}
	\caption{(Color online) The first arrow shows absorbing tensors $R^{[1]}$, $T$, and $R^{[3]}$ to renew tensors $C^{[1]}$, $R^{[2]}$, and $C^{[2]}$ in left operation. The second arrow shows the truncation of the enlarged bond of $\tilde{C}^{[1]}$, $\tilde{R}^{[2]}$ and $\tilde{C}^{[2]}$. Inset is the acquisition of the truncation matrix $Z$.}
	\label{fig-3CTMRG}
\end{figure}

In the 1960s, the corner transfer matrix (CTM)\index{CTM} idea was developed originally by Baxter in Refs. \cite{B68Prototype, B78CTM} and a book \cite{B16statBook}. Such ideas and methods have been applied to various models, for example, the chiral Potts model \cite{B91CTMPotts, B93CTMPottsP, B93CTMPottsTL}, the 8-vertex model \cite{B71EightVertex, B76CTM8VertexI, B76CTM8VertexII}, and to the 3D Ising model \cite{BF84CTM3DIsing}. Combining CTM with DMRG, Nishino and Okunishi proposed the CTMRG\index{CTMRG} \cite{NO96CTMRG0} in 1996 and applied it to several models \cite{NO96CTMRG0,nishino2001two,nishino2000self,nishino1998density,okunishi2000kramers,nishino1997numerical,nishino1997corner,tsushima1998phase,okunishi1999universal,li2001critical,gendiar2002latent}. In 2009, Or\'us and Vidal further developed CTMRG to deal with TN's \cite{OV09CTMRG}. What they proposed to do is to put eight \textit{variational tensors} to be optimized in the algorithm, which are four corner transfer matrices $C^{[1]}, C^{[2]}, C^{[3]}, C^{[4]}$ and four row (column) tensors $R^{[1]}, R^{[2]}, R^{[3]}, R^{[4]}$, on the boundary, and then to contract the tensors in the TN to these variational tensors in a specific order shown in Fig. \ref{fig-3CTMRGcontract}. The TN contraction is considered to be solved with the variational tensors when they converge in this contraction process. Compared with the boundary-state methods in the last subsection, the tensors in CTMRG define the states on both the boundaries and corners.


\textbf{\textit{Contraction}}. In each iteration step of CTMRG\index{CTMRG}, one choses two corner matrices on the same side and the row tensor between them, e.g., $C^{[1]}$, $C^{[2]}$ and $R^{[2]}$. The update of these tensors (Fig.\ref{fig-3CTMRG}) follows
\begin{eqnarray}
\tilde{C}^{[1]}_{\tilde{b}_2b'_1} &\leftarrow& \sum_{b_1} C^{[1]}_{b_1b_2}R^{[1]}_{b_1a_1b'_1},\\
\tilde{R}^{[2]}_{\tilde{b}_2 a_4 \tilde{b_3}} &\leftarrow& \sum_{a_2}R^{[2]}_{b_2a_2b_3}T_{a_1a_2a_3a_4}, \label{eq-3CTMRGcontract}\\
\tilde{C}^{[2]}_{\tilde{b}_3b'_4} &\leftarrow& \sum_{b_4} C^{[2]}_{b_3b_4}R^{[3]}_{b_4a_3b'_4},
\end{eqnarray}
where $\tilde{b}_2=(b_2,a_1)$ and $\tilde{b}_3=(b_3,a_1)$.

After the contraction given above, it can be considered that one column of the TN\index{TN} (as well as the corresponding row tensors $R^{[1]}$ and $R^{[3]}$) are contracted. Then one chooses other corner matrices and row tensors (such as $\tilde{C}^{[1]}$, $C^{[4]}$ and $R^{[1]}$) and implement similar contractions. By iteratively doing so, the TN is contracted in the way shown in Fig. \ref{fig-3CTMRGcontract}.

Note that for a finite TN\index{TN}, the initial corner matrices and row tensors should be taken as the tensors locating on the boundary of the TN. For an infinite TN, they can be initialized randomly, and the contraction should be iterated until the preset convergence is reached.

CTMRG\index{CTMRG} can be regarded as a \textit{polynomial contraction scheme}. One can see that the number of tensors that are contracted at each step is determined by the length of the boundary of the TN\index{TN} at each iteration time. When contracting a 2D TN defined on a $(L\times L)$ square lattice as an example, the length of each side is $L-2t$ at the $t$-th step. The boundary length of the TN (i.e., the number of tensors contracted at the $t$-th step) bears a linear relation with $t$ as $4(L-2t)-4$. For a 3D TN such as cubic TN, the boundary length scales as $6(L-2t)^2-12(L-2t)+8$, thus the CTMRG for a 3D TN (if exists) gives a polynomial contraction.

\textbf{\textit{Truncation}}. One can see that after the contraction in each iteration step, the bond dimensions of the variational tensors increase. Truncations are then in need to prevent the excessive growth of the bond dimensions. In Ref. \cite{OV09CTMRG}, the truncation is obtained by inserting a pair of isometries $V$ and $V^{\dagger}$ in the enlarged bonds. A reasonable (but not the only choice) of $V$ for translational invariant TN is to consider the eigenvalue decomposition on the sum of corner transfer matrices as
\begin{eqnarray}
\sum_{b}\tilde{C}^{[1]\dagger}_{\tilde{b}b}{\tilde{C}^{[1]}}_{\tilde{b}'b} + \sum_{b}\tilde{C}^{[2]\dagger}_{\tilde{b}b}{\tilde{C}^{[1]}}_{\tilde{b}'b}
\simeq \sum_{b=0}^{\chi-1} V_{\tilde{b}b} \Lambda_{b} V^{*}_{\tilde{b}'b}.
\label{eq-3CTMRGtrun}
\end{eqnarray}
Only the $\chi$ largest eigenvalues are preserved. Therefore, $V$ is a matrix of the dimension $D\chi \times \chi$, where $D$ is the bond dimension of $T$ and $\chi$ is the dimension cut-off. We then truncate $\tilde{C}^{[1]}$, $\tilde{R}^{[2]}$, and $\tilde{C}^{[2]}$ using $V$ as
\begin{eqnarray}
C^{[1]}_{b'_1b_2} &=& \sum_{\tilde{b}_2}\tilde{C}^{[1]}_{\tilde{b}_2b'_1}V^{*}_{\tilde{b}_2b_2},\\
R^{[2]}_{b_2a_4b_3} &=& \sum_{\tilde{b}_2,\tilde{b}_3}\tilde{R}^{[2]}_{\tilde{b}_2a_4\tilde{b}_3} V_{\tilde{b}_2b_2} V^{*}_{\tilde{b}_3b_3},\label{eq-3CTMRGtruncate} \\
C^{[2]}_{b_3b'_4} &=&
\sum_{\tilde{b}_3}\tilde{C}^{[2]}_{\tilde{b}_3b'_4} V_{\tilde{b}_3b_3}.
\end{eqnarray}


\textbf{\textit{Error and environment}}. Same as TRG\index{TRG} or TEBD\index{TEBD}, the truncations are obtained by the matrix decompositions of certain tensors that define the environment. From Eq. (\ref{eq-3CTMRGtrun}), the environment in CTMRG is the loop formed by the corner matrices and row tensors. Note that symmetries might be considered to accelerate the computation. For example, one may take $C^{[1]}=C^{[2]}=C^{[3]}=C^{[4]}$ and $R^{[1]}=R^{[2]}=R^{[3]}=R^{[4]}$ when the TN has rotational and reflection symmetries ($T_{a_1a_2a_3a_4} = T_{a_1'a_2'a_3'a_4'}$ after any permutation of the indexes).

\section{Time-evolving block decimation: linearized contraction and boundary-state methods}
\label{iTEBD}

The TEBD\index{TEBD} algorithm by Vidal was developed originally for simulating the time evolution of 1D quantum models \cite{V03TEBD,V04TEBD,V07iTEBD}. The (finite and infinite) TEBD algorithm has been widely applied to varieties of issues, such as criticality in quantum many body systems (e.g., \cite{TOIL08EntScaling, PMTM09EntScaling, PM10EScrit}), the topological phases \cite{PT12SPT}. the many-body localization \cite{DSK2013mbl, BPF2012unbounded, PPH2015mbl} and the thermodynamic property of quantum many-body systems \cite{PMJ2010entanglement, PMW2014correlations, BPG2009relaxation, FCE2014relaxation, BPG2010quantum, EF2016quench, LRGZX+11LTRG}.

\begin{figure}[tbp]
	\centering
	\includegraphics[angle=0,width=\linewidth]{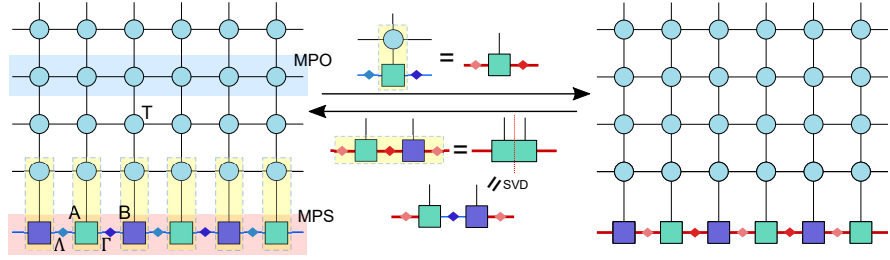}
	\caption{(Color online) The illustration of the contraction and truncation of the iTEBD\index{iTEBD} algorithm. In each iteration step, a row of tensors in the TN are contracted to the MPS, and truncations by SVD\index{SVD} are implemented so that the bond dimensions of the MPS\index{MPS} keep unchanged.}
	\label{fig-3iTEBD}
\end{figure}

In the language of TN\index{TN}, TEBD solves the TN contraction problems in a linearized manner, and the truncation is calculated in the context of an MPS. In the following, let us explain the infinite TEBD (iTEBD)\index{iTEBD} algorithm \cite{V07iTEBD} (Fig. \ref{fig-3iTEBD}) by still taking the infinite square TN formed by the copies of a forth-order tensor $T$ as an example. In each step, a row of tensors (which can be regarded as an MPO\index{MPO}) are contracted to an MPS\index{MPS} $|\psi\rangle$. Inevitably, the bond dimensions of the tensors in the MPS will increase exponentially as the contractions proceed. Therefore, truncations are necessary to prevent the bond dimensions diverging. The truncations are determined by minimizing the distance between the MPS's before and after the truncation. After the MPS $|\psi\rangle$ converges, the TN contraction becomes $\langle\psi |\psi\rangle$, which can be exactly and easily computed.

\textbf{\textit{Contraction}}. We use is two-site translational invariant MPS, which is formed by the tensors $A$ and $B$ on the sites and the spectrum $\Lambda$ and $\Gamma$ on the bonds as
\begin{eqnarray}
\sum_{\{a\}} \cdots \Lambda_{a_{n-1}} A_{s_{n-1}, a_{n-1} a_{n}} \Gamma_{a_{n}} B_{s_{n}, a_{n} a_{n+1}} \Lambda_{a_{n+1}} \cdots.
\end{eqnarray}
In each step of iTEBD, the contraction is given by
\begin{eqnarray}
A_{s, \tilde{a} \tilde{a}'} \leftarrow \sum_{s'} T_{s b s' b'} A_{s', a a'}, \ \ B_{s, \tilde{a} \tilde{a}'} \leftarrow \sum_{s'} T_{s b s' b'} B_{s', a a'},
\label{eq-3iTEBDEvolve}
\end{eqnarray}
where the new virtual bonds are entangled, satisfying $\tilde{a} = (b,a)$ and $\tilde{a}' = (b',a')$. Meanwhile, the spectrum are also updated as
\begin{eqnarray}
\Lambda_{\tilde{a}} \leftarrow  \Lambda_{a} \textbf{1}_{b},  \ \ \Gamma_{\tilde{a}'} \leftarrow \Gamma_{a'} \textbf{1}_{b'},
\end{eqnarray}
where $\textbf{1}$ is a vector with $\textbf{1}_b=1$ for any $b$.

It is readily to see that the number of tensors in iTEBD will be reduced linearly as $tN$, with $t$ the number of the contraction-and-truncation steps and $N \to \infty$ the number of the columns of the TN. Therefore, iTEBD (also finite TEBD) can be considered as a \textit{linearized contraction algorithm}, in contrast to the exponential contraction algorithm like TRG\index{TRG}.

\textbf{\textit{Truncation}}. Truncations are needed when the dimensions of the virtual bonds exceed the preset dimension cut-off $\chi$. In the original version of iTEBD \cite{V07iTEBD}\index{iTEBD}, the truncations are done by local SVD's\index{SVD}. To truncate the virtual bond $\tilde{a}$ for example, one defines a matrix by contracting the tensors and spectrum connected to the target bond as
\begin{eqnarray}
M_{s_1 \tilde{a}_1,s_{2} \tilde{a}_2} = \sum_{\tilde{a}} \Lambda_{\tilde{a}_1} A_{s_1, \tilde{a}_1 \tilde{a}} \Gamma_{\tilde{a}} B_{s_2, \tilde{a} \tilde{a}_2} \Lambda_{\tilde{a}_2}.
\end{eqnarray}

Then, perform SVD on $M$, keeping only the $\chi$-largest singular values and the corresponding basis as
\begin{eqnarray}
M_{s_1 \tilde{a}_1,s_2 \tilde{a}_2} = \sum_{a=0}^{\chi-1} U_{s_1, \tilde{a}_1 a} \Gamma_{a} V_{s_2, a \tilde{a}_2}.
\label{eq-3iTEBDenv}
\end{eqnarray}
The spectrum $\Gamma$ is updated by the singular values of the above SVD. The tensors $A$ and $B$ are also updated as
\begin{eqnarray}
A_{s_1, \tilde{a} a} = (\Lambda_{\tilde{a}})^{-1} U_{s_1, \tilde{a} a}, \ \
B_{s_2, a \tilde{a}} = V_{s_2, a \tilde{a}} (\Lambda_{\tilde{a}})^{-1}.
\label{eq-3iTEBDtruncate}
\end{eqnarray}
Till now, the truncation of the spectrum $\Gamma$ and the corresponding virtual bond have been completed. Any spectra and virtual bonds can be truncated similarly.

\textbf{\textit{Error and environment}}. Similar to TRG\index{TRG} and SRG\index{SRG}, the environment of the original iTEBD is $M$ in Eq. (\ref{eq-3iTEBDenv}), and the error is measured by the discarded singular values of $M$. Thus iTEBD\index{iTEBD} seems to only use local information to optimize the truncations. What is amazing is that when the MPO\index{MPO} is unitary or near unitary, the MPS converges to a so-called \textit{canonical form} \cite{PVWC07MPSRev, OV08canonical}. The truncations are then optimal by taking the whole MPS as the environment. If the MPO is far from being unitary, Or\'us and Vidal proposed the \textit{canonicalization} algorithm \cite{OV08canonical} to transform the MPS\index{MPS} into the canonical form before truncating. We will talk about this issue in detail in the next section.

\textbf{\textit{Boundary-state methods: density matrix renormalization group and variational matrix product state}}. The iTEBD\index{iTEBD} can be understood as a boundary-state method. One may consider one row of tensors in the TN\index{TN} as an MPO\index{MPO} (see Sec. \ref{sec-MPO} and Fig. \ref{fig-2MPO}), where the vertical bonds are the ``physical'' indexes and the bonds shared by two adjacent tensors are the geometrical indexes. This MPO is also called the \textit{transfer operator} or \textit{transfer MPO} of the TN. The converged MPS\index{MPS} is in fact the dominant eigenstate of the MPO \footnote{For simplicity, we assume the MPO gives an Hermitian operator so that its eigenstates and eigenvalues are well-defined.}. While the MPO represents a physical Hamiltonian or the imaginary-time evolution operator (see Sec. \ref{sec-phys2TN}), the MPS is the ground state. For more general situations, e.g., the TN represents a 2D partition function or the inner product of two 2D PEPS's, the MPS can be understood as the \textit{boundary state} of the TN (or the PEPS\index{PEPS}) \cite{CPSV17MPO, SPCP13boundary, CPSV11boundaryMPS}. The contraction of the 2D infinite TN becomes computing the boundary state, i.e., the dominant eigenstate (and eigenvalue) of the transfer MPO.

The boundary-state scheme gives several non-trivial physical and algorithmic implications \cite{ RPLLS17Scaling2D,CPSV17MPO, YLPVV+14boundary,SPCP13boundary, CPSV11boundaryMPS}, including the underlying resemblance between iTEBD\index{iTEBD} and the famous infinite DMRG (iDMRG)\index{iDMRG} \cite{M08iDMRGarxiv}. DMRG \cite{W92DMRG, W93DMRG} follows the idea of Wilson's NRG\index{NRG} \cite{W75NRGRev}, and solves the ground states and low-lying excitations of 1D or quasi-1D Hamiltonians (see several reviews \cite{S11DMRGRev, SW12DMRG2DRev, S05DMRGrev, CS11DMRGrevChem}); originally it has no direct relations to TN contraction problems. After the MPS and MPO become well understood, DMRG\index{DMRG} was re-interpreted in a manner that is more close to TN (see a review by Schollw\"ock \cite{S11DMRGRev}). In particular for simulating the ground states of infinite-size 1D systems, the underlying connections between the iDMRG and iTEBD were discussed by McCulloch \cite{M08iDMRGarxiv}. As argued above, the contraction of a TN can be computed by solving the the dominant eigenstate of its transfer MPO. The eigenstates reached by iDMRG and iTEBD are the same state up to a gauge transformation (note the gauge degrees of freedom of MPS will be discussed in Sec. \ref{sec.gauge}). Considering that DMRG mostly is not used to compute TN contractions and there are already several understanding reviews, we skip the technical details of the DMRG algorithms here. One may refer to the papers mentioned above if interested. However, later we will revisit iDMRG in the clue of multi-linear algebra.

Variational matrix-product-state (VMPS)\index{VMPS} method is a variational version of DMRG\index{DMRG} for (but not limited to) calculating the ground states of 1D systems with periodic boundary condition \cite{VPC04DMRGQinfo}. Compared with DMRG, VMPS is more directly related to TN\index{TN} contraction problems. In the following, we explain VMPS by solving the contraction of the infinite square TN. As discussed above, it is equivalent to solve the dominant eigenvector (denoted by $|\psi \rangle$) of the infinite MPO\index{MPO} (denoted by $\hat{rho}$) that is formed by a row of tensors in the TN. The task is to minimize $\langle \psi| \hat{\rho} |\psi \rangle$ under the constraint $\langle \psi| \psi \rangle = 1$. The eigenstate $|\psi \rangle$ written in the form of an MPS\index{MPS}.

\begin{figure}[tbp]
	\centering
	\includegraphics[angle=0,width=\linewidth]{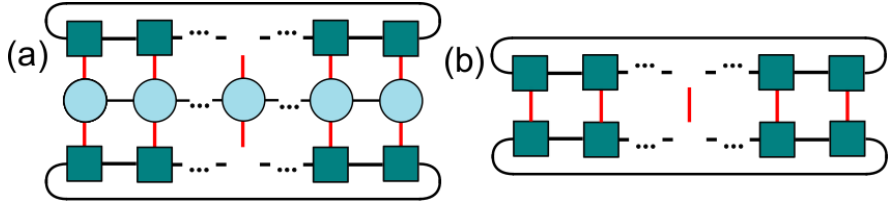}
	\caption{(Color online) The illustration of (a) $\hat{H}^{eff}$ and (b) $\hat{N}^{eff}$ in the variational matrix product state method}
	\label{fig-3vMPS}
\end{figure}

The tensors in $|\psi \rangle$ are optimized on by one. For instance, to optimize the $n$-th tensor, all other tensors are kept unchanged and considered as constants. Such a local minimization problem becomes $\hat{H}^{eff} |T_n\rangle = \mathcal{E} \hat{N}^{eff} |T_n\rangle$ with $\mathcal{E}$ the eigenvalue. $\hat{H}^{eff}$ is given by a 6-th order tensor defined by contracting all tensors in $\langle \psi| \hat{\rho} |\psi \rangle$ except for the $n$-th tensor and its conjugate [Fig. \ref{fig-3vMPS} (a)]. Similarly, $\hat{N}^{eff}$ is also given by a 6-th order tensor defined by contracting all tensors in $\langle \psi| \psi \rangle$ except for the $n$-th tensor and its conjugate [Fig. \ref{fig-3vMPS} (b)]. Again, the VMPS is different from the MPS obtained by TEBD\index{TEBD} only up to a gauge transformation.

Note that the boundary-state methods are not limited to solving TN contractions. An example is the time-dependent variational principle (TDVP)\index{TDVP}. The basic idea of TDVP was proposed by Dirac in 1930 \cite{dirac1930TDVP}, and then it was cooperated with the formulation of Hamiltonian \cite{KERMAN1976TDVP} and action function \cite{JACKIW1979TDVP}. For more details, one could refer to a review by Langhoff \textit{et al} \cite{LEK72TDVP}. In 2011, TDVP was developed to simulate the time evolution of many-body systems with the help of MPS \cite{HCOPVV11TDVP}. Since TDVP (and some other algorithms) concerns directly a quantum Hamiltonian instead of the TN\index{TN} contraction, we skip giving more details of these methods in this paper.

\section{Transverse contraction and folding trick}
\label{transverse}

\begin{figure}[tbp] 
	\centering
	\includegraphics[width=0.9\columnwidth]{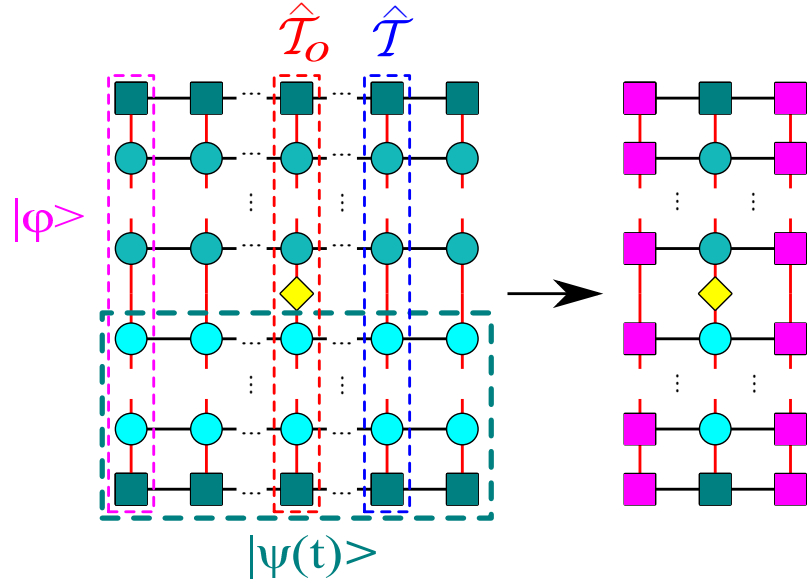}
	\caption{Transverse contraction of the TN\index{TN} for a local expectation value $\langle O(t) \rangle$.}
	\label{fig:folding}
\end{figure}

For the boundary-state methods introduced above, the boundary states are defined in the real space. Taking iTEBD\index{iTEBD} for the real-time evolution as an example, the contraction is implemented along the time direction, which is to do the time evolution in an explicit way. It is quite natural to consider implementing the contraction along the other direction. In the following, we will introduce the transverse contraction and the folding trick proposed and investigated in Refs. \cite{BHVC09folding, MCB12folding, HM15folding}. The motivation of transverse contraction is to avoid the explicit simulation of the time-dependent state $|\psi(t)\rangle$ that might be difficult to capture due to the fast growth of its entanglement.

\textbf{\textit{Transverse contraction}}. Let us consider to calculate the average of a one-body operator $o(t) = \langle \psi(t)| \hat{o} |\psi(t)\rangle$ with $|\psi(t)\rangle$ that is a quantum state of infinite size evolved to the time $t$. The TN\index{TN} representing $o(t)$ is given in the left part of Fig. \ref{fig:folding}, where the green squares give the initial MPS $|\psi(0)\rangle$ and its conjugate, the yellow diamond is $\hat{o}$, and the TN formed by the green circles represents the evolution operator $e^{it\hat{H}}$ and its conjugate (see how to define the TN in Sec. \ref{sec.2.evolution}). 

To perform the transverse contraction, we treat each column of the TN\index{TN} as an MPO\index{MPO} $\hat{\mathcal{T}}$. Then as shown in the right part of Fig. \ref{fig:folding}, the main task of computing $o(t)$ becomes to solve the dominant eigenstate $|\phi \rangle$ (normalized) of $\hat{\mathcal{T}}$, which is an MPS\index{MPS} illustrated by the purple squares. One may solve this eigenstate problems by any of the boundary-state methods (TEBD\index{TEBD}, DMRG\index{DMRG}, etc.). With $|\phi\rangle$, $o(t)$ can be exactly and efficiently calculated as
\begin{equation}
o(t) = \frac{\langle \psi (t) | \hat{o} | \psi (t) \rangle }{\langle \psi (t) | \psi (t) \rangle} = \frac{\langle \phi | \hat{\mathcal{T}}_o | \phi \rangle }{\langle \phi | \hat{\mathcal{T}} | \phi \rangle},
\end{equation}
with $\hat{\mathcal{T}}_o$ is the column that contains the operator $\hat{o}$. Note that the length of $|\phi\rangle$ (i.e., the number of tensors in the MPS\index{MPS}) is proportional to the time $t$, thus one should use the finite-size versions of the boundary-state methods. It should also be noted that $\hat{\mathcal{T}}$ may not be Hermitian. In this case, one should not use $|\phi \rangle$ and its conjugate, but compute the left and right eigenstates of $\hat{\mathcal{T}}$ instead.

Interestingly, similar ideas of the transverse contraction appeared long before the concept of TN emerged. For instance, transfer matrix renormalization group (TMRG)\index{TMRG} \cite{BXG96TMRG, WX97TMRG, S97TMRG, N95TMRG2Dclassic} can be used to simulate the finite-temperature properties of a 1D system. The idea of TMRG is to utilize DMRG\index{DMRG} to calculate the dominant eigenstate of the transfer matrix (similar to $\mathcal{T}$). In correspondence with the TN\index{TN} terminology, it is to use DMRG to compute $|\phi \rangle$ from the TN that defines the imaginary-time evolution. We will skip of the details of TMRG since it is not directly related to TN. One may refer the related references if interested.

\begin{figure}[tbp] 
	\centering
	\includegraphics[width=1\linewidth]{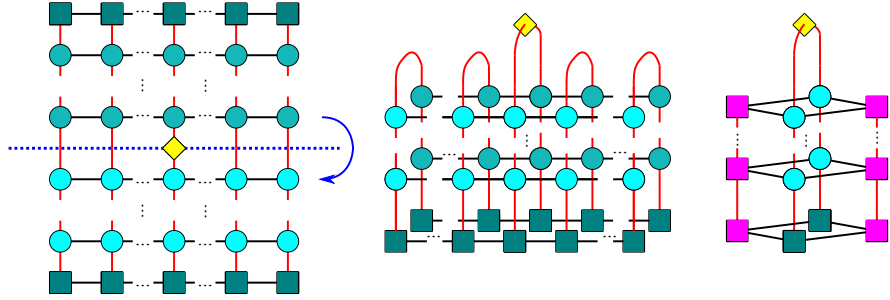}
	\caption{The illustration of the folding trick.}
	\label{fig:folding2}
\end{figure}

\textbf{\textit{Folding trick}}. The main bottleneck of a boundary-state method concerns the entanglement of the boundary state. In other words, the methods will become inefficient when the entanglement of the boundary state grows too large. One example is the real-time simulation of a 1D chain, where the entanglement entropy increases linearly with time. Solely with the transverse contraction, it will not essentially solve this problem. Taking the imaginary-time evolution as an example, it has been shown that with the dual symmetry of space and time, the boundary states in the space and time directions possess the same entanglement \cite{HM15folding, TTLR18tMPS}. 

In Ref. \cite{BHVC09folding}, the folding trick was proposed. The idea is to ``fold'' the TN\index{TN} before the transverse contraction (Fig. \ref{fig:folding2}). In the folded TN, each tensor is the tensor product of the original tensor and its conjugate. The length of the folded TN in the time direction is half of the original TN, and so is the length of the boundary state.

The previous work (Ref. \cite{BHVC09folding}) on the dynamic simulations of 1D spin chains showed that the entanglement of the boundary state is in fact reduced compared with that of the boundary state without folding. This suggests that the folding trick provides a more efficient representation of the entanglement structure of the boundary state. The authors of Ref. \cite{BHVC09folding} suggested an intuitive picture to understand the folding trick. Consider a product state as the initial state at $t-0$ and a single localized excitation at the position $x$ that propagates freely with velocity $v$. By evolving for a time $t$, only $(x \pm vt)$ sites will become entangled. With the folding trick, the evolutions (that are unitary) besides the $(x \pm vt)$ sites will not take effects since they are folded with the conjugates and become identities. Thus the spins outside $(x \pm vt)$ will remain product state and will not contribute entanglement to the boundary state. In short, one key factor to consider here is the entanglement structure, i.e., the fact that the TN is formed by unitaries. The transverse contraction with the folding trick is a convincing example to show that the efficiency of contracting a TN can be improved by properly designing the contraction way according to the entanglement structure of the TN.

\section{Relations to exactly contractible tensor networks and entanglement renormalization}

\begin{figure}[tbp]
	\centering
	\includegraphics[angle=0,width=0.85\linewidth]{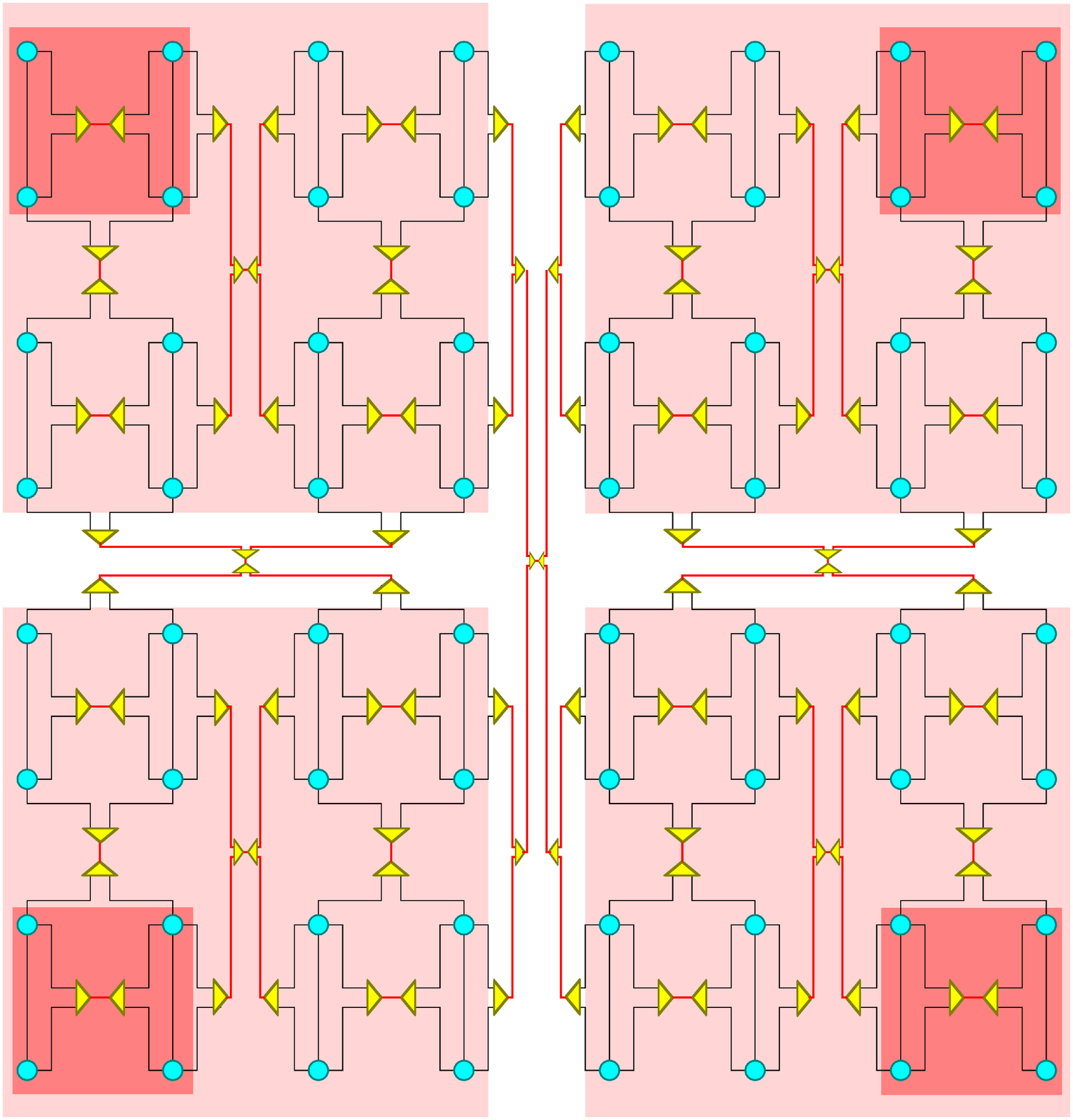}
	\caption{(Color online) The exactly contractible TN in the HOTRG algorithm.}
	\label{fig-3HOTRGexact}
\end{figure}

The TN algorithms explained above are aimed at dealing with contracting optimally the TN's that cannot be exactly contracted. Then a question rises: is a classical computer really able to handle these TN's? In the following, we show that by explicitly putting the isometries for truncations inside, the TN's that are contracted in these algorithms become eventually exactly contractible, dubbed as exactly contractible TN (ECTN)\index{ECTN}. Different algorithms lead to different ECTN. That means the algorithm will show a high performance if the TN can be accurately approximated by the corresponding ETNC.

Fig. \ref{fig-3HOTRGexact} shows the ECTN emerging in the plaquette renormalization \cite{WKS11TRG} or higher-order TRG (HOTRG)\index{HOTRG}\index{TRG} algorithms \cite{XCQZYX12HOSRG}. Take the contraction of a TN\index{TN} (formed by the copies of tensor $T$) on square lattice as an example. In each iteration step, four nearest-neighbor $T$'s in a square are contracted together, which leads to a new square TN formed by tensors ($T^{(1)}$) with larger bond dimensions. Then, isometries (yellow triangles) are inserted in the TN to truncate the bond dimensions (the truncations are in the same spirit of those in CTMRG\index{CTMRG}, see Fig. \ref{fig-3CTMRG}). Let us not contract the isometries with the tensors, but leave them there inside the TN. Still, we can move on to the next iteration, where we contract four $T^{(1)}$'s (each of which is formed by four $T$ and the isometries, see the darked-red plaques in Fig. \ref{fig-3HOTRGexact}) and obtain more isometries for truncating the bond dimensions of $T^{(1)}$. By repeating this process for several times, one can see that tree TN's appear on the boundaries of the coarse-grained plaques. Inside the 4-by-4 plaques (light red shadow), we have the two-layer tree TN's formed by three isometries. In the 8-by-8 plaques, the tree TN has three layers with seven isometries. These tree TN's separate the original TN into different plaques, so that it can be exactly contracted, similar to the fractal TN's introduced in Sec. \ref{Sec-exactTN}.

\begin{figure}[tbp]
	\centering
	\includegraphics[angle=0,width=0.7\linewidth]{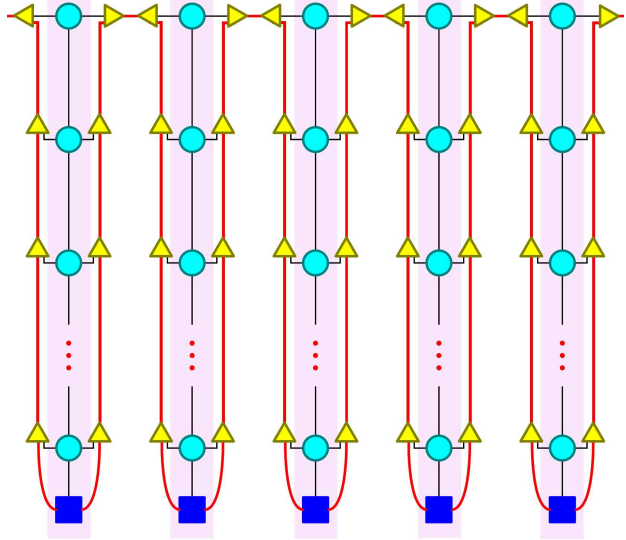}
	\caption{(Color online) The exactly contractible TN in the iTEBD algorithm.}
	\label{fig-3iTEBDexact}
\end{figure}

In the iTEBD\index{iTEBD} algorithm \cite{V03TEBD,V04TEBD,V07iTEBD,OV08canonical} (Fig. \ref{fig-3iTEBDexact}), one starts with an initial MPS\index{MPS} (dark blue squares). In each iteration, one tensor (light blue circles) in the TN\index{TN} is contracted with the tensor in the MPS and then the bonds are truncated by isometries (yellow triangles). Globally seeing, the isometries separate the TN into many ``tubes'' (red shadow) that are connected only at the top. The length of the tubes equals to the number of the iteration steps in iTEBD. Obviously, this TN is exactly contractible. Such a tube-like structure also appears in the contraction algorithms based on PEPS.

\begin{figure}[tbp]
	\centering
	\includegraphics[angle=0,width=0.7\linewidth]{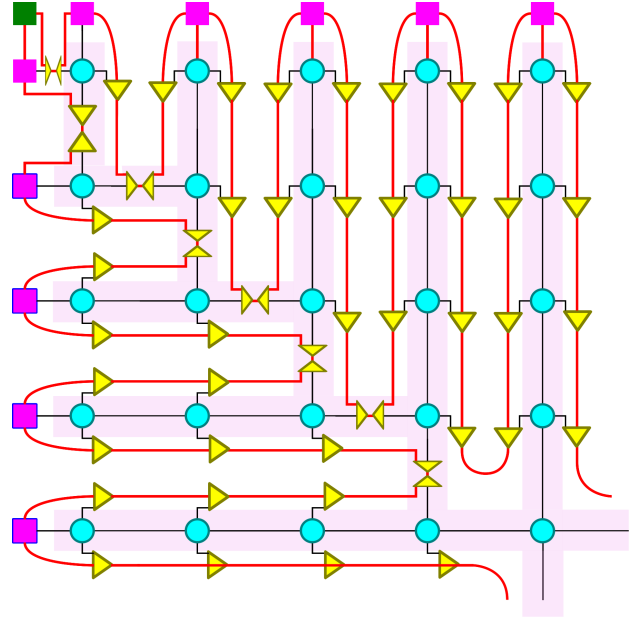}
	\caption{(Color online) A part of the exactly contractible TN in the CTMRG algorithm.}
	\label{fig-3CTMRGexact}
\end{figure}

For the CTMRG\index{CTMRG} algorithm \cite{OV09CTMRG}, the corresponding ECTN\index{ECTN} is a little bit complicated (see one quarter of it in Fig. \ref{fig-3CTMRGexact}). The initial row (column) tensors and the corner transfer matrices are represented by the pink and green squares. In each iteration step, the tensors (light blue circles) located most outside are contracted to the row (column) tensors and the corner transfer matrices, and isometries are introduced to truncate the bond dimensions. Globally seeing the picture, the isometries separate the TN into a tree-like structure (red shadow), which is exactly contractible.

For these three algorithms, each of them gives an ECTN that is formed by two part: the tensors in the original TN and the isometries that make the TN exactly contractible. After optimizing the isometries, the original TN is approximated by the ECTN. The structure of the ECTN depends mainly on the contraction strategy and the way of optimizing the isometries depend on the chosen environment.

The ECTN picture shows us explicitly how the correlations and entanglement are approximated in different algorithms. Roughly speaking, the correlation properties can be read from the minimal distance of the path in the ECTN that connects two certain sites, and the (bipartite) entanglement can be read from the number of bonds that cross the boundary of the bipartition. How well the structure suits the correlations and entanglement should be a key factor of the performance of a TN contraction algorithm. Meanwhile, this picture can assist us to develop new algorithms by designing the ECTN and taking the whole ECTN as the environment for optimizing the isometries. These issues still need further investigations.

The unification of the TN contraction and the ECTN\index{ECTN} has been explicitly utilized in the TN renormalization (TNR)\index{TNR} algorithm \cite{EV15TNR,EV15TNRmera}, where both isometries and unitaries (called \textit{disentangler}) are put into the TN\index{TN} to make it exactly contractible. Then instead of tree TN's or MPS's\index{MPS}, one will have MERA's\index{MERA} (see Fig. \ref{fig-1TTNS} (c) for example) inside which can better capture the entanglement of critical systems.

\section{A shot summary}

In this section, we have discussed about several contraction approaches for dealing with 2D TN's. Applying these algorithms, many challenging problems can be efficiently solved, including the ground-state and finite-temperature simulations of 1D quantum systems, and the simulations of 2D classical statistic models. Such algorithms consist of two key ingredients: contractions (local operations of tensors) and truncations. The local contraction determines the way how the TN is contracted step by step, or in other words, how the entanglement information is kept according to the ECTN\index{ECTN} structure. Different (local or global) contractions may lead to different computational costs, thus optimizing the contraction sequence is necessary in many cases \cite{BHVC09folding,EP14TNalgo,PHV14TNCorder}. The truncation is the approximation to discard less important basis so that the computational costs are properly bounded. One essential concept in the truncations is ``environment'', which plays the role of the reference when determining the weights of the basis. Thus, the choice of environment concerns the balance between the accuracy and efficiency of a TN\index{TN} algorithm.

\chapter{Tensor network approaches for higher-dimensional quantum lattice models}
\label{sec4}

\abstract{In this section, we will show several representative TN\index{TN} approaches for simulating the quantum lattice models in ($d>1$) dimensions. We will mainly use the language of TN contractions. One may refer to several existing reviews \cite{S11DMRGRev, VMC08MPSPEPSRev, CV09TNSRev, O14TNSRev, HV17TMTNrev} for more exhaustive understanding on the TN simulations for quantum problems.  We will focus on the algorithms based on PEPS\index{PEPS}, and show the key roles that the 2D TN contraction algorithms presented in Sec. \ref{sec3} play in the higher-dimensional cases.}

\section{Variational approaches of projected-entangled pair state}

\begin{figure}[tbp]
	\centering
	\includegraphics[angle=0,width=0.6\linewidth]{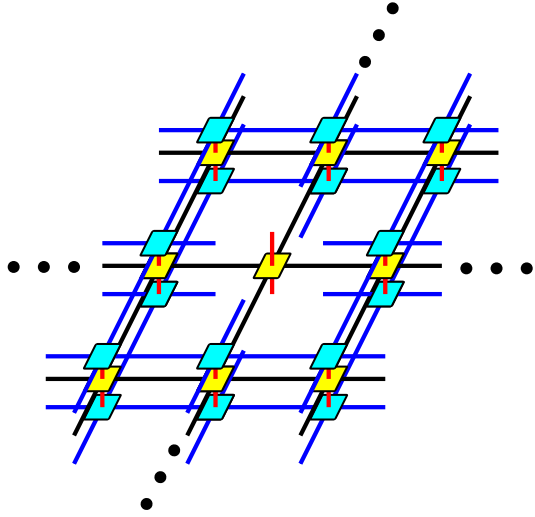}
	\caption{(Color online) The illustration of $\hat{H}^{eff}$ in Eq. (\ref{eq-4Heff}).}
	\label{fig-variational_PEPS}
\end{figure}

Without losing generality, we consider a 2D quantum system with nearest-neighbor coupling on an infinite square lattice as an example. The ground state can be represented by an iPEPS\index{iPEPS} (see Sec. \ref{sec.peps}). Similar to MPS\index{MPS} (Sec. \ref{sec.2.evolution}), the central task is to minimize the energy
\begin{eqnarray}
E= \frac{\langle \psi | \hat{H} | \psi \rangle}{ \langle \psi | \psi \rangle}.
\label{eq-4Eg}
\end{eqnarray}
There are in general two ways to do the minimization. One way proposed firstly by Verstraete and Cirac \cite{VC06PEPSArxiv} is considering the elements in the tensors as variational parameters. The tensors in the TN are updated one by one. In a similar spirit as the boundary-state methods (see Sec. \ref{iTEBD}), the key of this approach is to transform the global minimization to local ones, where one tensor (say $P^{[i]}$, see the PEPS form in Eq. (\ref{eq-2PEPSsimple}), Sec. \ref{sec.peps}) is updated by a local minimization problem
\begin{eqnarray}
E = \frac{P^{[i]\dagger} \hat{H}^{eff} P^{[i]}}{P^{[i]\dagger} \hat{N}^{eff} P^{[i]}}.
\label{eq-4Heff}
\end{eqnarray}
$\hat{H}^{eff}$ is an ``effective'' Hamiltonian by computing $\langle \psi | \hat{H} | \psi \rangle$ but after taking $P^{[i]\dag}$ in $\langle \psi |$ and $P^{[i]}$ in $| \psi \rangle$ out. Fig. \ref{fig-variational_PEPS} depicts $\hat{H}^{eff}$ where $\hat{H}$ is written as an infinite PEPO\index{PEPO} (iPEPO\index{iPEPO}, also see Sec. \ref{sec-MPO} for PEPO) for a better illustration. Similarly, $\hat{N}^{eff}$ is defined by computing $\langle \psi | \psi \rangle$ but after taking $P^{[i]\dag}$ and $P^{[i]}$ out.

Obviously, the computations of both $\hat{H}^{eff}$ and $\hat{N}^{eff}$ are in fact to contract the corresponding 2D TN's where the 2D TN\index{TN} contraction algorithms are needed. In \cite{C16vPEPS}, Corboz used CTMRG\index{CTMRG} (see \cite{NO96CTMRG0} or Sec. \ref{sec.CTMRG}) to compute the contractions. In \cite{VHCV16GrdPEPS}, Vanderstraeten \textit{et al} further developed this idea to a gradient method, where the gradient is calculated by implementing similar 2D TN contractions. The gradient is given as
\begin{eqnarray}
\frac{\partial E}{\partial P^{[i]\dag}} = \frac{\partial \langle \psi | \hat{H} | \psi \rangle / \langle \psi | \psi \rangle}{\partial P^{[i]\dag}} \nonumber = 2 \frac{\partial_{P^{[i]\dag}} \langle \psi | \hat{H} | \psi \rangle}{\langle \psi | \psi \rangle} - 2 \frac{\langle \psi | \hat{H} | \psi \rangle}{\langle \psi | \psi \rangle^2} \partial_{P^{[i]\dag}} \langle \psi | \psi \rangle.
\label{eq-4gradient}
\end{eqnarray}
By imposing the normalization condition $\langle \psi | \psi \rangle = 1$ and shifting the ground-state energy to zero by $\hat{H} \leftarrow \hat{H} - \langle \psi | \hat{H} | \psi \rangle$, the gradient is simplified as 
\begin{eqnarray}
\frac{\partial E}{\partial P^{[i]\dag}} = 2 \partial_{P^{[i]\dag}} \langle \psi | \hat{H} | \psi \rangle.
\label{eq-4gradient_simp}
\end{eqnarray}
Thus the gradient is computed by contracting the TN of $\langle \psi | \hat{H} | \psi \rangle$ after taking $P^{[i]\dag}$ out.

The gradient method is consistent with the effective Hamiltonian schemes. In fact, one has $\frac{\partial E}{\partial P^{[i]\dag}} = 2\hat{H}^{eff} P^{[i]}$. At the minimal point, the gradient should vanish $\frac{\partial E}{\partial P^{[i]\dag}} = 0$. It means $2\hat{H}^{eff} P^{[i]} = 0$, i.e., $P^{[i]}$ is the dominant eigenstate of $\hat{H}^{eff}$ with a zero eigenvalue. Considering the ground-state energy is shifted to zero, $P^{[i]}$ is the ground state of the effective Hamiltonian $\hat{H}^{eff}$.

Note that the infinite PEPO (iPEPO)\index{iPEPO} representation is not enforced to define $\hat{H}^{eff}$. In fact, it is not easy to obtain the iPEPO of an arbitrary 2D (or 3D) Hamiltonian. The usual way is to start from the summation form of the Hamiltonian $\hat{H} = \sum \hat{H}_{ij}$, and compute the contribution to $\hat{H}^{eff}$ from each $\hat{H}_{ij}$ separately \cite{C16vPEPS}. Each term is computed by contracting a 2D TN, where one can reuse the results to improve the efficiency.

Following the same clue (minimizing $E$), algorithms were proposed to combine TN with the QMC\index{QMC} methods \cite{SV09TNQMC, SWVC08QMCTN, SCSC10QMCTN, WPV11TNQMC, LDHGH17TNQMC}. Still let us focus on those based on PEPS\index{PEPS}. One may transform Eq. (\ref{eq-4Eg}) as
\begin{eqnarray}
E = \frac{\sum_{S,S'} W(S') \langle S'|\hat{H}| S\rangle W(S)}{\sum_{S} W(S)^2},
\end{eqnarray}
where $S = (s_1, s_2, \cdots)$ goes through all spin configurations and $W(S) = \langle s_1 s_2 \cdots| \psi \rangle$ is the coefficient of the iPEPS for the given configuration. QMC sampling can be implemented by defining the weight function as $W(S)^2$ and the estimator $E(S)$ as
\begin{eqnarray}
E(S) = \sum_{S'}  \frac{W(S')}{W(S)} \langle S'|\hat{H}| S\rangle,
\end{eqnarray}
so that the energy becomes
\begin{eqnarray}
E = \langle E(S) \rangle = \sum_{S} W(S)^2 E(S).
\end{eqnarray}
It is easy to see that the normalization condition of the weights $\sum_S W(S)^2 = 1$ is satisfied.

The task becomes to compute $W(S)$ and $\langle S'|\hat{H}| S\rangle$ with different configurations. The computation of $\langle S'|\hat{H}| S\rangle$ is relatively easy since $|S\rangle$ and $|S'\rangle$ are just two product states. The computation of $W(S)$ is more tricky. When $|\psi \rangle$ is a PEPS on a square lattice, $W(S)$ is a 2D scalar TN\index{TN} by fixing all the physical indexes of the PEPS\index{PEPS} as
\begin{eqnarray}
W(S) = \text{tTr} \prod_{n} P^{[n]}_{s_n},
\label{eq-4TNQMCW}
\end{eqnarray}
where $P^{[n]}_{s_n}$ is a forth-order tensor that only has the geometrical index \footnote{One may refer to Eq. (\ref{eq-2PEPSsimple}) to better understand Eq. (\ref{eq-4TNQMCW})}. The $n$-th physical index is taken as $s_n$. Considering that most of the configurations are not translationally invariant, such QMC-TN\index{QMC} methods are usually applied to finite-size models. One may use the finite-TN version of the algorithms reviewed in Sec. \ref{sec3}.

\section{Imaginary-time evolution methods}
\label{sec.evo2D}

Another way to compute the ground-state iPEPS is to do imaginary-time evolution, analog to the MPS methods presented in Sec. \ref{sec.2.evolution}. For a $d$-dimensional quantum model, its ground-state simulation can be considered as computing the contraction of a ($d+1$)-dimensional TN. 

\begin{figure}[tbp]
	\centering
	\includegraphics[angle=0,width=0.8\linewidth]{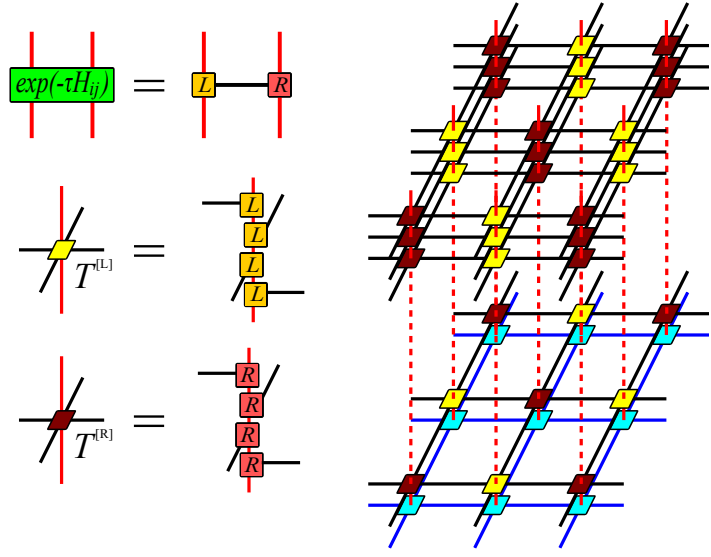}
	\caption{(Color online) The evolution of a PEPS\index{PEPS} can be mapped to the contraction of a 3D TN\index{TN}.}
	\label{fig-TPO_iEV}
\end{figure}

Firstly, let us show how the evolution operator for an infinitesimal imaginary-time step $\tau$ can be written as an iPEPO, which is in fact one layer of the 3D TN (Fig. \ref{fig-TPO_iEV}). The evolution of the iPEPS is put the iPEPS at the bottom and to contract the TN layer by layer to it.

To proceed, we divide the local Hamiltonians on the square lattice into four group: $\hat{H}_{e,e} = \sum_{\text{even i,j}} \hat{H}^{[i,j;i,j+1]} + \hat{H}^{[i.j;i+1,j]}$, $\hat{H}_{o,o} = \sum_{\text{odd i,j}} \hat{H}^{[i,j;i,j+1]} + \hat{H}^{[i.j;i+1,j]}$, $\hat{H}_{e,o} = \sum_{\text{even i, odd j}} \hat{H}^{[i,j;i,j+1]} + \hat{H}^{[i.j;i+1,j]}$ and $H_{o,e} = \sum_{\text{odd i, even j}} \hat{H}^{[i,j;i,j+1]} + \hat{H}^{[i.j;i+1,j]}$. One can see that each two terms in one group commute to each other. The evolution operator for an infinitesimal imaginary-time step ($\tau \to 0$) then can be written as
\begin{equation}
\hat{U} = \exp(-\tau \hat{H})
= \exp(-\tau \hat{H}^{[e,e]}) \exp(-\tau \hat{H}^{[o,o]}) \exp(-\tau \hat{H}^{[e,o]}) \exp(-\tau \hat{H}^{[o,e]}) + O(\tau^2).
\label{eq-4evolU}
\end{equation}

Let us assume translational invariance to the Hamiltonian, i.e., $\hat{H}^{[i,j]} = \hat{H}^{[two]}$. The element of two-body evolution operator is a forth-order tensor $U_{s_is_js_i's_j'} = \langle s_i's_j'| \exp(-\tau \hat{H}^{[two]}) |s_is_j \rangle$. Implement SVD\index{SVD} or QR\index{QR} decomposition on $U$ (\ref{fig-TPO_iEV}) as
\begin{eqnarray}
U_{s_is_js_i's_j'} = \sum_{\alpha} L_{s_is_i', a} R_{s_js_j',a}.
\end{eqnarray}
Then the two tensors $T^{[L]}$ and $T^{[R]}$ that form the iPEPO\index{iPEPO} of $\hat{U}$ is obtained as
\begin{eqnarray}
\begin{aligned}
T^{[L]}_{ss', a_1a_2a_3a_4} = \sum_{s_1s_2s_3} L_{ss_1,a_1} L_{s_1s_2,a_2} L_{s_2s_3, a_3} L_{s_3s',a_4}, \\
T^{[R]}_{ss', a_1a_2a_3a_4} = \sum_{s_1s_2s_3} R_{ss_1,a_1} R_{s_1s_2,a_2} R_{s_2s_3, a_3} R_{s_3s',a_4}.
\label{eq-4PEPSevolve}
\end{aligned}
\end{eqnarray}
The four $L$'s (or $R$'s) in $T^{[L]}$ (or $T^{[R]}$) correspond to the evolution operators of the two-body terms in $\hat{H}^{[e,e]}$, $\hat{H}^{[o,o]}$, $\hat{H}^{[e,o]}$, and $\hat{H}^{[o,e]}$ in Eq. (\ref{eq-4evolU}), respectively (see the left part of Fig. \ref{fig-TPO_iEV}).

While the TN for the imaginary-time evolution with the iPEPO\index{iPEPO} is a cubic TN\index{TN}, one may directly use the tensor $U$, which also gives a 3D but not cubic TN. Without losing generality, we in the following will use the iPEPO to present the algorithms for contraction a cubit TN. The algorithm can be readily applied to deal with the statistic models on cubic lattice or other problems that can be written as the contraction of a cubic TN.

The evolution $\hat{U} |\psi \rangle$ is to contract the iPEPO (one layer of the tensors) to the iPEPS\index{iPEPS}. In accordance to the translational invariance of the iPEPO, the iPEPS is also formed by two inequivalent tensors (denoted by $P^{[L]}$ and $P^{[R]}$). Locally, the tensors in the evolved iPEPS are given as
\begin{eqnarray}
\tilde{P}^{[L]}_{s, \tilde{\alpha}_1\tilde{\alpha}_2\tilde{\alpha}_3\tilde{\alpha}_4} = \sum_{s'}  T^{[L]}_{ss', a_1a_2a_3a_4} P^{[L]}_{s', \alpha_1\alpha_2\alpha_3\alpha_4}, \\
\tilde{P}^{[R]}_{s, \tilde{\alpha}_1\tilde{\alpha}_2\tilde{\alpha}_3\tilde{\alpha}_4} = \sum_{s'}  T^{[R]}_{ss', a_1a_2a_3a_4} P^{[R]}_{s', \alpha_1\alpha_2\alpha_3\alpha_4},
\end{eqnarray}
with the composite indexes $\tilde{\alpha}_x = (a_x, \alpha_x)$ ($x = 1, 2, 3, 4$). Obviously, the bond dimensions of the new tensors are increased by $\dim(a_x)$ times. It is necessary to preset a dimension cut-off $\chi$: when the bond dimensions become larger than $\chi$, approximations will be introduced to reduce the dimensions back to $\chi$. One then can iterate the evolution of the iPEPS with bounded computational cost. After the iPEPS converges, it is considered that the ground state is reached. Therefore, one key step in the imaginary-time schemes (as well as the similar contraction schemes of 3D TN's\index{TN}) is to find the optimal truncations of the enlarged bonds. In the following, we will concentrate on the truncation of bond dimensions, and present three kinds of scheme known as \textit{full}, \textit{simple}, and \textit{cluster} updates according to which environment the truncations are optimized \cite{LCB14PEPScontract} \footnote{The definition of the update schemes also apply to finite-size PEPS and the variational methods; for example, the variational methods which contract the whole TN to update the (i)PEPS are also called full update.}.

\section{Full, simple, and cluster update schemes}
\label{sec.updateschemes}

For truncating the dimensions of the geometrical bonds of an iPEPS\index{iPEPS}, the task is to minimize the distance between the iPEPS's before and after the truncation, i.e.,
\begin{equation}
\mathcal{\varepsilon} = || \tilde{\psi} \rangle - |\psi \rangle|.
\end{equation}
With the normalization condition of the iPEPS's, the problem can be reduced to the maximization of the fidelity
\begin{equation}
\mathcal{Z} = \langle \tilde{\psi} |\psi \rangle.
\label{eq-4fidtrun}
\end{equation}
As discussed in Sec. \ref{sec.QobTN}, $\mathcal{Z}$ is in fact a scalar TN\index{TN}.

\textbf{Full update.} Among the three kinds of update schemes, full update seems to be the most natural and reasonable, in which the truncation is optimized referring to the whole iPEPS \cite{XCQZYX12HOSRG, OV09CTMRG, XJCWX09SRG, LCB14PEPScontract, O12CTMRG, JOVVC08PEPS, PBTCO15FastFullUpdate}. Les us consider a translationally invariant iPEPS. For square lattice, the iPEPS is formed by the infinite copies of two tensors $P^{[L]}$ and $P^{[R]}$ located on the two sub-lattices, respectively. Their evolution is given by Eq. (\ref{eq-4PEPSevolve}). We use $\tilde{P}^{[L]}$ and $\tilde{P}^{[R]}$ to denote the tensors with enlarged bond dimensions. Below, we follow Ref. \cite{XJCWX09SRG} to explain the truncation process. To truncate the forth bond $\tilde{\alpha}_4$ of the tensor for example, one firstly defines the tensor $M$ by contracting a pair of $\tilde{P}^{[L]}$ and $\tilde{P}^{[R]}$ as
\begin{equation}
M_{s_1\tilde{\alpha}_1 \tilde{\alpha}_2 \tilde{\alpha}_3, s_2  \tilde{\alpha}_1' \tilde{\alpha}_2' \tilde{\alpha}_3'} = \sum_{\tilde{\alpha}_4} \tilde{P}^{[L]}_{s_1, \tilde{\alpha}_1 \tilde{\alpha}_2 \tilde{\alpha}_3 \tilde{\alpha}_4} \tilde{P}^{[R]}_{s_2, \tilde{\alpha}_1' \tilde{\alpha}_2' \tilde{\alpha}_3' \tilde{\alpha}_4}.
\end{equation}
Note that $P^{[L]}$ and $P^{[R]}$ share the bond $\tilde{\alpha}_4$ that is to be truncated. Compute the \textit{environment tensor} $M^e$ by contracting the TN of $\mathcal{Z}$ after taking a pair of $\tilde{P}^{[L]}$ and $\tilde{P}^{[R]}$ out from the TN. $\mathcal{M}$ is in fact an eighth-order tensor of the same dimensions as $M$. Decompose $M^e$ by SVD as
\begin{equation}
M^e_{s_1\tilde{\alpha}_1 \tilde{\alpha}_2 \tilde{\alpha}_3, s_2  \tilde{\alpha}_1' \tilde{\alpha}_2' \tilde{\alpha}_3'} = \sum_{\alpha} V^{[L]}_{s_1 \tilde{\alpha}_1 \tilde{\alpha}_2 \tilde{\alpha}_3, \alpha} \Lambda_{\alpha} V^{[R]}_{s_2  \tilde{\alpha}_1' \tilde{\alpha}_2' \tilde{\alpha}_3', \alpha}.
\label{eq-4MSVD}
\end{equation}

Define a new matrix as $\tilde{M} = \Lambda^{1/2} V^{[R]} M V^{[L]} \Lambda^{1/2}$ and decompose it by SVD as $\tilde{M} \simeq \tilde{V}^{[L]} \tilde{\Lambda} \tilde{V}^{[R]}$ by taking only the $\chi$-largest singular values and singular vectors. Finally, two tensors are updated by $P^{[L]} = \tilde{\Lambda}^{1/2} \tilde{V}^{[L]T} \Lambda^{-1/2} V^{[R]T}$ and $P^{[R]} = \tilde{\Lambda}^{1/2} \tilde{V}^{[R]} \Lambda^{-1/2} V^{[L]}$. One can check
\begin{equation}
M_{s_1\tilde{\alpha}_1 \tilde{\alpha}_2 \tilde{\alpha}_3, s_2  \tilde{\alpha}_1' \tilde{\alpha}_2' \tilde{\alpha}_3'} \simeq \sum_{\alpha_4} P^{[L]}_{s_1, \tilde{\alpha}_1 \tilde{\alpha}_2 \tilde{\alpha}_3 \alpha_4} P^{[R]}_{s_2, \tilde{\alpha}_1' \tilde{\alpha}_2' \tilde{\alpha}_3' \alpha_4},
\label{eq-4recoverM}
\end{equation}
with the dimension of the shared bond $\dim(\alpha_4) = \chi$. We shall stress that Eq. (\ref{eq-4recoverM}) is not the SVD\index{SVD} of $M$; the decomposition and truncation are optimized by the SVD of $M^e$, hence is a non-local optimization.

With the formula given above, the task is to compute the environment tensor $M^e$ by the contraction algorithms of 2D TN's\index{TN}. In Ref. \cite{XJCWX09SRG}, the authors developed the SRG\index{SRG}, where $M^e$ is computed by a modified version of TRG\index{TRG} algorithm \cite{LN07TRG}. Other options include iTEBD\index{iTEBD} \cite{JOVVC08PEPS}, CTMRG\index{CTMRG} \cite{OV09CTMRG} and etc. Note that how to define the environment as well as how to truncate by the environment may have subtle differences in different works. The spirit is the same, which is to minimize the fidelity in Eq. (\ref{eq-4fidtrun}) referring to the whole iPEPS\index{iPEPS}.

\textbf{Simple update.} A much more efficient way known as the simple update was proposed by Jiang \textit{et al} \cite{JWX08SimpleUpdate}; it uses local environment to determine the truncations, providing an extremely efficient algorithm to simulate the 2D ground states. As shown in Fig. \ref{fig-1PEPSgraph} (c), the iPEPS\index{iPEPS} used in the simple update is formed by the tensors on the site and the spectra on the bonds: two tensors $P^{[L]}$ and $P^{[R]}$ located on the two sub-lattices, and $\lambda^{[1]}$, $\lambda^{[2]}$, $\lambda^{[3]}$, and $\lambda^{[4]}$ on the four inequivalent geometrical bonds of each tensor. The evolution of the tensors in such an iPEPS is given by Eq. (\ref{eq-4PEPSevolve}). $\lambda^{[i]}$ should be simultaneously evolved as $\tilde{\lambda}^{[i]}_{(a_i, \alpha_i)} = I_{a_i} \lambda_{\alpha_i}$ with $I_{a_i} = 1$.

To truncate the forth geometrical bond of $P^{[L]}$ (and $P^{[R]}$), for example, we construct a new tensor by contracting $P^{[L]}$ and $P^{[R]}$ and the adjacent spectra as
\begin{equation}
M_{s_1\tilde{\alpha}_1 \tilde{\alpha}_2 \tilde{\alpha}_3, s_2  \tilde{\alpha}_1' \tilde{\alpha}_2' \tilde{\alpha}_3'} = \sum_{\tilde{\alpha}_4} \tilde{P}^{[L]}_{s_1, \tilde{\alpha}_1 \tilde{\alpha}_2 \tilde{\alpha}_3 \tilde{\alpha}_4} \tilde{P}^{[R]}_{s_2, \tilde{\alpha}_1' \tilde{\alpha}_2' \tilde{\alpha}_3' \tilde{\alpha}_4} \tilde{\lambda}^{[1]}_{\tilde{\alpha}_1} \tilde{\lambda}^{[2]}_{\tilde{\alpha}_2} \tilde{\lambda}^{[3]}_{\tilde{\alpha}_3} \tilde{\lambda}^{[1]\prime}_{\tilde{\alpha}_1'} \tilde{\lambda}^{[2]\prime}_{\tilde{\alpha}_2} \tilde{\lambda}^{[3]\prime}_{\tilde{\alpha}_3} \tilde{\lambda}^{[4]}_{\tilde{\alpha}_4}.
\end{equation}
Then implement SVD\index{SVD} on $M$ as
\begin{equation}
M_{s_1\tilde{\alpha}_1 \tilde{\alpha}_2 \tilde{\alpha}_3, s_2  \tilde{\alpha}_1' \tilde{\alpha}_2' \tilde{\alpha}_3'} \simeq \sum_{\alpha=1}^{\chi} U^{[L]}_{s_1 \tilde{\alpha}_1 \tilde{\alpha}_2 \tilde{\alpha}_3, \alpha} \lambda_{\alpha} U^{[R]}_{s_2  \tilde{\alpha}_1' \tilde{\alpha}_2' \tilde{\alpha}_3', \alpha}.
\end{equation}
where one takes only the $\chi$-largest singular values and the basis. $P^{[L]}$ and $P^{[R]}$ are updated as
\begin{equation}
\begin{aligned}
P^{[L]}_{s_1, \tilde{\alpha}_1 \tilde{\alpha}_2 \tilde{\alpha}_3 \alpha} = U^{[L]}_{s_1 \tilde{\alpha}_1 \tilde{\alpha}_2 \tilde{\alpha}_3, \alpha} (\tilde{\lambda}^{[1]}_{\tilde{\alpha}_1})^{-1} (\tilde{\lambda}^{[2]}_{\tilde{\alpha}_2})^{-1} (\tilde{\lambda}^{[3]}_{\tilde{\alpha}_3})^{-1}, \\
P^{[R]}_{s_2, \tilde{\alpha}_1 \tilde{\alpha}_2 \tilde{\alpha}_3 \alpha} = U^{[R]}_{s_2 \tilde{\alpha}_1 \tilde{\alpha}_2 \tilde{\alpha}_3, \alpha} (\tilde{\lambda}^{[1]}_{\tilde{\alpha}_1})^{-1} (\tilde{\lambda}^{[2]}_{\tilde{\alpha}_2})^{-1} (\tilde{\lambda}^{[3]}_{\tilde{\alpha}_3})^{-1}.
\end{aligned}
\end{equation}
The spectrum $\tilde{\lambda}_4$ is updated by $\lambda$ in the SVD.

The above procedure truncates $\dim(\tilde{\alpha}_4)$ to the dimension cut-off $\chi$, which can be readily applied to truncate any other bonds. According to the discussion about SVD in Sec. \ref{sec.2SVD}, the environment is the two tensors and the adjacent spectra $\lambda$'s in $M$, where the $\lambda$'s play the role of an ``effective'' environment that approximate the true environment ($M^e$ in the full update). From this viewpoint, the simple update uses local environment. Later by borrowing from the idea of the orthogonal form of the iPEPS on Bethe lattices \cite{SDV06TTN, NFGSS08TreeMPS, TEV09TTN, MVLN10TTN, LDX12TTN, NC13TTN, PVK13TTN, MVSNL15TTN}, it was realized that the environment of the simple update is the iPEPS on the infinite trees \cite{RXLS13NCD, RLXZS12ODTNS, RPPSL17AOP3D}, not just several tensors. We will talk about this in detail in the next chapter from the perspective of the multi-linear algebra.

\textbf{Cluster update.} By keeping the same dimension cut-off, the simple update is much more efficient than the full update. On the other hand, obviously, the full update possesses higher accuracy than the simple update by considering better the environment. The cluster update is between the simple and full updates, which is more flexible to balance between the efficiency and accuracy \cite{RXLS13NCD, LCB14PEPScontract, WV11PEPSclusterArxiv}.

One way is to choose a finite cluster of the infinite TN\index{TN} and define the environment tensor by contracting the finite TN after taking a pair of $\tilde{P}^{[L]}$ and $\tilde{P}^{[R]}$ out. One can consider to firstly use the simple update to obtain the spectra and put them on the boundary of the cluster \cite{WV11PEPSclusterArxiv}. This is equivalent to using a new boundary condition \cite{RXLS13NCD, RPPSL17AOP3D}, different from the open or periodic boundary conditions of a finite cluster. Surely, the bigger the cluster becomes, more accurate but more consuming the computation will be. One may also consider an infinite-size cluster, which is formed by a certain number of rows of the tensors in the TN \cite{LCB14PEPScontract}. Again, both the accuracy and computational cost will in general increase with the number of rows. With infinite rows, such a cluster update naturally becomes the full update. Despite the progresses, there are still many open questions, for example, how to best balance the efficiency and accuracy in the cluster update.

\section{Summary of the tensor network algorithms in higher dimensions}

In this section, we mainly focused on the iPEPS\index{iPEPS} algorithm that simulate the ground states of 2D lattice models. The key step is to compute the environment tensor, which is to contract the corresponding TN\index{TN}. For several special cases such as trees and fractal lattices, the environment tensor corresponds to an exactly contractible TN, and thus can be computed efficiently (see Sec. \ref{Sec-exactTN}). For the regular lattices such as square lattice, the environment tensor is computed by the TN contraction algorithms, which is normally the most consuming step in the iPEPS approaches. 

The key concepts and ideas, such as environment, (simple, cluster, and full) update schemes, and the use of SVD\index{SVD}, can be similarly applied to finite-size cases \cite{PWV11tPEPS, LCB14fPEPS}, the finite-temperature simulations \cite{CCD12FTPEPS, RXLS13NCD, RLXZS12ODTNS, CD15TPO, CD15FTPEPS, CDO16TPO, CRD16TPO, CDO17TPOQMC, KREO18PEPOthermal}, and real-time simulations \cite{CDC19TPO, PWV11tPEPS} in two dimensions. The computational cost of the TN approaches is quite sensitive to the spatial dimensions of the system. The simulations of 3D quantum systems are much more consuming than the 2D cases, where the task become to contract the 4D TN. The 4D TN contraction is extremely consuming, one may consider to generalize the simple update \cite{RPPSL17AOP3D, JO18graphPEPS}, or to construct finite-size effective Hamiltonians that mimic the infinite 3D quantum models \cite{RPPSL17AOP3D, RXPS+18FTQES}

Many technical details of the approaches can be flexibly modified according to the problems under consideration. For example, the iPEPO\index{iPEPO} formulation is very useful when computing a 3D statistic model, which is to contract the corresponding 3D TN. To simulating the imaginary-time evolution, to directly use the two-body evolution operators (see, e.g., \cite{JWX08SimpleUpdate, OV09CTMRG}) is normally more efficient than to use the iPEPO.  The environment is not necessarily defined by the tensors; it can be defined by contracting everything of the TN\index{TN} except for the aimed geometrical bond \cite{CCD12FTPEPS, RLXZS12ODTNS}. The contraction order also significantly affects the efficiency and accuracy. One may consider to use the ``single-layer'' picture \cite{LCB14PEPScontract, PWV11tPEPS}, or an ``intersected'' optimized contraction scheme \cite{XLHX+19TNcon}.

\chapter{Tensor network contraction and multi-linear algebra}

\abstract{This chapter is aimed at understanding TN algorithms from the perspective of MLA\index{MLA}. In Sec. \ref{sec5-eig}, we start from a simple example with a 1D TN\index{TN} stripe, which can be ``contracted'' by solving the eigenvalue decomposition of matrices. This relates to several important MPS techniques such as canonicalization \cite{OV08canonical} that enables to implement optimal truncations of the bond dimensions of MPS's\index{MPS} (Sec. \ref{sec5-canon}). In Sec. \ref{sec5-SO}, we discuss about super-orthogonalization \cite{RLXZS12ODTNS} inspired by Tucker decomposition \cite{DDV00HOSVD} in MLA, which is also a higher-dimensional generalization of canonicalization; it is proposed to implement optimal truncations of the iPEPS's defined on trees. In Sec. \ref{sec5-rank1}, we explain based on the rank-1 decomposition  \cite{LMV00Rank1}, that super-orthogonalization in fact provides the ``loopless approximation'' of the iPEPS's\index{iPEPS} on regular lattices \cite{RXLS13NCD}; it explains how the approximations in the simple update algorithm works for the ground-state simulations on 2D regular lattices \cite{JWX08SimpleUpdate}. In Sec. \ref{sec5-TRD}, we will discuss tensor ring decomposition (TRD)\index{TRD} \cite{R16AOP}, which is a rank-$N$ generalization of the rank-1 decomposition. TRD naturally provides a unified description of iDMRG\cite{W92DMRG,W93DMRG,M08iDMRGarxiv}\index{iDMRG}, iTEBD\cite{V07iTEBD}\index{iTEBD}, and CTMRG\index{CTMRG} \cite{OV09CTMRG,FVZHV17fastCTMRG} when considering the contractions of 2D TN's\index{TN}.}

\section{A simple example of solving tensor network contraction by eigenvalue decomposition}

\label{sec5-eig}

As discussed in the previous sections, the TN algorithms are understood mostly based on the linear algebra, such as eigenvalue and singular-value decompositions. Since the elementary building block of a TN is a tensor, it is very natural to think about using the MLA to understand and develop TN algorithms. MLA is also known as tensor decompositions or tensor algebra \cite{KB09MLA}. It is a highly inter-disciplinary subject. One of its tasks is to generalize the techniques in the linear algebra to higher order tensors. For instance, one key question is how to define the rank of a tensor and how to determine its optimal lower-rank approximation. This is exactly what we need in the TN algorithms.

Let us begin with a trivial example by simply considering the trace of the product of $N$ number of ($\chi \times \chi$) matrices $M$ as
\begin{eqnarray}
\text{Tr} \mathcal{M} = \text{Tr}(M^{[1]} M^{[2]} \cdots M^{[N]}) = \text{Tr} \prod_{n=1}^{N} M^{[n]},
\label{eq-4Z1D}
\end{eqnarray}
with $M^{[n]} = M$. In the language of TN\index{TN}, this can be regarded as a 1D TN with periodic boundary condition. For simplicity, we assume that the dominant eigenstate of $M$ is unique.

Allow us to firstly use a clumsy way to do the calculation: contract the shared bonds one by one from left to right. For each contraction, the computational cost is $O(\chi^3)$, thus the total cost is $O(N\chi^3)$.

Now let us be smarter by using the eigenvalue decomposition (assume it exists for $M$) in the linear algebra, which reads
\begin{eqnarray}
M = U \Lambda U^{\dagger},
\label{eq-4eig}
\end{eqnarray}
where $\Lambda$ are diagonal and $U$ is unitary satisfying $UU^{\dagger} = U^{\dagger} U= I$. Substituting Eq. (\ref{eq-4eig}) into Eq. (\ref{eq-4Z1D}), we can readily have the contraction as
\begin{eqnarray}
\text{Tr} \mathcal{M} = \text{Tr} (U \Lambda U^{\dagger} U \Lambda U^{\dagger} \cdots U \Lambda U^{\dagger}) = \text{Tr} (U \Lambda^N U^{\dagger}) = \sum_{a=0}^{\chi-1} \Lambda_a^N.
\end{eqnarray}
The dominant computational cost is around $O(\chi^3)$.

In the limit of $N \to \infty$, things become even easier, where we have
\begin{eqnarray}
\text{Tr} \mathcal{M} = \lim_{N \to \infty} \Lambda_0^N \sum_{a=0}^{\chi-1} (\frac{\Lambda_a}{\Lambda_0})^N = \Lambda_0^N,
\end{eqnarray}
where $\Lambda_0$ is the largest eigenvalue, and we have $\lim_{N \to \infty} (\frac{\Lambda_a}{\Lambda_0})^N = 0$ for $a>0$. It means all the contributions except for the dominant eigenvalue vanish when the TN is infinitely long. What we should do is just to compute the dominant eigenvalue. The efficiency can be further improved by numerous more mature techniques (such as Lanczos algorithm).


\subsection{Canonicalization of matrix product state}
\label{sec5-canon}

Before considering a 2D TN, let us take some more advantages of the eigenvalue decomposition on the 1D TN's, which is closely related to the \textit{canonicalization} of MPS proposed by Or\'us and Vidal for non-unitary evolution of MPS\index{MPS} \cite{OV08canonical}. The utilization of canonicalization are mainly in two aspects: locating optimal truncations of the MPS, and fixing the gauge degrees of freedom of the MPS for better stability and efficiency.

\subsection{Canonical form and globally optimal truncations of MPS}

\begin{figure}[tbp]
	\centering
	\includegraphics[angle=0,width=0.9\linewidth]{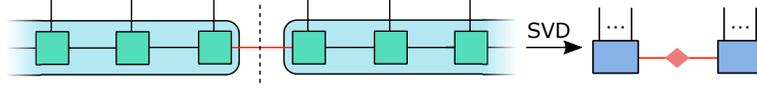}
	\caption{(Color online) An impractical scheme to get the global optimal truncation of the virtual bond (red). First, the MPS\index{MPS} is cut into two parts. All the indexes on each side of the cut are grouped into one big index. Then by contracting the virtual bond and doing the SVD\index{SVD}, the virtual bond dimension is optimally reduced to $\chi$ by only taking the $\chi$-largest singular values and the corresponding vectors.}
	\label{fig-4MPSbipar}
\end{figure}

As discussed in the above chapter, when using iTEBD\index{iTEBD} to contract a TN\index{TN}, one needs to find the optimal truncations of the virtual bonds of the MPS\index{MPS}. In other words, the problem is how to optimally reduce the dimension of an MPS.

The globally optimal truncation can be down in the following expensive way. Let us divide the MPS into two parts by cutting the bond that is to be truncated (Fig. \ref{fig-4MPSbipar}). Then, if we contract all the virtual bonds on the left hand side and reshape all the physical indexes there into one index, we will obtain a large matrix denoted as $L_{\cdots s_n,\alpha_n}$ that has one big physical and one virtual index. Another matrix denoted as $R_{s_{n+1}\cdots,\alpha_n}^{\ast}$ can be obtained by doing the same thing on the right hand side. The conjugate of $R$ is taken there to obey some conventions.

Then, by contracting the virtual bond and doing SVD\index{SVD} as
\begin{eqnarray}
\sum_{a_n} L_{\cdots s_n,a_n} R_{s_{n+1}\cdots,a_n}^{\ast} = \sum_{a_n'} \tilde{L}_{\cdots s_n,a_n'} \lambda_{a_n'} \tilde{R}_{s_{n+1}\cdots,a_n'}^{\ast},
\label{eq-4wholeMPSsvd}
\end{eqnarray}
the virtual bond dimension is optimally reduced to $\chi$ by only taking the $\chi$-largest singular values and the corresponding vectors. The truncation error that is minimized is the distance between the MPS before and after the truncation. Therefore, the truncation is optimal globally concerning the whole MPS as the environment.

In practice, we do not implement the SVD above. It is actually the decomposition of the whole wave function, which is exponentially expensive. Canonicalization provides an efficient way to realize the SVD through only local operations.

\begin{figure}[tbp]
	\centering
	\includegraphics[angle=0,width=0.75\linewidth]{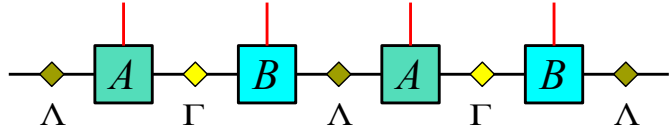}
	\caption{(Color online) The MPS\index{MPS} with two-site translational invariance.}
	\label{fig-4MPS2}
\end{figure}

Considering an infinite MPS with two-site translational invariance (Fig. \ref{fig-4MPS2}); it is formed by the tensors $A$ and $B$ as well as the diagonal matrices $\Lambda$ and $\Gamma$ as
\begin{eqnarray}
\sum_{\{a\}} \cdots \Lambda_{a_{n-1}} A_{s_{n-1}, a_{n-1} a_{n}} \Gamma_{a_{n}} B_{s_{n}, a_{n} a_{n+1}} \Lambda_{a_{n+1}} \cdots = \text{tTr} (\cdots \Lambda A \Gamma B \Lambda \cdots).
\end{eqnarray}
This is the MPS\index{MPS} used in the iTEBD\index{iTEBD} algorithm (see Chap.\ref{iTEBD} and Fig.\ref{fig-3iTEBD}). Note that all argument can be readily generalized to the infinite MPS's with $n$-site translational invariance, or even to the finite MPS's.

\begin{figure}[tbp]
	\centering
	\includegraphics[angle=0,width=0.8\linewidth]{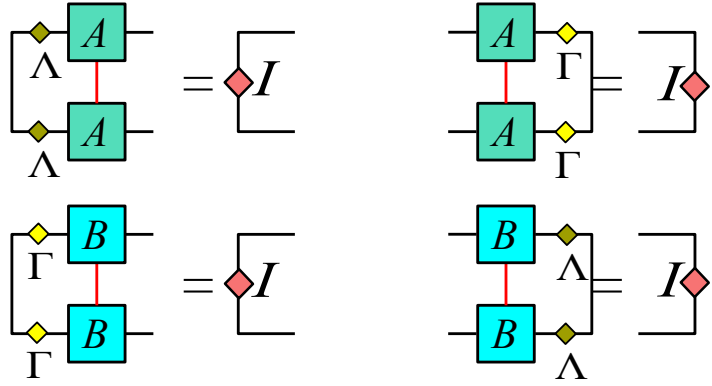}
	\caption{(Color online) Four canonical conditions of an MPS.}
	\label{fig-4CanonCond}
\end{figure}

An MPS is in the \textit{canonical form} if the tensors satisfy
\begin{eqnarray}
\sum_{sa} \Lambda_{a} A_{s, a a'} \Lambda^{\ast}_{a} A^{\ast}_{s, a a''} = I_{a' a''}, \label{eq-4Canonical1} \\
\sum_{sa} A_{s, a' a} \Gamma_{a} A^{\ast}_{s, a'' a} \Gamma^{\ast}_{a} = I_{a' a''}, \label{eq-4Canonical2} \\
\sum_{sa} \Gamma_{a} B_{s, a a'} \Gamma^{\ast}_{a} B^{\ast}_{s, a a''} = I_{a' a''}, \label{eq-4Canonical3} \\
\sum_{sa} B_{s, a' a} \Lambda_{a} B^{\ast}_{s, a'' a} \Lambda^{\ast}_{a} = I_{a' a''}, \label{eq-4Canonical4}
\end{eqnarray}
where $\Lambda$ and $\Gamma$ are positive-defined (Fig. \ref{fig-4CanonCond}). Eqs. (\ref{eq-4Canonical1}) -  (\ref{eq-4Canonical4}) are called the \textit{canonical conditions} of the MPS\index{MPS}. Note there will be $2n$ equations with $n$-site translational invariance, meaning that each inequivalent tensor will obey to two (left and right) conditions.


In the canonical form, $\Lambda$ or $\Gamma$ directly give the singular values by cutting the MPS on the corresponding bond. To see this, let us calculate Eq. (\ref{eq-4wholeMPSsvd}) from a canonical MPS. From the canonical conditions, matrices $L$ and $R$ are unitary, satisfying $L^{\dagger}L=I$ and $R^{\dagger}R=I$ (the physical indexes are contracted). Meanwhile, $\Lambda$ (or $\Gamma$) is positive-defined, thus $L$, $\Lambda$ (or $\Gamma$) and $R$ of a canonical MPS directly define the SVD, and $\Lambda$ or $\Gamma$ is indeed the singular value spectrum. Then the optimal truncations of the virtual bonds are reached by simply keeping $\chi$-largest values of $\Lambda$ and the corresponding basis of the neighboring tensors. This is true when cutting any one of the bonds of the MPS. From the uniqueness of SVD\index{SVD}, Eqs. (\ref{eq-4Canonical1}) and (\ref{eq-4Canonical2}) leads to a unique MPS representation, thus such a form is called ``\textit{canonical}''. In other words, the canonicalization fixes the gauge degrees of freedom of the MPS.

For any finite MPS\index{MPS}, the uniqueness is robust. For an infinite MPS, there will be some additional complexity. Let us define the left and right \textit{transfer matrices} $M^L$ and $M^R$ of as
\begin{eqnarray}
M^L_{a_1 a_1' a_2 a_2'} = \sum_{s} \Lambda_{a_1} A_{s, a_1 a_2} \Lambda^{\ast}_{a_1'} A^{\ast}_{s, a_1' a_2'}, \\
M^R_{a_1 a_1' a_2 a_2'} = \sum_{s} A_{s, a_1 a_2} \Gamma_{a_1} A^{\ast}_{s, a_1' a_2'} \Gamma_{a_1'}^{\ast}.
\label{eq-4LTM}
\end{eqnarray}
Then the canonical conditions [Eq. (\ref{eq-4Canonical1})] say that the identity is the left (right) eigenvector of $M^L$ ($M^R$), satisfying
\begin{eqnarray}
\sum_{a_1 a_1'} I_{a_1 a_1'} M^L_{a_1 a_1' a_2 a_2'} = \lambda^L I_{a_2 a_2'}, \\
\sum_{a_1 a_1'} I_{a_2 a_2'} M^R_{a_1 a_1' a_2 a_2'} = \lambda^R I_{a_1 a_1'},
\end{eqnarray}
with $\lambda^L$ ($\lambda^R$) the eigenvalue.

Similar eigenvalue equations can be obtained from the canonical conditions associated to the tensor $B$, where we have the transfer matrices as
\begin{eqnarray}
N^L_{a_1 a_1' a_2 a_2'} = \sum_{s} \Gamma_{a_1} B_{s, a_1 a_2} \Gamma^{\ast}_{a_1'} B^{\ast}_{s, a_1' a_2'}, \\
N^R_{a_1 a_1' a_2 a_2'} = \sum_{s} B_{s, a_1 a_2} \Lambda_{a_1} B^{\ast}_{s, a_1' a_2'} \Lambda_{a_1'}^{\ast}.
\label{eq-4RTM}
\end{eqnarray}
Now the canonical conditions are given by four eigenvalue equations and can be reinterpreted as the following: with an infinite MPS formed by $A$, $B$, $\Lambda$ and $\Gamma$, it is canonical when the identity is the eigenvector of its transfer matrices.

Simply from the canonical conditions, it does not require the ``identity'' to be dominant eigenvector. However, if the identity is not the dominant one, the canonical conditions will become unstable under an arbitrarily small noise. Below, we will show that the canonicalization algorithm assures that the identity is the leading eigenvector, since it transforms the leading eigenvector to an identity. In addition, if the dominant eigenvector of $M^L$ and $M^R$ (also $N^L$ and $N^R$) is degenerate, the canonical form will not be unique. See Ref. \cite{OV08canonical} for more details.

\begin{figure}[tbp]
	\centering
	\includegraphics[angle=0,width=1\linewidth]{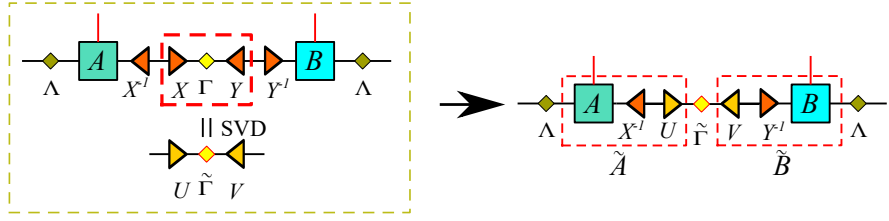}
	\caption{(Color online) The illustration of the canonical transformations.}
	\label{fig-4Canon}
\end{figure}

\subsection{Canonicalization algorithm and some related topics}
\label{sec5.canon}

Considering the iTEBD\index{iTEBD} algorithm \cite{V07iTEBD} (see Sec. \ref{iTEBD}), while the MPO represents an unitary operator, the canonical form of the MPS\index{MPS} will be reserved by the evolution (contraction). For the imaginary-time evolution, the MPO is near-unitary. For the Trotter step $\tau \to 0$, the MPO\index{MPO} approaches to be an identity. It turns out that in this case, the MPS will be canonicalized by the evolution in the standard iTEBD algorithm. When the MPO is non-unitary (e.g., when contracting the TN of a 2D statistic model) \cite{OV08canonical}, the MPS will not be canonical, and the canonicalization might be needed to better truncate the bond dimensions of the MPS.

\textbf{Canonicalization algorithm}. An algorithm to canonicalize an arbitrary MPS was proposed by Or\'us and Vidal \cite{OV08canonical}. The idea is to compute the first eigenvectors of the transfer matrices, and introduce proper gauge transformations on the virtual bonds that map the leading eigenvector to identity.

Let us take the gauge transformations on the virtual bonds between $A$ and $B$ as an example. Firstly, compute the dominant left eigenvector $v^L$ of the matrix $N^L M^L$, and similarly the dominant right eigenvector $v^R$ of the matrix $N^R M^R$. Then, reshape $v^L$ and $v^R$ as two matrices and decompose them symmetrically as
\begin{eqnarray}
v^R_{a_1 a_1'} &=& \sum_{a_1''} X_{a_1 a_1''}X^{\ast}_{a_1' a_1''}, \\
v^L_{a_1 a_1'} &=& \sum_{a_1''} Y_{a_1 a_1''}Y^{\ast}_{a_1' a_1''}.
\end{eqnarray}
$X$ and $Y$ can be calculated using eigenvalue decomposition, i.e., $v^R=WDW^{\dagger}$ with $X=W\sqrt{D}$.



Insert the identities $X^{-1}X$ and $Y Y^{-1}$ on the virtual bond as shown in Fig. \ref{fig-4Canon}, then we get a new matrix $\mathcal{M} = X \Gamma Y$ on this bond. Apply SVD\index{SVD} on on $\mathcal{M}$ as $\mathcal{M} = U \tilde{\Gamma} V^{\dagger}$, where we have the updated spectrum $\tilde{\Gamma}$ on this bond. Meanwhile, we obtain the gauge transformations to update $A$ and $B$ as $\mathcal{U}=X^{-1}U$ and $\mathcal{V}=V^{\dagger} Y^{-1}$, where the transformations are implemented as
\begin{eqnarray}
A_{s_1,a_1 a_2} \leftarrow \sum_{a} A_{s_1,a_1 a} \mathcal{U}_{a a_2},\\
B_{s_1,a_1 a_2} \leftarrow \sum_{a} B_{s_1,a a_2} \mathcal{V}_{a_1 a}.
\end{eqnarray}
Implement the same steps given above on the virtual bonds between $B$ and $A$, then the MPS is transformed to the canonical form.



\textbf{\textit{Variants of the canonical form}}. From the canonical form of an MPS, one can define the \textit{left or right canonical forms}. Define the follow tensors
\begin{eqnarray}
A^L_{s, a a'} &=& \Lambda_{a} A_{s, a a'}, \\
A^R_{s, a a'} &=& A_{s, a a'} \Gamma_{a'}, \\
B^L_{s, a a'} &=& \Gamma_{a} B_{s, a a'}, \\
B^R_{s, a a'} &=& B_{s, a a'} \Lambda_{a'}, \\
A^M_{s, a a'} &=& \Lambda_{a} A_{s, a a'} \Gamma_{a'}.
\end{eqnarray}
The left-canonical MPS is defined by $A^L$ and $B^L$ as
\begin{eqnarray}
\text{tTr} (\cdots A^L B^L  A^L B^L \cdots).
\end{eqnarray}
Similarly, the right-canonical MPS\index{MPS} is defined by $A^R$ and $B^R$ as
\begin{eqnarray}
\text{tTr} (\cdots A^R B^R  A^R B^R \cdots).
\end{eqnarray}
The central orthogonal MPS is defined as
\begin{eqnarray}
\text{tTr} (\cdots A^L B^L A^M B^R A^R \cdots).
\end{eqnarray}
One can easily check that these MPS's and the canonical MPS can be transformed to each other by gauge transformations.

From the canonical conditions, $A^L$, $A^R$, $B^L$, and $B^R$ are non-square orthogonal matrices (e.g., $\sum_{s a} A^L_{s, a a'} A^{L\ast}_{s, a a''} = I_{a' a''}$), called \textit{isometries}. $A^M$ is called the \textit{central tensor} of the central-orthogonal MPS. This MPS\index{MPS} form is the state ansatz behind the DMRG\index{DMRG} algorithm \cite{W92DMRG,W93DMRG}, and is very useful in TN-based methods (see for example the works of McCulloch \cite{M07DMRGsymme,M08iDMRGarxiv}). For instance, when applying DMRG to solve 1D quantum model, the tensors $A^L$ and $B^L$ define a left-to-right RG\index{RG} flow that optimally compresses the Hilbert space of the left part of the chain. $A^R$ and $B^R$ define a right-to-left RG flow similarly. The central tensor between these two RG flows is in fact the ground state of the effective Hamiltonian given by the RG flows of DMRG. Note that the canonical MPS is also called the \textit{central canonical form}, where the directions of the RG flows can be switched arbitrarily by gauge transformations, thus there is no need to define the directions of the flows or a specific center.

\textbf{\textit{Relations to tensor train decomposition}}. It is worth mentioning the TTD\index{TTD} \cite{O11TTD} proposed in the field of MLA\index{MLA}. As argued in Chap. 2, one advantage of MPS\index{MPS} is it lowers the number of parameters from an exponential size dependence to a polynomial one. Let us consider a similar problem: for a $N$-th order tensor that has $d^N$ parameters, how to find its optimal MPS\index{MPS} representation, where there are only $[2d\chi+(N-2)d\chi^2]$ parameters? TTD was proposed for this aim: by decomposing a tensor into a tensor-train form that is similar to a finite open MPS, the number of parameters becomes linearly relying to the order of the original tensor. The TTD algorithm shares many similar ideas with MPS and the related algorithms (especially DMRG\index{DMRG} which was proposed about two decades earlier). The aim of TTD\index{TTD} is also similar to the truncation tasks in the TN\index{TN} algorithms, which is to compress the number of parameters.

\section{Super-orthogonalization and Tucker decomposition}
\label{sec5-SO}

\begin{figure}[tbp]
	\centering
	\includegraphics[angle=0,width=1\linewidth]{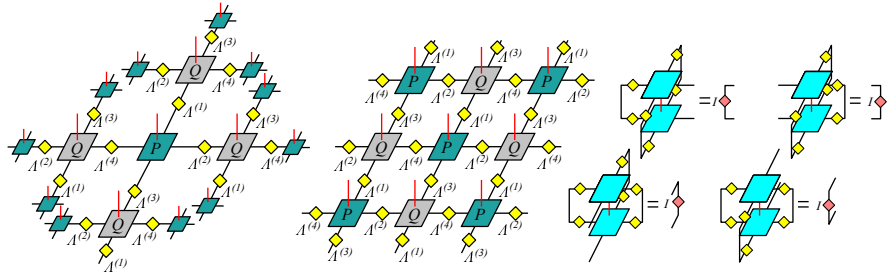}
	\caption{(Color online) The first two figures show the iPEPS\index{iPEPS} on tree and square lattices, with two-site translational invariance. The last one shows the super-orthogonal conditions.}
	\label{fig-4SO}
\end{figure}

As discussed in the above section, the canonical form of an MPS brings a lot of advantages, such as determining the entanglement and the optimal truncations of the virtual bond dimensions by local transformations. The canonical form can be readily generalized to the iPEPS's\index{iPEPS} on trees. Can we also define the canonical form for the iPEPS's in higher-dimensional regular lattices, such as square lattice (Fig. \ref{fig-4SO})? If this can be done, we would known how to find the globally optimal transformations that reduces the bond dimensions of the iPEPS, just like what we can do with an MPS. Due to the complexity of 2d TN's\index{TN}, unfortunately, there is no such a form in general. In the following, we explain the \textit{super-orthogonal form} of iPEPS proposed in 2012 \cite{RLXZS12ODTNS}, which applies the canonical form of tree iPEPS to the iPEPS on regular lattices. The super-orthogonalization is a generalization of the Tucker decomposition (a higher-order generalization of matrix SVD\index{SVD}) \cite{DDV00HOSVD}, providing a zero-loop approximation scheme \cite{RXLS13NCD} to define the entanglement and truncate the bond dimensions. 


\subsection{Super-orthogonalization}

Let us start from the iPEPS\index{iPEPS} on the (infinite) Bethe lattice with the coordination number $z=4$. It is formed by two tensors $P$ and $Q$ on the sites as well as four spectra $\Lambda^{(k)}$ ($k=1,2,3,4$) on the bonds, as illustrated in Fig. \ref{fig-4SO}. Here, we still take the two-site translational invariance for simplicity.

There are eight \textit{super-orthogonal conditions}, of which four associate to the tensor $P$ and four to $Q$. For $P$, the conditions are
\begin{eqnarray}
\sum_s \sum_{\cdots a_{k-1} a_{k+1} \cdots} P_{s,\cdots a_k \cdots} P_{s,\cdots a_k' \cdots}^{\ast} \prod_{n \neq k} \Lambda^{(n)}_{a_n} \Lambda^{(n)\ast}_{a_n} = I_{a_{k} a_{k}'}, \ \ (\forall \ k),
\label{eq-4SOconditions}
\end{eqnarray}
where all the bonds along with the corresponding spectra are contracted except for $a_k$. It means that by putting $a_k$ as one index and all the rest as another, the \textit{$k$-rectangular matrix} $S^{(k)}$ defined as
\begin{eqnarray}
S^{(k)}_{s\cdots a_{k-1} a_{k+1} \cdots, a_k} = P_{s,\cdots a_k \cdots} \prod_{n \neq k} \Lambda^{(n)}_{a_n},
\label{eq-4RCmatrix}
\end{eqnarray}
is an isometry, satisfying $S^{(k)\dagger} S^{(k)} = I$. The super-orthogonal conditions of the tensor $Q$ are defined in the same way. $\Lambda^{(k)}$ is dubbed \textit{super-orthogonal spectrum} when the super-orthogonal conditions are fulfilled.

\begin{figure}[tbp]
	\centering
	\includegraphics[angle=0,width=1\linewidth]{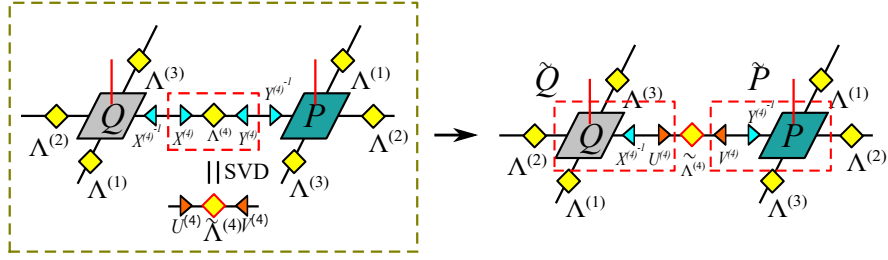}
	\caption{(Color online) The illustrations of gauge transformations in the super-orthogonalization algorithm.}
	\label{fig-4SOalgo}
\end{figure}

In the canonicalization of MPS\index{MPS}, the vectors on the virtual bonds give the bipartite entanglement defined by Eq. (\ref{eq-4wholeMPSsvd}). Meanwhile, the bond dimensions can be optimally reduced by discarding certain smallest elements of the spectrum. In the super-orthogonalization, this is not always true for iPEPS's. For example, given a translational invariant iPEPS\index{iPEPS} defined on a tree (or called Bethe lattice, see Fig. \ref{fig-4SO} (a)) \cite{SDV06TTN,NFGSS08TreeMPS,TEV09TTN,MVLN10TTN,LDX12TTN,NC13TTN,PVK13TTN,MVSNL15TTN},  the super-orthogonal spectrum indeed  gives the bipartite entanglement spectrum by cutting the system at the corresponding place. However, when considering loopy lattices, such as the iPEPS defined on a square lattice (Fig. \ref{fig-4SO} (b)), this will no longer be true. Instead, the super-orthogonal spectrum provides an approximation of the entanglement of the iPEPS by optimally ignoring the loops. One can still truncate the bond dimensions according to the super-orthogonal spectrum, giving in fact the simple update (see Sec. \ref{sec.updateschemes}). We will discuss the loopless approximation in detail in Sec. \ref{sec.looplessTNrank-1} using the rank-1 decomposition.

\subsection{Super-orthogonalization algorithm}

Any PEPS\index{PEPS} can be transformed to the super-orthogonal form by iteratively implementing proper gauge transformations on the virtual bonds \cite{RLXZS12ODTNS}. The algorithm consists of two steps. Firstly, compute the reduced matrix $\mathcal{M}^{(k)}$ of the $k$-rectangular matrix of the tensor $P$ [Eq. (\ref{eq-4RCmatrix})] as
\begin{eqnarray}
\mathcal{M}^{(k)}_{a_{k} a_{k}'}=\sum_s \sum_{\cdots a_{k-1} a_{k+1} \cdots} S^{(k)}_{s\cdots a_{k-1} a_{k+1} \cdots, a_k} S^{(k)\ast}_{s\cdots a_{k-1} a_{k+1} \cdots, a_k'}.
\label{eq-4bondRM}
\end{eqnarray}
Compared with the super-orthogonal conditions in Eq. (\ref{eq-4SOconditions}), one can see that $\mathcal{M}^{(k)}=I$ when the PEPS is super-orthogonal. Similarly, we define the reduced matrix $\mathcal{N}^{(k)}$ of the tensor $Q$.

When the PEPS is not super-orthogonal, $\mathcal{M}^{(k)}$ and $\mathcal{N}^{(k)}$ are not identities but Hermitian matrices. Decompose them as $\mathcal{M}^{(k)}=X^{(k)}X^{(k)\dagger}$ and $\mathcal{N}^{(k)}=Y^{(k)} Y^{(k)\dagger}$. Then, insert the identities $X^{(k)} [X^{(k)}]^{-1}$ and $Y^{(k)} [Y^{(k)}]^{-1}$ on the virtual bonds to perform gauge transformations along four directions as shown in Fig. \ref{fig-4SOalgo}. Then, we can use SVD\index{SVD} to renew the four spectra by $X^{(k)}\Lambda^{(k)}Y^{(k)T} = U^{(k)}\tilde{\Lambda}^{(k)}V^{(k)\dagger}$. Meanwhile, we transform the tensors as
\begin{eqnarray}
P_{s,\cdots a_k \cdots} \leftarrow \sum_{a_k'a_k''} P_{s,\cdots a_k' \cdots} [X^{(k)}]^{-1}_{a_k'a_k''}  U^{(k)}_{a_k'' a_k},\\
Q_{s,\cdots a_k \cdots} \leftarrow \sum_{a_k'a_k''} Q_{s,\cdots a_k' \cdots} [Y^{(k)}]^{-1}_{a_k'a_k''}  V^{(k)\ast}_{a_k'' a_k}.
\end{eqnarray}

Compared with the canonicalization algorithm of MPS\index{MPS}, one can see that the gauge transformations in the super-orthogonalization algorithm are quite similar. The difference is that one cannot transform a PEPS\index{PEPS} into the super-orthogonal form by a single step, since the transformation on one bond might cause some deviation from obeying the super-orthogonal conditions on other bonds. Thus, the above procedure should be iterated until all the tensors and spectra converge.

\subsection{Super-orthogonalization and dimension reduction by Tucker decomposition}

Such an iterative scheme is closely related to the Tucker decomposition in MLA\index{MLA} \cite{DDV00HOSVD}. Tucker decomposition is considered as a generalization of (matrix) SVD to higher-order tensors, thus it is also called higher-order or multi-linear SVD. The aim is to find the optimal reductions of the bond dimensions for a single tensor.

\begin{figure}[tbp]
	\centering
	\includegraphics[angle=0,width=0.6\linewidth]{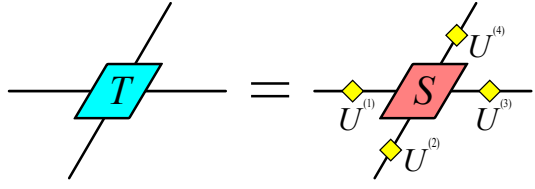}
	\caption{(Color online) The illustrations of Tucker decomposition [Eq. (\ref{eq-4tucker})].}
	\label{fig-4tucker}
\end{figure}

Let us define the $k$-reduced matrix of a tensor $T$ as
\begin{eqnarray}
M^{(k)}_{a_k a_k'} = \sum_{a_1 \cdots a_{k-1} a_{k+1} \cdots} T_{a_1 \cdots a_{k-1} a_k a_{k+1} \cdots} T_{a_1 \cdots a_{k-1} a_k' a_{k+1} \cdots}^{\ast},
\end{eqnarray}
where all except the $k$-th indexes are contracted. The Tucker decomposition (Fig. \ref{fig-4tucker}) of a tensor $T$ has the form as
\begin{eqnarray}
T_{a_1 a_2 \cdots} = \sum_{a_1a_2 \cdots} S_{b_1b_2 \cdots} \prod_{k} U^{(k)}_{a_kb_k},
\label{eq-4tucker}
\end{eqnarray}
where the following conditions should be satisfied:
\begin{itemize}	
	\item \textit{Unitarity}. $U^{(k)}$ are unitary matrices satisfying $U^{(k)} U^{(k)\dagger} = I$.
	\item \textit{All-orthogonality}. For any $k$, the $k$-reduced matrix $M^{(k)}$ of the tensor $S$ is diagonal, satisfying
	\begin{eqnarray}
	M^{(k)}_{a_k a_k'} = \Gamma^{(k)}_{a_k} I_{a_k a_k'}.
	\end{eqnarray}
	\item \textit{Ordering}. For any $k$, the elements of $\Gamma^{(k)}$ in the $k$-reduced matrix are positive-defined and in the descending order, satisfying $\Gamma_0 > \Gamma_1 > \cdots$.
\end{itemize}

From these conditions, one can see that the tensor $T$ is decomposed as the contraction of another tensor $S$ with several unitary matrices. $S$ is called the \textit{core tensor}. In other words, the optimal lower-rank approximation of the tensor can be simply obtained by
\begin{eqnarray}
T_{a_1 a_2 \cdots} \simeq \sum_{a_1a_2 \cdots = 0}^{\chi-1} S_{b_1b_2 \cdots} \prod_{k} U^{(k)}_{a_kb_k},
\end{eqnarray}
where we only take the first $\chi$ terms in the summation of each index.

Such an approximations can be understood in terms of the SVD\index{SVD} of matrices. Applying the conditions of the $k$-reduced matrix of $T$, we have
\begin{eqnarray}
M^{(k)}_{a_k a_k'} = \sum_{b_k} U^{(k)}_{a_kb_k} \Gamma^{(k)}_{b_k} U^{(k)\dagger}_{a_k'b_k}.
\end{eqnarray}
Since $U^{(k)}$ is unitary and $\Gamma^{(k)}$ is positive-defined and in the descending order, the above equation is exactly the eigenvalue decomposition of $M^{(k)}$. From the relation between the SVD of a matrix and the eigenvalue decomposition of its reduced matrix, we can see that $U^{(k)}$ and $\Gamma^{(k)}$ in fact give the SVD of the matrix $T_{a_1 \cdots a_{k-1} a_{k+1} \cdots, a_{k}}$ as
\begin{eqnarray}
T_{a_1 \cdots a_{k-1} a_{k+1} \cdots, a_{k}} = \sum_{b_k} \mathcal{S}_{a_1 \cdots a_{k-1} a_{k+1} \cdots, b_{k}} \sqrt{\Gamma^{(k)}_{b_k}} U^{(k)}_{a_kb_k}.
\end{eqnarray}
Then, The optimal truncation of the rank of each index is reached by the corresponding SVD\index{SVD}. The truncation error is obviously the distance defined as
\begin{eqnarray}
\varepsilon^{(k)} = |T_{a_1 \cdots a_{k-1} a_{k+1} \cdots, a_{k}} - \sum_{b_k=1}^{\chi} \mathcal{S}_{a_1 \cdots a_{k-1} a_{k+1} \cdots} \Gamma^{(k)}_{b_k} U^{(k)}_{a_kb_k}|,
\end{eqnarray}
which is minimized in this SVD.

For the algorithms of Tucker decomposition, one simple way is to do the eigenvalue decomposition of each $k$-reduced matrix, or the SVD of each $k$-rectangular. Then for a $K$-th ordered tensor, $K$ SVD's will give us the Tucker decomposition and a lower-rank approximation. This algorithm is often called \textit{higher-order SVD} (HOSVD)\index{HOSVD}, which has been successfully applied to implement truncations in the TRG\index{TRG} algorithm \cite{XCQZYX12HOSRG}. The accuracy of HOSVD can be improved. Since the truncation on one index will definitely affect the truncations on other indexes, there will be some ``interactions'' among different indexes (modes) of the tensor. The truncations in HOSVD are calculated independently, thus such ``interactions''are ignored. One improved way is the \textit{high-order orthogonal iteration} (HOOI)\index{HOOI}, where the interactions among different modes are considered by iteratively doing SVD's until reaching the convergence. See more details in Ref. \cite{DDV00HOSVD}.

Compared with the conditions of Tucker decomposition, let us redefine the super-orthogonal conditions of a PEPS\index{PEPS} as
\begin{itemize}	
	\item \textit{Super-orthogonality}. For any $k$, the reduced matrix of the $k$-rectangular matrix $\mathcal{M}^{(k)}$ [Eq. (\ref{eq-4bondRM})] is diagonal, satisfying
	\begin{eqnarray}
	\mathcal{M}^{(k)}_{a_k a_k'} = \Gamma^{(k)}_{a_k} I_{a_k a_k'}.
	\end{eqnarray}
	\item \textit{Ordering}. For any $k$, the elements of $\Gamma^{(k)}$ are positive-defined and in the descending order, satisfying $\Gamma_0 > \Gamma_1 > \cdots$.
\end{itemize}
Note that the condition ``unitary'' (first one in Tucker decomposition) is hidden in the fact that we use gauge transformations to transform the PEPS into the super-orthogonal form. Therefore, the super-orthogonalization is also called \textit{network Tucker decomposition} (NTD)\index{NTD}.

In the Tucker decomposition, the ``all-orthogonality'' and ``ordering'' lead to an SVD\index{SVD} associated to a single tensor, which explains how the optimal truncations work from the decompositions in linear algebra. In the NTD, the SVD picture is generalized from a single tensor to a non-local PEPS. Thus, the truncations are optimized in a non-local way.

Let us consider a finite-size PEPS\index{PEPS} and arbitrarily choose one geometrical bond (say $a$). If the PEPS is on a tree, we can cut the bond and separate the TN\index{TN} into three disconnecting parts: the spectrum ($\Lambda$) on this bond and two tree brunches stretching to the two sides of the bond. Specifically speaking, each brunch contains one virtual bond and all the physical bonds on the corresponding side, formally denoted as $\Psi^L_{i_1i_2 \cdots,a}$ (and $\Psi^R_{j_1j_2 \cdots,a}$ on the other side). Then the state given by the iPEPS can be written as
\begin{eqnarray}
\sum_a \Psi^L_{i_1i_2 \cdots,a} \Lambda_a \Psi^R_{j_1j_2 \cdots,a}.
\label{eq-4TreeSVD}
\end{eqnarray}

To get the SVD\index{SVD} picture, we need to prove that $\Psi^L$ and $\Psi^R$ in the above equation are isometries, satisfying the orthogonal conditions as
\begin{eqnarray}
\begin{aligned}
\sum_{i_1i_2 \cdots} \Psi^L_{i_1i_2 \cdots,a} \Psi^L_{i_1i_2 \cdots,a'} = I_{aa'},\\
\sum_{j_1j_2 \cdots} \Psi^R_{j_1j_2 \cdots,a} \Psi^R_{j_1j_2 \cdots,a'} = I_{aa'}.
\end{aligned}
\label{eq-4OrtPhiLR}
\end{eqnarray}
Note that the spectrum $\Lambda$ is already positive-defined according to the algorithm. To this end, we construct the TN of $\sum_{i_1i_2 \cdots} \Psi^{L(R)}_{i_1i_2 \cdots,a} \Psi^{L(R)}_{i_1i_2 \cdots,a'}$ from its boundary. If the PEPS\index{PEPS} is super-orthogonal, the spectra must be on the boundary of the TN\index{TN} because the super-orthogonal conditions are satisfied everywhere \footnote{With the open boundary condition, one may consider the geometrical bond dimensions as one, and define the spectra by the one-dimensional vector $[1]$.}. Then the contractions of the tensors on the boundary satisfy Eq. (\ref{eq-4SOconditions}), which gives identities. Then we have on the new boundary again the spectra to iterate the contractions. All tensors can be contracted by iteratively using the super-orthogonal conditions, which in the end gives identities as Eq. (\ref{eq-4OrtPhiLR}). Thus, $\Psi^L$ and $\Psi^R$ are indeed isometries and Eq. (\ref{eq-4TreeSVD}) indeed gives the SVD\index{SVD} of the whole wavefunction. The truncations of the bond dimensions is globally optimized by taking the whole tree PEPS as the environment.

For an iPEPS\index{iPEPS}, it can be similarly proven that $\Psi^L$ and $\Psi^R$ are isometries. One way is to put any non-zero spectra on the boundary and iterate the contraction by Eq. (\ref{eq-4bondRM}). While the spectra on the boundary can be arbitrary, the results of the contractions by Eq. (\ref{eq-4bondRM}) converge to identities quickly \cite{RLXZS12ODTNS}. Then the rest of the contractions are exactly given by the super-orthogonal conditions [Eq. (\ref{eq-4SOconditions})]. In other words, the identity is a stable fixed point of the above iterations. Once the fixed point is reached, it can be considered that the contraction is from infinitely far away, meaning from the ``boundary'' of the iPEPS. In this way, one proves $\Psi^L$ and $\Psi^R$ are isometries, i.e., $\Psi^{L\dagger} \Psi^L = I$ and $\Psi^{R\dagger} \Psi^R = I$.

\section{Zero-loop approximation on regular lattices and rank-1 decomposition}
\label{sec.looplessTNrank-1}

\subsection{Super-orthogonalization works well for truncating the PEPS on regular lattice: some intuitive discussions}
\label{sec5-rank1}

From the discussions above, we can see that the ``canonical'' form of a TN\index{TN} state is strongly desired, because it is expected to give the entanglement and the optimal truncations of the bond dimensions. Recall that to contract a TN that cannot be contracted exactly, truncations are inevitable, and locating the optimal truncations is one of the main tasks in the computations. The super-orthogonal form provides a robust way to optimally truncate the bond dimensions of the PEPS\index{PEPS} defined on a tree, analog to the canonicalization of MPS\index{MPS}. 

Interestingly, the super-orthogonal form does not require the tree structure. For an iPEPS\index{iPEPS} defined on a regular lattice, for example the square lattice, one can still super-orthogonalize it using the same algorithm. What is different is that the SVD\index{SVD} picture of the wave function (generally, see Eq. (\ref{eq-4wholeMPSsvd})) will not rigorously hold, as well as the robustness of the optimal truncations. In other words, the super-orthogonal spectrum does not exactly give the entanglement. A question rises: can we still truncate iPEPS defined on a square lattice according to the super-orthogonal spectrum?

Surprisingly, numeric simulations show that the accuracy by truncating according to the super-orthogonal spectrum is still good in many cases. Let us take the ground-state simulation of a 2D system by imaginary-time evolution as an example. As discussed in Sec. \ref{sec.evo2D}, the simulation becomes the contraction of a 3D TN\index{TN}. One usual way to compute this contraction is to contract layer by layer to an iPEPS\index{iPEPS} (see, e.g., \cite{JOVVC08PEPS,JWX08SimpleUpdate}). The contraction will enlarge the virtual bond dimensions, and truncations are needed. When the ground state is gapped (see, e.g., \cite{RLXZS12ODTNS,JWX08SimpleUpdate}), the truncations produce accurate results, which means the super-orthogonal spectrum approximates the true entanglement quite well.

It has been realized that using the simple update algorithm \cite{JWX08SimpleUpdate}, the iPEPS will converge to the super-orthogonal form for a vanishing Trotter step $\tau \to 0$. The success of the simple update suggests that the optimal truncation method on trees still works well for regular lattices. Intuitively, this can be understood in the following way. Comparing a regular lattice with a tree, if it has the same coordination number, the two lattices look exactly the same if we only inspect locally on one site and its nearest neighbors. The difference appears when one goes round the closed loops on the regular lattice, since there are no loop in the tree. Thus, the error applying the optimal truncation schemes (such as super-orthogonalization) of a tree to a regular lattice should be characterized by some non-local features associated to the loops. This explains in a descriptive way why the simple update works well for gapped states, where the physics is dominated by short-range correlations. For the systems that possess small gaps or are gapless, simple update is not sufficiently accurate\cite{XCYKNX14PESS}, particularly for the non-local physical properties such as the correlation functions.

\subsection{Rank-1 decomposition and algorithm}
\label{sec.rank1algo}

\begin{figure}[tbp]
	\centering
	\includegraphics[angle=0,width=0.5\linewidth]{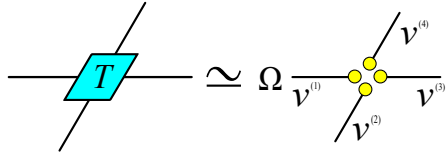}
	\caption{(Color online) The illustrations of rank-1 decomposition [Eq. (\ref{eq-4rank1})].}
	\label{fig-4rank1}
\end{figure}

\textit{Rank-1 decomposition} in MLA\index{MLA} \cite{LMV00Rank1} provides a more mathematic and rigorous way to understand the approximation by super-orthogonalization (simple update) to truncate PEPS\index{PEPS} on regular lattices \cite{RXLS13NCD}. For a tensor $T$, its rank-1 decomposition (Fig. \ref{fig-4rank1}) is defined as
\begin{eqnarray}
T_{a_1a_2\cdots a_K} \simeq \Omega \prod_{k=1}^K v^{(k)}_{a_k},
\label{eq-4rank1}
\end{eqnarray}
where $v^{(k)}$ are normalized vectors and $\Omega$ is a constant that satisfies
\begin{eqnarray}
\Omega = \sum_{a_1a_2\cdots a_K} T_{a_1a_2\cdots a_K} \prod_{k=1}^K v^{(k)\ast}_{a_k}.
\label{eq-4rank1Const}
\end{eqnarray}
Rank-1 decomposition provides an approximation of $T$, where the distance between $T$ and its rank-1 approximation is minimized, i.e.,
\begin{eqnarray}
\min_{|v^{(k)}_{a_k}|=1} |T_{a_1a_2\cdots a_K} - \Omega \prod_{k=1}^K v^{(k)}_{a_k}|.
\label{eq-4rank1Min}
\end{eqnarray}

\begin{figure}[tbp]
	\centering
	\includegraphics[angle=0,width=0.8\linewidth]{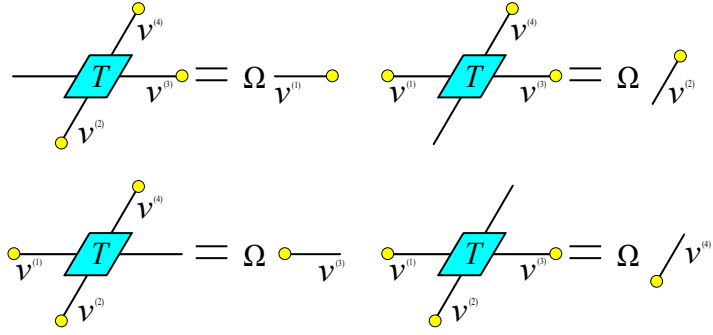}
	\caption{(Color online) The illustrations of self-consistent conditions for the rank-1 decomposition [Eq. (\ref{eq-4rank1Self})].}
	\label{fig-4rank1cond}
\end{figure}

The rank-1 decomposition is given by the fixed point of a set of self-consistent equations (Fig. \ref{fig-4rank1cond}), which are
\begin{eqnarray}
\sum_{\text{all except } a_k} T_{a_1a_2\cdots a_K} \prod_{j \neq k} v^{(j)}_{a_j} = \Omega v^{(k)}_{a_k} \ \ (\forall \ k).
\label{eq-4rank1Self}
\end{eqnarray}
It means that $v^{(k)}$ is obtained by contracting all other vectors with the tensor. This property provides us an algorithm to compute rank-1 decomposition: one arbitrarily initialize the vectors $\{v^{(k)}\}$ of norm-1 and recursively update each vector by the tensor and the rest vectors using Eq. (\ref{eq-4rank1Self}) until all vectors converge. 

Apart from some very special cases, such an optimization problem is concave, thus rank-1 decomposition is unique {\footnote{In fact, the uniqueness of rank-1 decomposition has not been rigorously proven.}}. Furthermore, if one arbitrarily chooses a set of norm-1 vectors, they will converge to the fixed point exponentially fast with the iterations. To the best of our knowledge, the exponential convergence has not been proved rigorously, but observed in most cases.

\subsection{Rank-1 decomposition, super-orthogonalization, and zero-loop approximation}
\label{sec.zeroloop}

Let us still consider an translational invariant square TN that is formed by infinite copies of the 4th-order tensor $T$ (Fig. \ref{fig-3TNcontract}). The rank-1 decomposition of $T$ provides an approximative scheme to compute the contraction of the TN, which is called the \textit{theory of network contractor dynamics} (NCD)\index{NCD} \cite{RXLS13NCD}.

\begin{figure}[tbp]
	\centering
	\includegraphics[angle=0,width=0.8\linewidth]{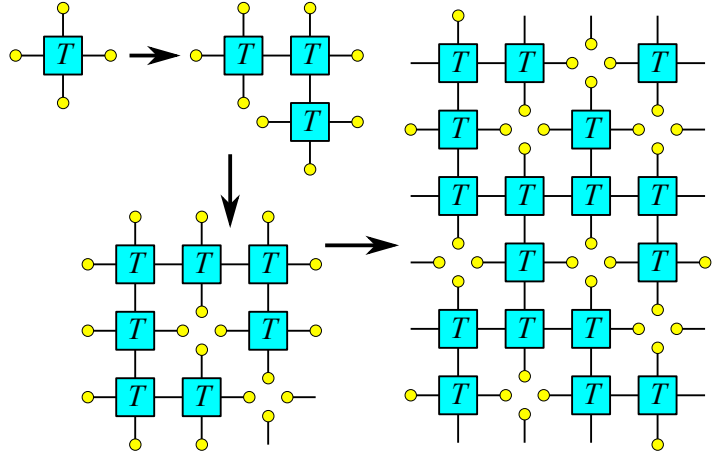}
	\caption{(Color online) Using the self-consistent conditions of the rank-1 decomposition, a tree TN\index{TN} with no loops can grow to cover the infinite square lattice. The four vectors gathering in a same site give the rank-1 approximation of the original tensor.}
	\label{fig-4rank1encoding}
\end{figure}

The picture of NCD\index{NCD} can be understood in an opposite way to contraction, but by iteratively using the self-consistent conditions [Eq. (\ref{eq-4rank1Self})] to ``grow''  a tree TN\index{TN} that covers the whole square lattice (Fig. \ref{fig-4rank1encoding}). Let us start from Eq. (\ref{eq-4rank1Const}) of $\Omega$. Using Eq. (\ref{eq-4rank1Self}), we substitute each of the four vectors by the contraction of $T$ with the other three vectors. After doing so, Eq. (\ref{eq-4rank1Const}) becomes the contraction of more than one $T$'s with the vectors on the boundary. In other words, we ``grow'' the local TN contraction from one tensor plus four vectors to that with more tensors and vectors.

By repeating the substitution, the TN\index{TN} can be grown to cover the whole square lattice, where each site is allowed to put maximally one $T$. Inevitably, some sites will not have $T$, but four vectors instead. These vectors (also called \textit{contractors}) give the rank-1 decomposition of $T$ as Eq. (\ref{eq-4rank1}). This is to say that some tensors in the square TN are replaced by its rank-1 approximation, so that all loops are destructed and the TN becomes a loopless tree covering the square lattice. In this way, the square TN is approximated by such a tree TN on square lattice, so that its contraction is simply computed by Eq. (\ref{eq-4rank1Const}).

The growing process as well as the optimal tree TN is only to understand the zero-loop approximation with rank-1 decomposition. There is no need to practically implement such a process. Thus, it does not matter how the TN is grown or where the rank-1 tensors are put to destroy the loops. All information we need is given by the rank-1 decomposition. In other words, the zero-loop approximation of the TN is encoded in the rank-1 decomposition.

For growing the TN\index{TN}, we shall remark that using the contraction of one $T$ with several vectors to substitute one vector is certainly not unique. However, the aim of ``growing'' is to reconstruct the TN formed by $T$. Thus, if $T$ has to appear in the substitution, the vectors should be uniquely chosen as those given in the rank-1 decomposition due to the uniqueness of rank-1 decomposition. Secondly, there are hidden conditions when covering the lattice by ``growing''. A stronger version is
\begin{eqnarray}
T_{a_1a_2a_3a_4} = T_{a_3a_2a_1a_4}^{\ast} = T_{a_1a_4a_3a_2}^{\ast} = T_{a_3a_4a_1a_2}.
\label{eq-4NCDcond1}
\end{eqnarray}
And a weaker one only requires the vectors to be conjugate to each other as
\begin{eqnarray}
v^{(1)} = v^{(3)\dagger}, \ \ v^{(2)} = v^{(4)\dagger}.
\label{eq-4NCDcond2}
\end{eqnarray}
These conditions assure that the self-consistent equations encodes the correct tree that optimally in the rank-1 sense approximates the square TN.

Comparing with Eqs. (\ref{eq-4SOconditions}) and (\ref{eq-4rank1Self}), the super-orthogonal conditions are actually equivalent to the above self-consistent equations of rank-1 decomposition by defining the tensor $T$ and vector $v$ as
\begin{eqnarray}
T_{a_1a_2 \cdots a_K} &=& \sum_s P_{s,a_1' a_2' \cdots a_K'} P_{s,a_1'' a_2'' \cdots a_K''}^{\ast} \prod_{k=1}^K \sqrt{\Lambda^{(k)}_{a_k'} \Lambda^{(k)\ast}_{a_k''}}, \label{eq-4DoubleT}\\
v^{(k)}_{a_k} &=& \sqrt{\Lambda^{(k)}_{a_k'} \Lambda^{(k)\ast}_{a_k''}},
\label{eq-4SOtensor}
\end{eqnarray}
with $a_k= (a_k',a_k'')$. Thus, the super-orthogonal spectrum provides an optimal approximation for the truncations of the bond dimensions in the zero-loop level. This provides a direct connection between the simple update scheme and rank-1 decomposition.

\subsection{Error of zero-loop approximation and tree-expansion theory based on rank-decomposition.}

The error of NCD\index{NCD} (and simple update) is an important issue. From the first glance, the error seems to be the error of rank-1 decomposition $\varepsilon = |T - \prod_k v^{(k)}|$. This would be true if we replaced all tensors in the square TN\index{TN} by the rank-1 version. In this case, the PEPS\index{PEPS} is approximated by a product state with zero entanglement. In the NCD\index{NCD} scheme, however, we only replace a part of the tensors to destruct the loops. The corresponding approximative PEPS is an entanglement state with a tree structure. Therefore, the error of rank-1 decomposition cannot properly characterize the error of simple update.

\begin{figure}[tbp]
	\centering
	\includegraphics[angle=0,width=0.6\linewidth]{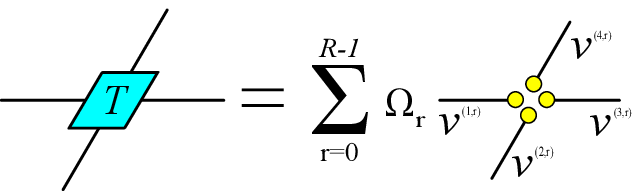}
	\caption{(Color online) The illustrations of rank-1 decomposition [Eq. (\ref{eq-4rank})].}
	\label{fig-4rank}
\end{figure}

To control the error, let us introduce the \textit{rank decomposition} (also called CANDECOMP/PARAFAC decomposition)\index{CANDECOMP/PARAFAC} of $T$ in MLA\index{MLA} (Fig. \ref{fig-4rank}) that reads
\begin{eqnarray}
T_{a_1a_2 \cdots} = \sum_{r=0}^{R-1} \Omega_r \prod_k v^{(k,r)}_{a_{k}},
\label{eq-4rank}
\end{eqnarray}
where $v^{(k,r)}$ are normalized vectors. The idea of rank decomposition \cite{H1927rankorigin,H1928rankorigin1} is to expand $T$ into the summation of $R$ number of rank-1 tensors with $R$ called the \textit{tensor rank}. The elements of the vector $\Omega$ can always be in the descending order according to the absolute values. Then the leading term $\Omega_0 \prod_k v^{(k,0)}$ gives exactly the rank-1 decomposition of $T$, and the error of the rank-1 decomposition becomes $|\sum_{r=1}^{R-1} \Omega_r \prod_k v^{(k,r)}_{a_{k}}|$.

In the optimal tree TN\index{TN}, let us replace the rank-1 tensors back by the full rank tensor in Eq. (\ref{eq-4rank}). We suppose the rank decomposition is exact, thus we will recover the original TN by doing so. The TN contraction becomes the summation of $R^{\tilde{N}}$ terms with $\tilde{N}$ the number of rank-1 tensors in the zero-loop TN. Each term is the contraction of a tree TN, which is the same as the optimal tree TN except that certain vectors are changed to $v^{(k,r)}$ instead of the rank-1 term $v^{(k,0)}$. Note that in all terms, we use the same tree structure; the leading term in the summation is the zero-loop TN in the NCD\index{NCD} scheme. It means with rank decomposition, we expand the contraction of the square TN by the summation of the contractions of many tree TN's.

Let us order the summation referring to the contributions of different terms. For simplicity, we assume $R=2$, meaning $T$ can be exactly decomposed as the summation of two rank-1 tensors, which are the leading term given by the rank-1 decomposition, and the next-leading term denoted as $T_1 = \Omega_1 \prod_k v^{(k,1)}$. We dub as the next-leading term as the \textit{impurity tensor}. Defining $\tilde{n}$ as the number of the impurity tensors appearing in one of the tree TN in the summation, the expansion can be written as
\begin{eqnarray}
Z = \Omega_0^{\tilde{N}} \sum_{\tilde{n}=0}^{\tilde{N}} (\frac{\Omega_1}{\Omega_0})^{\tilde{n}} \sum_{\mathcal{C} \in \mathcal{C}(\tilde{n})} Z_{\mathcal{C}}.
\label{eq-4RankExpansion}
\end{eqnarray}
We use $\mathcal{C}(\tilde{n})$ to denote the set of all possible \textit{configurations} of $\tilde{n}$ number of the impurity tensors, where there are $\tilde{n}$ of $T_1$'s located in different positions in the tree. Then $Z_{\mathcal{C}}$ denotes the contraction of such a tree TN with a specific configuration of $T_1$'s. In general, the contribution is determined by the order of $|\Omega_1/\Omega_0|$ since $|\Omega_1/\Omega_0|<1$.

\begin{figure}[tbp]
	\centering
	\includegraphics[angle=0,width=1\linewidth]{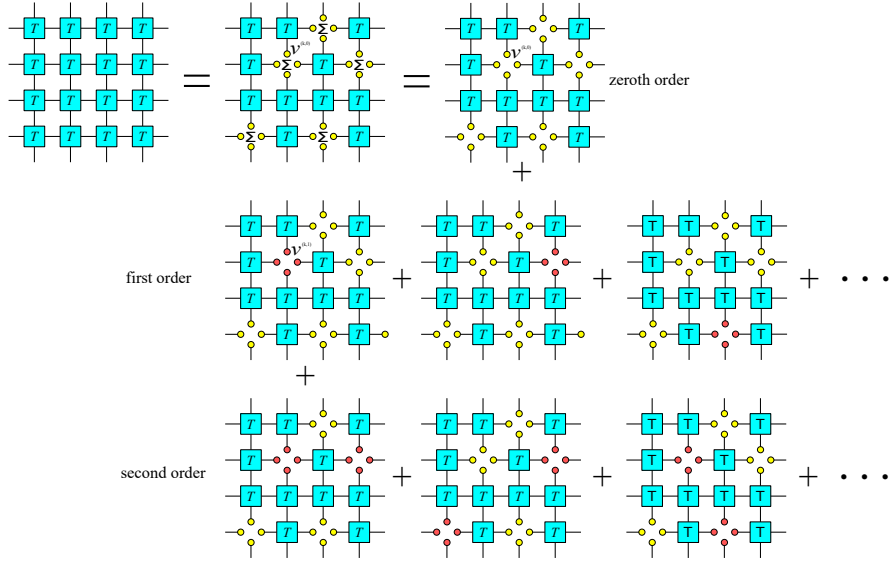}
	\caption{(Color online) The illustrations of the expansion with rank decomposition. The yellow and red circles stand for $v^{(k,0)}_{a_k}$ (zeroth order terms in the rank decomposition) and $v^{(k,1)}_{a_k}$ (first order terms), respectively. Here, we consider the tensor rank $R=2$ for simplicity.}
	\label{fig-4rankexpansion}
\end{figure}

To proceed, we choose one tensor in the tree as the original point, and always contract the tree TN\index{TN} by ending at this tensor. Then the distance $\mathcal{D}$ of a vector is defined as the number of tensors in the path that connects this vector to the original point. Note that one impurity tensor is the tensor product of several vectors, and each vector may have different distance to the original point. For simplicity, we take the shortest one to define the distance of an impurity tensor.

Now, let us utilize the exponential convergence of the rank-1 decomposition. After contracting any vectors with the tensor in the tree, the resulting vector approaches to the fixed point (the vectors in the rank-1 decomposition) in an exponential speed. Define $\mathcal{D}_0$ as the average number of the contractions that will project any vectors to the fixed point with a tolerable difference. Consider any impurity tensors with the distance $\mathcal{D}>\mathcal{D}_0$, their contributions to the contraction are approximately the same, since after $\mathcal{D}_0$ contractions, the vectors have already been projected to the fixed point.

From the above argument, we can see that the error is related not only to the error of the rank-1 decomposition, but also to the speed of the convergence to the rank-1 component. The smaller $\mathcal{D}_0$ is, the smaller the error (the total contribution from the non-dominant terms) will be. Calculations show that the convergence speed is related to the correlation length (or gap) of the physical system, but their rigorous relations have not been established yet. Meanwhile, the expansion theory of the TN\index{TN} contraction given above requires the rank decomposition, which, however is not uniquely defined of an arbitrarily given tensor.

\section{iDMRG, iTEBD, and CTMRG revisited by tensor ring decomposition}
\label{sec5-TRD}

We have shown that the rank-1 decomposition solves the contraction of infinite-size tree TN\index{TN} and provides a mathematic explanation of the approximation made in the simple update. Then, it is natural to think: can we generalize this scheme beyond being only rank-1, in order to have better update schemes? In the following, we will show that besides the rank decomposition, the \textit{tensor ring decomposition} (TRD)\index{TRD} \cite{R16AOP} was suggested as another rank-N generalization for solving TN contraction problems. 

TRD is defined by a set of self-consistent eigenvalue equations (SEE's)\index{SEE} with certain constraints. The original proposal of TRD requires all eigenvalue equations to be Hermitian \cite{R16AOP}. Later, a generalize version was proposed \cite{TLR16tMPSArxiv} that provides an unified description of the iDMRG\index{iDMRG} \cite{W92DMRG,W93DMRG,M08iDMRGarxiv}, iTEBD\index{iTEBD} \cite{V07iTEBD}, and CTMRG\index{CTMRG} \cite{OV09CTMRG} algorithms. We will concentrate on this version in the following.

\subsection{Revisiting iDMRG, iTEBD, and CTMRG: a unified description with tensor ring decomposition}

\begin{figure}[tbp]
	\centering
	\includegraphics[angle=0,width=\linewidth]{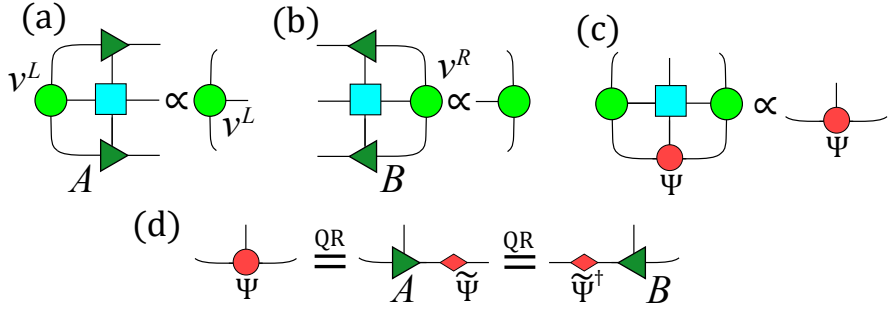}
	\caption{(Color online) The (a), (b) and (c) show the three local eigenvalue equations given by Eqs. (\ref{eq-4vR}) and (\ref{eq-4HeffDMRG}). The isometries $A$ and $B$ are obtained by the QR decompositions of $\Psi$ in two different ways in Eq. (\ref{eqs-QRPsi}), as shown in (d).}
	\label{fig-EigEqsiDMRG}
\end{figure}

Let us start from the iDMRG\index{iDMRG} algorithm. The TN contraction can be solved using the iDMRG \cite{W92DMRG,W93DMRG,M08iDMRGarxiv} by considering an infinite-size row of tensors in the TN as an MPO \cite{VGC04MPDO,ZV04MPO,LRGZXY+11LTRG,BKTMWH17FT1D,GIK17FT1D} (also see some related discussions in Sec. \ref{iTEBD}). We introduce three third-order variational tensors denoted by $v^L$, $v^R$ (dubbed as the \textit{boundary} or \textit{environmental tensors}) and $\Psi$ (dubbed as the \textit{central tensor}). These tensors are the fixed-point solution of the a set of eigenvalue equations. $v^L$ and $v^R$ are, respectively, the left and right dominant eigenvector of the following matrices (Fig. \ref{fig-EigEqsiDMRG} (a) and (b))
\begin{eqnarray}
M^L_{c'b_1'b_1,cb_2'b_2} = \sum_{aa'} T_{a'c'ac} A^{\ast}_{a'b_1'b_2'} A_{a b_1b_2}, \label{eqs-MLiDMRG} \label{eq-4vL} \\
M^R_{c'b_1'b_1,cb_2'b_2} = \sum_{aa'} T_{a'c'ac} B^{\ast}_{a'b_1'b_2'} B_{a b_1b_2}, \label{eq-4vR}
\end{eqnarray}
where $A$ and $B$ are the left and right orthogonal parts obtained by the QR\index{QR} decomposition (or SVD\index{SVD}) of $\Psi$ (Fig. \ref{fig-EigEqsiDMRG} (d)) as
\begin{eqnarray}
\Psi_{abb'} = \sum_{b''} A_{abb''} \tilde{\Psi}_{b''b'} = \sum_{b''} \tilde{\Psi}^{\dagger}_{bb''} B_{ab''b'}.  \label{eqs-QRPsi}
\end{eqnarray}
$\Psi$ is the dominant eigenvector of the Hermitian matrix (Fig. \ref{fig-EigEqsiDMRG} (c)) that satisfies
\begin{eqnarray}
\mathcal{H}_{a'b_1'b_2',ab_1b_2} = \sum_{cc'} T_{a'c'ac} v^{L}_{c'b_1'b_1} v^R_{cb_2'b_2}. \label{eq-4HeffDMRG}
\end{eqnarray}

One can see that each of the eigenvalue problems is parametrized by the solutions of others, thus we solve them in a recursive way. First, we initialize arbitrarily the central tensors $\Psi$ and get $A$ and $B$ by Eq. (\ref{eqs-QRPsi}). Note that a good initial guess can make the simulations faster and more stable. Then we update $v^L$ and $v^R$ by multiplying with $M^L$ and $M^R$ as Eqs. (\ref{eq-4vL}) and (\ref{eq-4vR}). Then we have the new $\Psi$ by solving the dominant eigenvector of $\mathcal{H}$ in Eq. (\ref{eq-4HeffDMRG}) that is defined by the new $v^L$ and $v^R$. We iterate such a process until all variational tensors converge.

\begin{figure}[tbp]
	\centering
	\includegraphics[angle=0,width=0.55\linewidth]{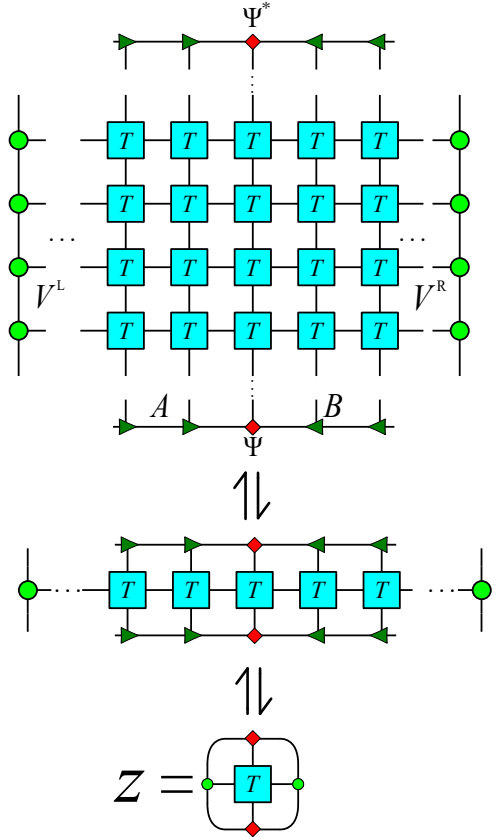}
	\caption{(Color online) The eigenvalue equations as illustrated ``encode'' the infinite TN.}
	\label{fig-4TNE}
\end{figure}

Let us rephrase the iDMRG\index{iDMRG} algorithm given above in the language of TN\index{TN} contraction/reconstruction. When the variational tensors give the fixed point, the eigenvalue equations ``encodes'' the infinite TN, i.e., the TN can be reconstructed from the equations. To do so, we start from a local representation of $Z$ (Fig. \ref{fig-4TNE}) written as
\begin{eqnarray}
Z \leftrightharpoons \sum T_{a'c'ac} \Psi^{\ast}_{a'b_1b_2} \Psi_{ab_3b_4} v^{L}_{c'b_1b_3} v^R_{c b_2b_4},
\label{eq-4LocalTNE}
\end{eqnarray}
where the summation goes through all indexes. According to the fact that $\Psi$ is the leading eigenvector of Eq. (\ref{eq-4HeffDMRG}), $Z$ is maximized with fixed $v^L$ and $v^R$. We here and below use the symbol ``$\leftrightharpoons$'' to represent the contraction relation up to a difference of a constant factor.

Then, we use the eigenvalue equations of $v^L$ and $v^R$ to add one $M^L$ and one $M^R$ [Eq. (\ref{eq-4vL}) and (\ref{eq-4vR})] in the contraction, i.e., we substitute $v^L$ by $v^LM^L$ and $v^R$ by $M^Rv^R$. After doing so for one time, a finite central orthogonal MPS\index{MPS} appears, formed by $A$, $B$ and $\Psi$. Such substitutions can be repeated for infinite times, and then we will have an infinite central-orthogonal MPS formed by $\Psi$, $A$ and $B$ as
\begin{equation}
\Phi_{\cdots a_n \cdots} = \sum_{\{b\}} \cdots A_{a_{n-2}b_{n-2}b_{n-1}} A_{a_{n-1}b_{n-1}b_{n}} \Psi_{a_n b_nb_{n+1}} B_{a_{n+1}b_{n+1}b_{n+2}} B_{a_{n+2}b_{n+2}b_{n+3}} \cdots.
\label{eq-4iMPS}
\end{equation}
One can see that the bond dimension of $b_n$ is in fact the dimension cut-off of the MPS.

Now, we have
\begin{equation}
Z \leftrightharpoons \Phi^{\dagger} \rho \Phi,
\label{eq-4ZMPO}
\end{equation}
where $\rho$ is an infinite-dimensional matrix that has the form of an MPO (middle of Fig. \ref{fig-4TNE}) as
\begin{eqnarray}
\rho_{\cdots a_n' \cdots, \cdots a_n \cdots} = \sum_{\{c\}} \cdots T_{a_{n}'c_{n} a_{n} c_{n+1}} T_{a_{n+1}'c_{n+1} a_{n+1} c_{n+2}} \cdots.
\label{eq-4iMPO}
\end{eqnarray}
$\rho$ is in fact one raw of the TN\index{TN}. Compared with Eq. (\ref{eq-4LocalTNE}), the difference of $Z$ is only a constant factor that can be given by the dominant eigenvalues of $M^L$ and $M^R$.

After the substitutions from Eq. (\ref{eq-4LocalTNE}) to (\ref{eq-4ZMPO}), $Z$ is still maximized by the given $\Phi$, since $v^L$ and $v^R$ are the dominant eigenvectors. Note that such a maximization is reached under the assumption that the dominant eigenvector $\Phi$ can be well represented in an MPS\index{MPS} with finite bond dimensions. Meanwhile, one can easily see that the MPS is normalized $|\Phi_{\cdots a_n \cdots}|=1$, thanks to the orthogonality of $A$ and $B$. Then we come to a conclusion that $\Phi$ is the optimal MPS that gives the dominant eigenvector of $\rho$, satisfying $\Phi \leftrightharpoons \rho \Phi$ \footnote{For an Hermitian matrix $M$, $v$ is its dominant eigenvector if $|v| = 1$ and $v^{\dagger} M v$ is maximized.}. Then, we can rewrite the TN\index{TN} contraction as $Z \leftrightharpoons \lim_{K \to \infty} \Phi^{\dagger} \rho^K \Phi$, where the infinite TN appears as $\rho^K$ (Fig. \ref{fig-4TNE}).

\begin{figure}[tbp]
	\centering
	\includegraphics[angle=0,width=0.5\linewidth]{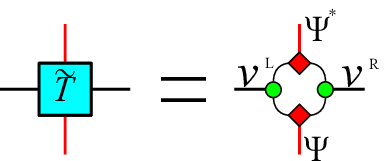}
	\caption{(Color online) The illustrations of the TRD\index{TRD} in Eq. (\ref{eq-4TRD}).}
	\label{fig-4TRD}
\end{figure}

Now we define the \textit{tensor ring decomposition} (TRD)\index{TRD} \footnote{The definition of TRD given here is from Ref. \cite{R16AOP}, which is completely different from the tensor ring decomposition proposed in Ref. \cite{ZZXZ+16tensorring}. While their TRD provides an approximation of a single tensor, the TRD discussed in this paper is more like an encoding scheme of an infinite-size TN\index{TN}.}: with the following conditions
\begin{itemize}
	\item $Z$ [Eq. (\ref{eq-4LocalTNE})] is maximized under the constraint that $v^L$ and $v^R$ are normalized,
	\item $\Phi^{\dagger} \rho \Phi$ is maximized under the constraint that $\Phi$ is normalized,
\end{itemize}
the TRD (Fig. \ref{fig-4TRD}) of $T$ is defined by $\tilde{T}$ as
\begin{eqnarray}
\tilde{T}_{a'c'ac} = \sum_{b_1b_2b_3b_4} \Psi^{\ast}_{a'b_1b_2} \Psi_{ab_3b_4} v^{L}_{c'b_1b_3} v^R_{c b_2b_4},
\label{eq-4TRD}
\end{eqnarray}
so that the TN contraction $Z = \sum T_{a'c'ac} \tilde{T}_{a'c'ac}$ [Eq. (\ref{eq-4LocalTNE})] is maximized. Like the NTD\index{NTD} and rank-1 decomposition, TRD belongs to the decompositions that encode infinite TN's, i.e., an infinite TN can be reconstructed from the self-consistent equations (note the rank decomposition does not encode any TN's). Comparing with Eq. (\ref{eq-4rank1Self}), TRD is reduced to the rank-1 decomposition by taking the dimensions of $\{b\}$ as one.

It was shown that for the same system, the ground state obtained by iDMRG\index{iDMRG} is equivalent to the ground state by iTEBD\index{iTEBD}, up to a gauge transformation \cite{M08iDMRGarxiv, HLOVV16TDVPMPS}. Different from this connection, TRD further unifies iDMRG and iTEBD. For iTEBD, after combining the contraction and truncation given by Eqs. (\ref{eq-3iTEBDEvolve}) and (\ref{eq-3iTEBDtruncate}), we have the equation for updating the tensor there as
\begin{eqnarray}
A_{s,cc'} \leftrightharpoons \sum_{s'aba'b'} T_{sbs'b'} A_{s',aa'} X_{ab,c} Y_{a'b',c'}.
\label{eq-4iTEBDupdate}
\end{eqnarray}
Looking at Eqs. (\ref{eq-4vR}) and (\ref{eq-4vR}), Eq. (\ref{eq-4iTEBDupdate}) is just the eigenvalue equation for updating $v^{[L(R)]}$ by $v^{[L(R)]} \leftrightharpoons M^{[L(R)]} v^{[L(R)]}$; the QR decomposition in Eq. (\ref{eqs-QRPsi}) guaranties that the ``truncations in iTEBD'' are implemented by isometries. In other words, one can consider another MPS\index{MPS} defined in the vertical direction, which is formed by $v^{[L(R)]}$ and updated by the iTEBD algorithm. It means that while implementing iDMRG in the parallel direction of the TN, one is in fact simultaneously implementing iTEBD to update another MPS along the vertical direction.

Particularly, when one uses iDMRG to solve the ground state of a 1D system, the MPS formed by $v^{[L(R)]}$ in the imaginary-time direction satisfies the continuous structure \cite{HM15folding, TLR16tMPSArxiv} that was originally proposed for continuous field theories \cite{VC10cMPS}. Such an iTEBD calculation can also be considered as the transverse contraction of the TN \cite{BHVC09folding, MCB12folding, HM15folding}.

CTMRG\index{CTMRG} \cite{OV09CTMRG, FVZHV17fastCTMRG} is also closely related to the scheme given above, which leads to the CTMRG without the corners. The tensors $\Psi$, $v^L$ and $v^R$ correspond to the row and column tensors, and the equations for updating these tensors are the same to the equations of updating the row and column tensors in CTMRG (see Eqs. (\ref{eq-3CTMRGcontract}) and (\ref{eq-3CTMRGtruncate})). Such a relation becomes more explicit in the rank-1 case, when corners become simply scalars. The difference is that in the original CTMRG by Or\'us \textit{et al} \cite{OV09CTMRG}, the tensors are updated with a power method, i.e., $\Psi \leftarrow \mathcal{H} \Psi$ and $v^{[L(R)]} \leftarrow M^{[L(R)]} v^{[L(R)]}$. Recently, eigen-solvers instead of power method were suggested in CTMRG (\cite{FVZHV17fastCTMRG} and a related review \cite{HV17TMTNrev}), where the eigenvalue equations of the row and column tensors are the same to those given in TRD\index{TRD}. The efficiency was shown to be largely improved with this modification.

\subsection{Extracting the information of tensor networks from eigenvalue equations: two examples}
\label{sec5.extract}

In the following, we present how to extract the properties of the TN\index{TN} by taking the \textit{free energy} and \textit{correlation length} as two example related to the eigenvalue equations. Note that these quantities correspond to the properties of the physical model and have been employed in many places (see, e.g., a review \cite{O14TNSRev}). In the following, we treated these two quantities as the properties of the TN itself. When the TN is used to represent different physical models, these quantities will be interpreted accordingly to different physical properties.

For an infinite TN, the contraction usually gives a divergent or vanishing value. The \textit{free energy} per tensor of the TN is defined to measure the contraction as
\begin{eqnarray}
f = -\lim_{N \to \infty}  \frac{\ln \mathcal{Z}}{N},
\label{eq-4freeE}
\end{eqnarray}
with $\mathcal{Z}$ the value of the contraction in theory and $N$ denoting the number of tensors. Such a definition is closely related to some physical quantities, such as the free energy of classical models and the average fidelity of TN states \cite{ZOV08fidTN}. Meanwhile, $f$ can enable us to compare the values of the contractions of two TN's without actually computing $\mathcal{Z}$.

The free energy is given by the dominant eigenvalues of $M^L$ and $M^R$. Let us reverse the above reconstructing process to show this. Firstly, we use the MPS in Eq. (\ref{eq-4iMPS}) to contract the TN\index{TN} in one direction, and have $\mathcal{Z} = (\lim_{K \to \infty} \eta^K) \Phi^{\dagger} \Phi = \lim_{K \to \infty} \eta^K$ with $\eta$ the dominant eigenvalue of $\rho$. The problem becomes getting $\eta$. By going from $\Phi^{\dagger} \rho \Phi$ to Eq. (\ref{eq-4LocalTNE}), we can see that the eigenvalue problem of $\Phi$ is transformed to that of $\mathcal{H}$ in Eq. (\ref{eq-4HeffDMRG}) multiplied by a constant $\lim_{\tilde{K} \to \infty} \kappa_0^{\tilde{K}}$ with $\kappa_0$ the dominant eigenvalue of $M^L$ and $M^R$ and $\tilde{K}$ the number of tensors in $\rho$. Thus, we have $\eta = \eta_0 \kappa_0^{\tilde{K}}$ with $\eta_0$ the dominant eigenvalue of $\mathcal{H}$. Finally, we have the TN contraction $\mathcal{Z} = [\eta_0 \kappa_0^{\tilde{K}}]^K = \eta_0^K \kappa_0^{N}$ with $K\tilde{K}=N$. By substituting into Eq. (\ref{eq-4freeE}), we have $f = -\ln \kappa_0 - \lim_{\tilde{K} \to \infty} (\ln \eta_0) /K_1 = -\ln \kappa_0$.

The second issue is about the correlations of the TN\index{TN}. The \textit{correlation function} of a TN can be defined as
\begin{eqnarray}
F(\tilde{T}^{[\bf{r}_1]}, \tilde{T}^{[\bf{r}_2]}) = \mathcal{Z}(\tilde{T}^{[\bf{r}_1]}, \tilde{T}^{[\bf{r}_2]})/\mathcal{Z} - \mathcal{Z}(\tilde{T}^{[\bf{r}_1]}, T^{[\bf{r}_2]}) \mathcal{Z}(T^{[\bf{r}_1]}, \tilde{T}^{[\bf{r}_2]}) / \mathcal{Z}^2,
\label{eq-4CorrF}
\end{eqnarray}
where $\mathcal{Z}(\tilde{T}^{[\bf{r}_1]}, \tilde{T}^{[\bf{r}_2]})$ denotes the contraction of the TN\index{TN} after substituting the original tensors in the positions $\bf{r_1}$ and $\bf{r_2}$ by two different tensors $\tilde{T}^{[\bf{r}_1]}$ and $\tilde{T}^{[\bf{r}_2]}$. $T^{[\bf{r}]}$ denotes the original tensor at the position $\bf{r}$. 

Though the correlation functions depend on the tensors that are substituted with, and can be defined in many different ways, the long-range behavior share some universal properties. For a sufficiently large distance ($|\mathbf{r_1}-\mathbf{r_2}| \gg 1$), if $\tilde{T}^{[\bf{r}_1]}$ and $\tilde{T}^{[\bf{r}_2]}$ are in a same column,  $F$ satisfies
\begin{eqnarray}
F \sim e^{-|\mathbf{r_1}-\mathbf{r_2}|/\xi}.
\label{eq-4CorrelationExp}
\end{eqnarray}
One has the correlation length
\begin{eqnarray}
\xi = 1/(\ln \eta_0 - \ln \eta_1),
\label{eq-5Xiv}
\end{eqnarray}
with $\eta_0$ and $\eta_1$ the two dominant eigenvalues of $\mathcal{H}$. If $\tilde{T}^{[\bf{r}_1]}$ and $\tilde{T}^{[\bf{r}_2]}$ are in a same row, one has
\begin{eqnarray}
\xi = 1/(\ln \kappa_0 - \ln \kappa_1),
\label{eq-4CorrelationL}
\end{eqnarray}
with $\kappa_0$ and $\kappa_1$ the two dominant eigenvalues of $M^{L(R)}$.

To prove Eq. (\ref{eq-5Xiv}), we rewrite $\mathcal{Z}(\tilde{T}^{[\bf{r}_1]}, \tilde{T}^{[\bf{r}_2]})/\mathcal{Z}$ as
\begin{eqnarray}
\mathcal{Z}(\tilde{T}^{[\bf{r}_1]}, \tilde{T}^{[\bf{r}_2]})/\mathcal{Z} = [\Phi^{\dagger} \rho(\tilde{T}^{[\bf{r}_1]}, \tilde{T}^{[\bf{r}_2]}) \Phi]/(\Phi^{\dagger} \rho \Phi).
\end{eqnarray}
Then, introduce the transfer matrix $M$ of $\Phi^{\dagger} \rho \Phi$, i.e., $\Phi^{\dagger} \rho \Phi = {\bf Tr} M^{\tilde{K}}$ with $\tilde{K} \to \infty$. With the eigenvalue decomposition of $\mathcal{M} = \sum_{j=0}^{D-1} \eta_j v_j v_j^{\dagger}$ with $D$ the matrix dimension and $v_j$ the $j$-th eigenvectors, one can further simply the equation as
\begin{eqnarray}
\mathcal{Z}(\tilde{T}^{[\bf{r}_1]}, \tilde{T}^{[\bf{r}_2]})/\mathcal{Z} =
\sum_{j=0}^{D-1} (\eta_j/\eta_0)^{|\bf{r}_1 - \bf{r}_2|} v_0^{\dagger} \mathcal{M}(\tilde{T}^{[\bf{r}_1]}) v_{j} v_{j}^{\dagger} \mathcal{M}(\tilde{T}^{[\bf{r}_1]}) v_0,
\end{eqnarray}
with $\mathcal{M}(\tilde{T}^{[\bf{r}]})$ the transfer matrix after substituting the original tensor at $\bf{r}$ with $\tilde{T}^{[\bf{r}]}$. Similarly, one has
\begin{eqnarray}
\mathcal{Z}(\tilde{T}^{[\bf{r}_1]}, T)/\mathcal{Z} = v_0^{\dagger} \mathcal{M}(\tilde{T}^{[\bf{r}_1]}) v_0, \\
\mathcal{Z}(T,\tilde{T}^{[\bf{r}_2]})/\mathcal{Z} = v_0^{\dagger} \mathcal{M}(\tilde{T}^{[\bf{r}_2]}) v_0.
\end{eqnarray}
Note that one could transform the MPS into a translationally invariant form (e.g., the canonical form) to uniquely define the transfer matrix of $\Phi^{\dagger} \rho \Phi$. Substituting the equations above in Eq. (\ref{eq-4CorrF}), one has
\begin{eqnarray}
F(\tilde{T}^{[\bf{r}_1]}, \tilde{T}^{[\bf{r}_2]}) =
\sum_{j=1}^{D-1} (\eta_j/\eta_0)^{|\bf{r}_1 - \bf{r}_2|} v_0^{\dagger} \mathcal{M}(\tilde{T}^{[\bf{r}_1]}) v_{j} v_{j}^{\dagger} \mathcal{M}(\tilde{T}^{[\bf{r}_1]}) v_0.
\end{eqnarray}
When the distance is sufficiently large, i.e., $|\mathbf{r}_1 - \mathbf{r}_2| \gg 1$, only the dominant term takes effects, which is
\begin{eqnarray}
F(\tilde{T}^{[\bf{r}_1]}, \tilde{T}^{[\bf{r}_2]}) \simeq
(\eta_1/\eta_0)^{|\bf{r}_1 - \bf{r}_2|} v_0^{\dagger} \mathcal{M}(\tilde{T}^{[\bf{r}_1]}) v_{1} v_{1}^{\dagger} \mathcal{M}(\tilde{T}^{[\bf{r}_1]}) v_0.
\end{eqnarray}
Compared with Eq. (\ref{eq-4CorrelationExp}), one has $\xi = 1/(\ln \eta_0 - \ln \eta_1)$. The second case can be proven similarly.

These two quantities are defined independently on specific physical models that the TN\index{TN} might represent, thus they can be considered as the mathematical properties of the TN. By introducing physical models, these quantities are closely related to the physical properties. For example, when the TN represents the the partition function of a classical lattice model, Eq. (\ref{eq-4freeE}) multiplied by the temperature is exactly the free energy. And the correlation lengths of the TN are also the physical correlation lengths of the model in two spatial directions. When the TN gives the imaginary time evolution of an infinite 1D quantum chain, the correlation lengths of the TN are the spatial and dynamical correlation length of the ground state.

It is a huge topic to investigate the properties of the TN's or TN states. Paradigm examples include injectivity and symmetries \cite{ZBWC15fPEPS,HVSC15TNgauge,  MOP16TNsymme,PSGWC10symme,  W12Qspace, SCP10PEPSsymme, SPV10TNsymme, SPV11symmeU1, O11advTN, BCOT11assyme, SV12SU2symme, TCL14TNsymme, RDS15PEPSZ2, JR15TNsymme, LH16TNsymme, O14inject}, statistics and fusion rules \cite{GLSW09StringTPS, BAV09StringTPS,  DBJC12TopoTNS, PCBA+10AnyonTN}.  These issues are beyond the scope of this lecture notes. One may refer to the related works if interested.

\chapter{Quantum entanglement simulation inspired by tensor network}

\abstract{This chapter is focused the quantum entanglement simulation approach \cite{RPPSL17AOP3D}. The idea is to use few-body models embedded in the ``entanglement bath'' to mimic the properties of large and infinite-size quantum systems. The few-body models are dubbed as quantum entanglement simulators. Generally speaking, the QES\index{QES} approach consists of three stages: first, determine the effective interactions that give the infinite boundary condition \cite{RPPSL17AOP3D, PVM12InfBound} by the MPS/PEPS\index{MPS}\index{PEPS} methods, such as iDMRG\index{iDMRG} in one dimension or zero-loop scheme in two and three dimensions; second, construct the simulators by surrounding a finite-size cluster of the targeted model with the effective interactions; third, simulate the properties of the quantum entanglement simulator by the finite-size algorithms or quantum hardware, and extract the properties of the targeted model within the bulk.} 


\section{Motivation and general ideas}

An impression one may have for the TN\index{TN} approaches of the quantum lattice models is that the algorithm (i.e., how to contract and/or truncate) will dramatically change when considering different models or lattices. This motivates us to look for more unified approaches. Considering that a huge number of problems can be transformed to TN contractions, one general question we may ask is: how can we reduce a nondeterministic-polynomial-hard TN contraction problem approximately to an effective one that can be computed exactly and efficiently by classical computers? We shall put some principles while considering this question: the effective problem should be as simple as possible, containing as few parameters to solve as possible. We would like to coin this principle for TN contractions as the \textit{ab-initio} optimization principle (AOP)\index{AOP} of TN \cite{R16AOP}. The term ``\textit{ab-inito}'' is taken here to pay respect to the famous \textit{ab-inito} principle approaches in condensed matter physics and quantum chemistry (see several recent reviews in \cite{S10DFTRev,B12DFT,B14DFT}). Here, ``\textit{ab-inito}'' means to think from the very beginning, with least prior knowledge of or assumptions to the problems.

One progress achieved in the spirit of AOP is the TRD\index{TRD} introduced in Sec. \ref{sec5-TRD}. Considering the TN on an infinite square lattice, its contraction is reduced to a set of self-consistent eigenvalue equations that can be efficiently solved by classical computers. The variational parameters are just two tensors. One advantage of TRD is that it connects the TN algorithms (iDMRG\index{iDMRG}, iTEBD\index{iTEBD}, CTMRG\index{CTMRG}), which are previously considered to be quite different, in a unified picture.

Another progress made in the AOP\index{AOP} spirit is called QES\index{QES} for simulating infinite-size physical models \cite{RPPSL17AOP3D, R16AOP, RXPS+18FTQES}. It is less dependent on the specific models; it also provides a natural way for designing quantum simulators and for hybridized-quantum-classical simulations of many-body systems. Hopefully in the future when people are aboe to readily realize the designed Hamiltonians on artificial quantum platforms, QES will enable us to design the Hamiltonians that will realized quantum many-body phenomena

\section{Simulating one-dimensional quantum lattice models}

Let us firstly take the ground-state simulation of the infinite-size 1D quantum system as an example. The Hamiltonian is the summation of two-body nearest-neighbor terms, which reads $\hat{H}_{Inf} = \sum_{n} \hat{H}_{n,n+1}$. The translational invariance is imposed. The first step is to choose a supercell (e.g., a finite part of the chain with $\tilde{N}$ sites). Then the Hamiltonian of the supercell is $\hat{H}_{B} = \sum_{n=1}^{\tilde{N}} \hat{H}_{n,n+1}$, and the Hamiltonian connecting the supercell to the rest part is $\hat{H}_{\partial} = \hat{H}_{n',n'+1}$ (note the interactions are nearest-neighbor).

\begin{figure}[tbp]
	\centering
	\includegraphics[angle=0,width=0.75\linewidth]{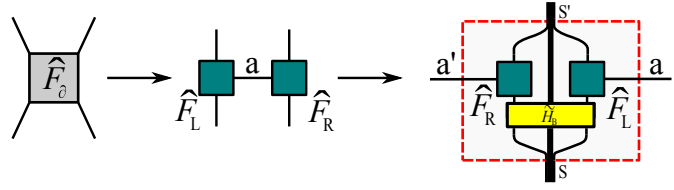}
	\caption{(Color online) Graphical representations of Eq.(\ref{eqs-Fpartial})-(\ref{eqs-cellT}).}
	\label{fig-4OperatorF}
\end{figure}

Define the operator $\hat{F}^{\partial}$ as
\begin{eqnarray}
\hat{F}_{\partial} = \hat{I} - \tau \hat{H}_{\partial},
\label{eqs-Fpartial}
\end{eqnarray}
with $\tau$ the Trotter-Suzuki step. This definition is to construct the Trotter-Suzuki decomposition \cite{SI87Trotter,IS88Trotter}. Instead of using the exponential form $e^{-\tau \hat{H}}$, we equivalently chose to shift $\hat{H}_{\partial}$ for algorithmic consideration. The errors of these two ways concerning the ground state are at the same level ($\mathcal{O}(\tau^2)$). Introduce an ancillary index $a$ and rewrite $\hat{F}_{\partial}$ as a sum of operators as
\begin{eqnarray}
\hat{F}_{\partial} = \sum_a \hat{F}_{L}(s)_a \hat{F}_{R}(s')_a,
\label{eqs-Fboundary}
\end{eqnarray}
where $\hat{F}_{L}(s)_a$ and $\hat{F}_{R}(s')_a$ are two sets of one-body operators (labeled by $a$) acting on the left and right one of the two spins ($s$ and $s'$) associated with $\hat{H}_{\partial}$, respectively (Fig. \ref{fig-4OperatorF}). Eq. (\ref{eqs-Fboundary}) can be easily achieved directly from the Hamiltonian or using eigenvalue decomposition. For example for the Heisenberg interaction with $\hat{H}_{\partial} = \sum_{\alpha=x,y,z} J_{\alpha} \hat{S}^{\alpha}(s)\hat{S}^{\alpha}(s')$ with $\hat{S}^{\alpha}(s)$ the spin operators. We have $\hat{F}_{\partial} = \sum_{a = 0,x,y,z} \tilde{J}_{a} \hat{S}^{a}(s) \hat{S}^{a}(s')$ with $\hat{S}^{0} = I$, $\tilde{J}_0 = 1$, $\tilde{J}_{a} = -\tau J_{a}$ ($\alpha = x, y, z$), hence we can define $\hat{F}_{L}(s)_a = \sqrt{|\tilde{J}_{a}|} \hat{S}^{\alpha}(s)$ and $\hat{F}_{R}(s)_a = \text{sign}(\tilde{J}_{a}) \sqrt{|\tilde{J}_{a}|} \hat{S}^{\alpha}(s')$.

Construct the operator $\hat{\mathcal{F}}(S)_{a'a}$, with $S=(s_1,\cdots,s_{\tilde{N}})$ representing the physical spins inside the super-cell, as
\begin{eqnarray}
\hat{\mathcal{F}}(S)_{a'a}= \hat{F}_{R}(s_1)_{a'}^{\dagger} \tilde{H}_B \hat{F}_{L}(s_{\tilde{N}})_a,
\label{eqs-getF}
\end{eqnarray}
with $\tilde{H}_B = \hat{I} - \tau \hat{H}_B$. $\hat{F}_{R}(s_1)_{a'}^{\dagger}$ and $\hat{F}_{L}(s_{\tilde{N}})_{a}$ act on the first and last sites of the super-cell, respectively. One can see that $\hat{\mathcal{F}}(S)_{a'a}$ represents a set of operators labeled by two indexes ($a'$ and $a$) that act on the supercell.

In the language of TN\index{TN}, the co-efficients of $\hat{\mathcal{F}}(S)_{a'a}$ in the local basis ($|S\rangle = |s_1\rangle \cdots |s_{\tilde{N}}\rangle$) is a forth-order cell tensor (Fig. \ref{fig-4OperatorF}) as
\begin{eqnarray}
T_{S'a'Sa} = \langle S'| \hat{\mathcal{F}}(S)_{a'a} | S\rangle.
\label{eqs-cellT}
\end{eqnarray}
On the left-hand-side, the order of the indexes are arranged to be consistent with the definition in the TN algorithm introduced in the precious sections. $T$ is the cell tensor, whose infinite copies form the TN of the imaginary-time evolution up to the first Trotter-Suzuki order. One may consider the second Trotter-Suzuki order by defining $\hat{\mathcal{F}}(S)_{a'a}$ as $\hat{\mathcal{F}}(S)_{a'a} = (\hat{I} - \tau \hat{H}_B /2) \hat{F}_{R}(s_1)_{a'}^{\dagger} \hat{F}_{L}(s_{\tilde{N}})_a (\hat{I} - \tau \hat{H}_B /2)$. With the cell tensor $T$, the ground-state properties can be solved using the TN algorithms (e.g., TRD) introduced above. The ground state is given by the MPS given by Eq. (\ref{eq-4iMPS}).

Let us consider one of the eigenvalue equations [also see Eq. (\ref{eq-4HeffDMRG})] in the TRD\index{TRD}
\begin{eqnarray}
\mathcal{H}_{S'b_1'b_2',Sb_1b_2} = \sum_{aa'} T_{S'a'Sa} v^{L}_{a'b_1'b_1} v^R_{ab_2'b_2}. \label{eq-4HeffDMRG1}
\end{eqnarray}
We define a new Hamiltonian $\hat{\mathcal{H}}$ by using $\mathcal{H}_{S'b_1'b_2',Sb_1b_2}$ as the coefficients
\begin{eqnarray}
\hat{\mathcal{H}} =\sum_{SS'} \sum_{b_1b_2b_1'b_2'} \mathcal{H}_{S'b_1'b_2',Sb_1b_2} |S'b_1'b_2'\rangle\langle Sb_1b_2|.
\label{eq-4HeffDMRGop}
\end{eqnarray}
$\hat{\mathcal{H}}$ is the effective Hamiltonian in iDMRG\index{iDMRG} \cite{W92DMRG,W93DMRG,M07DMRGsymme} or the methods which represent the RG\index{RG} of Hilbert space by MPS \index{MPS} \cite{ZVFVH18vuMPS, HLOVV16TDVPMPS}. The indexes $\{b\}$ are considered as virtual spins with basis $\{|b\rangle\}$. The virtual spins are called the \textit{entanglement bath sites} in the QES\index{QES}.

By substituting with the cell tensor $T$ [Eqs. (\ref{eqs-getF}) and (\ref{eqs-cellT})] inside the above equation, we have
\begin{eqnarray}
\mathcal{\hat{H}} = \mathcal{\hat{H}}_L \tilde{H}_B \mathcal{\hat{H}}_R,
\label{eqs-Heffect}
\end{eqnarray}
where the Hamiltonians $\mathcal{\hat{H}}_L$ and $\mathcal{\hat{H}}_R$ locate on the boundaries of $\mathcal{\hat{H}}$, whose coefficients satisfy
\begin{eqnarray}
\begin{aligned}
\langle b_1's_{1}'| \mathcal{\hat{H}}_L |b_1 s_{1}\rangle = \sum_{a} v^{L}_{ab_1'b_1} \langle s_1' |\hat{F}_{R}(s_1)_{a}^{\dagger} | s_1\rangle, \\
\langle s_{\tilde{N}}' b_2'| \mathcal{\hat{H}}_R |s_{\tilde{N}} b_2\rangle = \sum_a \langle s_{\tilde{N}}'| \hat{F}_{L}(s_{\tilde{N}})_a^{\dagger} | s_{\tilde{N}}\rangle v^{R}_{ab_2b_2'}.
\end{aligned}
\label{eqs-HBtwobody}
\end{eqnarray}
$\mathcal{\hat{H}}_L$ and $\mathcal{\hat{H}}_R$ are just two-body Hamiltonians, of which each acts on the bath site and its neighboring physical site on the boundary of the bulk; they define the infinite boundary condition for simulating the time evolution of 1D quantum systems \cite{PVM12InfBound}.

$\mathcal{\hat{H}}_L$ and $\mathcal{\hat{H}}_R$ can also be written in a shifted form as
\begin{equation}
\mathcal{\hat{H}}_{L(R)} = I-\tau \hat{H}_{L(R)}.
\label{eqs-HfbLR}
\end{equation}
This is because the tensor $v^L$ (and also $v^R$) satisfies a special form \cite{TLR16tMPSArxiv} as
\begin{eqnarray}
v^{L}_{0,bb'} = I_{bb'} - \tau Q_{bb'}, \\
v^{L}_{a,bb'} = \tau^{a/2} (R^a)_{bb'} \ \ (a > 0), \\
\end{eqnarray}
with $Q$ and $R$ two Hermitian matrices independent on $\tau$. In other words, the MPS\index{MPS} formed by infinite copies of $v^{L}$ or $v^{R}$ is a continuous MPS \cite{VC10cMPS}, which is known as the temporal MPS \cite{HM15folding}. Therefore,  $\hat{H}_{L(R)}$ is independent on $\tau$, called the \textit{physical-bath Hamiltonian}. Then $\mathcal{\hat{H}}$ can be written as the shift of a few-body Hamiltonian as $\mathcal{\hat{H}} = I- \tau \hat{H}_{FB}$, where $\hat{H}_{FB}$ has the standard summation form as
\begin{eqnarray}
\hat{H}_{FB} = \hat{H}_L + \sum_{n=1}^{L} \hat{H}_{n,n+1} + \hat{H}_R.
\label{eqs-Hfewbody}
\end{eqnarray}

For $\hat{H}_{L}$ and $\hat{H}_{R}$ with the bath dimension$\chi$, the coefficient matrix of $\hat{H}_{L(R)}$ is ($2\chi \times 2\chi$). Then $\hat{H}_{L(R)}$ can be generally expanded by $\hat{S}^{\alpha_1} \otimes \hat{\mathcal{S}}^{\alpha_2}$ with $\{\hat{\mathcal{S}}\}$ the generators of the SU($\chi$) group, and define the magnetic field and coupling constants associated to the entanglement bath
\begin{eqnarray}
\hat{H}_{L(R)} = \sum_{\alpha_1,\alpha_2} J^{\alpha_1\alpha_2}_{L(R)} \hat{\mathcal{S}}^{\alpha_1} \otimes \hat{S}^{\alpha_2},
\label{eqs-Constants}
\end{eqnarray}
with $\mathcal{S}$ denoting the SU($\chi$) spin operators and $\hat{S}$ the operators of the physical spin (with the identity included).

Let us take the bond dimension $\chi=2$ as an example, and $\hat{H}_{L(R)}$ just gives the Hamiltonian between two spin-$1/2$'s. Thus, it can be expanded by the spin (or Pauli) operators $\hat{S}^{\alpha_1} \otimes \hat{S}^{\alpha_2}$ as
\begin{eqnarray}
\hat{H}_{L(R)} = \sum_{\alpha_1,\alpha_2=0}^3 J^{\alpha_1\alpha_2}_{L(R)} \hat{S}^{\alpha_1} \otimes \hat{S}^{\alpha_2},
\label{eqs-ConstantsSU2}
\end{eqnarray}
where the spin-$1/2$ operators are labeled as $\hat{S}^0 = I$, $\hat{S}^1 = \hat{S}^x$, $\hat{S}^2 = \hat{S}^y$, and $\hat{S}^3 = \hat{S}^z$. Then with $\alpha_1 \neq 0$ and $\alpha_2 \neq 0$, we have $J^{\alpha_1\alpha_2}_{L(R)}$ as the coupling constants, and $J^{\alpha_1 0}_{L(R)}$ and $J^{0 \alpha_2}_{L(R)}$ the magnetic fields on the first and second sites, respectively. $J^{0 0}_{L(R)}$ only provides a constant shift of the Hamiltonian which does not change the eigenstates.

\begin{figure}[tbp]
	\centering
	\includegraphics[angle=0,width=1\linewidth]{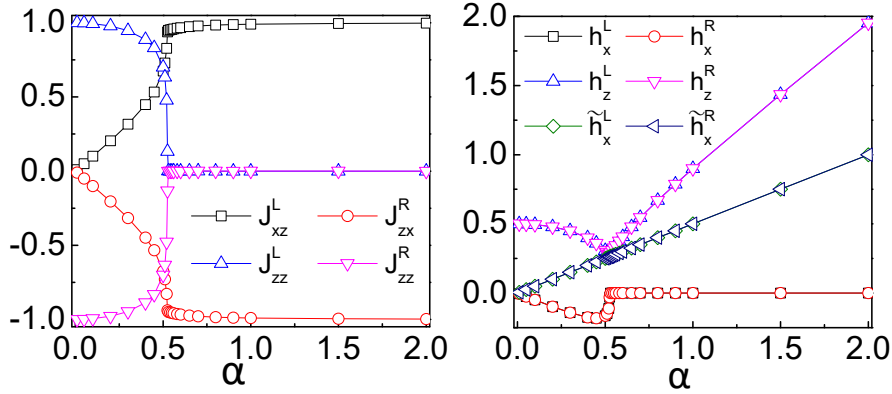}
	\caption{(Color online) The $\alpha$-dependence \cite{RPSL17CSW} of the coupling constants (left) and magnetic fields (right) of the few-body Hamiltonians [Eq. (\ref{eq-4HLR})]. Reused from \cite{RPSL17CSW} with permission.}
	\label{fig-4Hbath}
\end{figure}

As an example, we show the $\hat{H}_{L}$ and $\hat{H}_{R}$ for the infinite quantum Ising chain in a transverse field \cite{RPSL17CSW}. The original Hamiltonian reads $\hat{H}_{Ising} = \sum_n \hat{S}^z_n \hat{S}^z_{n+1} -\alpha \sum_{n} \hat{S}^x_n$, and $\hat{H}_{L}$ and $\hat{H}_{R}$ satisfies
\begin{eqnarray}
\begin{aligned}
&\hat{H}_{L} = J^L_{xz} \hat{S}^x_1 \hat{S}^z_2 + J^L_{zz} \hat{S}^z_1 \hat{S}^z_2 - h^L_{x} \hat{S}^x_1 - h^L_{z} \hat{S}^z_1 - \tilde{h}^L_{x} \hat{S}^x_2,\\
&\hat{H}_{R} = J^R_{zx} \hat{S}^z_{N-1} \hat{S}^x_N + J^R_{zz} \hat{S}^z_{N-1} \hat{S}^z_{N} - h^R_{x} \hat{S}^x_{N} - h^R_{z} \hat{S}^z_{N} - \tilde{h}^R_{x} \hat{S}^x_{N-1}.
\end{aligned}
\label{eq-4HLR}
\end{eqnarray}
The coupling constants and magnetic fields depend on the transverse field $\alpha$, as shown in Fig. \ref{fig-4Hbath}. The calculation shows that except the Ising interactions and the transverse field that originally appear in the infinite model, the $\hat{S}^x \hat{S}^z$ coupling and a vertical field emerge in $\hat{H}_{L}$ and $\hat{H}_{R}$. This is interesting, because the $\hat{S}^x \hat{S}^z$ interaction is the stabilizer on the open boundaries of the cluster state, a highly entangled state that has been widely used in quantum information sciences \cite{BR01EntgleState,RB01EntgleState}. More relations with the cluster state are to be further explored.

The physical information of the infinite-size model can be extracted from the ground state of $\hat{H}_{FB}$ (denoted by $|\Psi(Sb_1b_2)\rangle$) by tracing over the entanglement-bath degrees of freedom. To this aim, we calculate the reduced density matrix of the bulk as
\begin{eqnarray}
\hat{\rho}(S) = \text{Tr}_{b_1b_2} |\Psi(Sb_1b_2) \rangle \langle \Psi(Sb_1b_2)|.
\label{eq-4RDM}
\end{eqnarray}
Note $|\Psi(Sb_1b_2)\rangle = \sum_{Sb_1b_2} \Psi_{Sb_1b_2} |Sb_1b_2\rangle$ with $\Psi_{Sb_1b_2}$ the eigenvector of Eq. (\ref{eq-4HeffDMRG1}) or (\ref{eq-4HeffDMRG}). It is easy to see that $\Psi_{Sb_1b_2}$ is the central tensor in the central-orthogonal MPS\index{MPS} [Eq. (\ref{eq-4iMPS})], thus the $\hat{\rho}(S)$ is actually the reduced density matrix of the MPS. Since the MPS optimally gives the ground state of the infinite model, therefore, $\hat{\rho}(S)$ of the few-body ground state optimally gives the reduced density matrix of the original model.

In Eq. (\ref{eqs-Hfewbody}), the summation of the physical interactions are within the supercell that we choose to construct the cell tensor. To improve the accuracy to, e.g., capture longer correlations inside the bulk, one just needs to increase the supercell in $\hat{H}_{FB}$. In other words, $\hat{H}_L$ and $\hat{H}_R$ are obtained by TRD\index{TRD} from the supercell of a tolerable size $\tilde{N}$, and $\hat{H}_{FB}$ is constructed with a larger bulk as $\hat{H}_{FB} = \hat{H}_L + \sum_{n=1}^{\tilde{N}'} \hat{H}_{n,n+1} + \hat{H}_R$ with $\tilde{N}'>\tilde{N}$. Though $\hat{H}_{FB}$ becomes more expensive to solve, we have any well-established finite-size algorithms to compute its dominant eigenvector. We will show below that this way is extremely useful in higher dimensions.

\section{Simulating higher-dimensional quantum systems}

For ($D>1$)-dimensional quantum systems on, e.g., square lattice, one can use different update schemes to calculate the ground state. Here, we explain an alternative way by generalizing the above 1D simulation to higher dimensions \cite{RPPSL17AOP3D}. The idea is to optimize the physical-bath Hamiltonians by the zero-loop approximation (simple update, see Sec. \ref{sec.looplessTNrank-1}), e.g., iDMRG\index{iDMRG} on tree lattices \cite{LCP00TreeDMRG,NC13TTN}, and then construct the few-body Hamiltonian $\hat{H}_{FB}$ with larger bulks. The loops inside the bulk will be fully considered when solving the ground state of $\hat{H}_{FB}$, thus the precision will be significantly improved compared with the zero-loop approximation.

\begin{figure}[tbp]
	\centering
	\includegraphics[angle=0,width=0.6\linewidth]{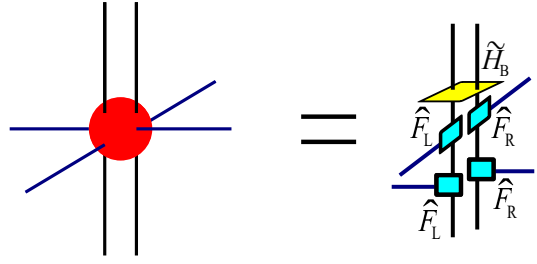}
	\caption{(Color online) Graphical representation of the cell tensor for 2D quantum systems [Eq. (\ref{eq-getF})].}
	\label{fig-4getF}
\end{figure}

The procedures are similar to those for 1D models. The first step is to contract the cell tensor, so that the ground-state simulation is transformed to a TN\index{TN} contraction problem. We choose the two sites connected by a parallel bond as the supercell, and construct the cell tensor that parametrizes the eigenvalue equations. The bulk interaction is simply the coupling between these two spins, i.e. $\hat{H}_B = \hat{H}_{i,j}$, and the interaction between two neighboring supercells is the same, i.e., $\hat{H}_{\partial} = \hat{H}_{i,j}$. By shifting $\hat{H}_{\partial}$, we define $\hat{F}_{\partial} = I-\tau \hat{H}_{\partial}$ and decompose it as
\begin{eqnarray}
\hat{F}_{\partial} = \sum_{a} \hat{F}_{L}(s)_a \otimes \hat{F}_{R}(s')_a.
\label{eq-Fsvd}
\end{eqnarray}
$\hat{F}_{L}(s)_a$ and $\hat{F}_{R}(s')_a$ are two sets of operators labeled by $a$ that act on the two spins ($s$ and $s'$) in the supercell, respectively (see the texts below Eq. (\ref{eqs-Fboundary}) for more detail).

Define a set of operators by the product of the (shifted) bulk Hamiltonian with $\hat{F}_{L}(s)_a$ and $\hat{F}_{R}(s)_a$ (Fig. \ref{fig-4getF}) as
\begin{equation}
\hat{\mathcal{F}}(S)_{a_1a_2a_3a_4}= \hat{F}_{R}(s)_{a_1} \hat{F}_{R}(s)_{a_2} \hat{F}_{L}(s')_{a_3} \hat{F}_{L}(s')_{a_4} \tilde{H}^B,
\label{eq-getF}
\end{equation}
with $S=(s,s')$ and $\tilde{H}^B = I- \tau \hat{H}^B$. The cell tensor that defines the TN\index{TN} is given by the coefficients of $\hat{\mathcal{F}}(S)_{a_1a_2a_3a_4}$ as
\begin{eqnarray}
T_{S'Sa_1a_2a_3a_4} = \langle S'| \hat{\mathcal{F}}(S)_{a_1a_2a_3a_4} |S\rangle.
\label{eq-4Tcell}
\end{eqnarray}
One can see that $T$ has six bonds, of which two ($S$ and $S'$) are physical and four ($a_1$, $a_2$, $a_3$, and $a_4$) are non-physical. For comparison, the tensor in the 1D quantum case has four bonds, where two are physical and two are non-physical [see Eq. (\ref{eqs-cellT})]. As discussed above in Sec. \ref{sec.evo2D}, the ground-state simulation becomes the contraction of a cubic TN formed by infinite copies of $T$. Each layer of the cubic TN gives the operator $\hat{\rho}(\tau) = I-\tau \hat{H}$, which is a PEPO\index{PEPO} defined on a square lattice. Infinite layers of the PEPO $\lim_{K \to \infty} \hat{\rho}(\tau)^K$ give the cubic TN\index{TN}. 

\begin{figure}[tbp]	\includegraphics[angle=0,width=0.8\linewidth]{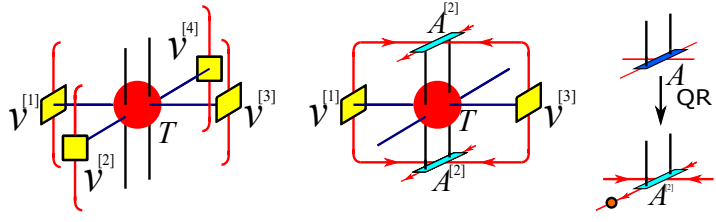}
	\centering
	\caption{(Color online) The left figure is the graphic representations of $\mathcal{H}_{S'b_1'b_2'b_3'b_4',Sb_1b_2b_3b_4}$ in Eq.(\ref{eq-effectH}), and we take Eq.(\ref{eq-M2}) from the self-consistent equations as an example shown in the middle. The QR\index{QR} decomposition in Eq.(\ref{eq-4QRPsi}) is shown in the right figure, where the arrows indicate the direction of orthogonality of $A^{[3]}$ in Eq.(\ref{eq-OrtA3}).}
	\label{fig-4SelfConsisEq}
\end{figure}

The next step is to solve the SEE's\index{SEE} of the zero-loop approximation. For the same model defined on the loopless Bethe lattice, the 3D TN is formed by infinite layers of PEPO $\hat{\rho}_{Bethe}(\tau)$ that is defined on the Bethe lattice. The cell tensor is defined exactly in the same way as Eq. (\ref{eq-4Tcell}). With the Bethe approximation, there are five variational tensors, which are $\Psi$ (central tensor) and $v^{[x]}$ ($x=1,2,3,4$, boundary tensors). Meanwhile, we have five self-consistent equations that encodes the 3D TN $\lim_{K \to \infty} \hat{\rho}_{Bethe}(\tau)^K$, which are given by five matrices as
\begin{eqnarray}
\mathcal{H}_{S'b_1'b_2'b_3'b_4',Sb_1b_2b_3b_4} = \sum_{a_1a_2a_3a_4} T_{S'Sa_1a_2a_3a_4} v^{[1]}_{a_1b_1 b_1'} v^{[2]}_{a_2b_2 b_2'} v^{[3]}_{a_3b_3 b_3'} v^{[4]}_{a_4b_4 b_4'}, \label{eq-effectH}\\
M^{[1]}_{a_1b_1b_1',a_3b_3b_3'} = \sum_{S'Sa_2a_4b_2b_2'b_4b_4'} T_{S'Sa_1a_2a_3a_4} A^{[1]\ast}_{S'b_1'b_2'b_3'b_4'} v^{[2]}_{a_2b_2 b_2'} A^{[1]}_{Sb_1b_2b_3b_4} v^{[4]}_{a_4b_4 b_4'},\label{eq-M1}\\
M^{[2]}_{a_2b_2b_2',a_4b_4b_4'} = \sum_{S'Sa_1a_3b_1b_1'b_3b_3'} T_{S'Sa_1a_2a_3a_4} A^{[2]\ast}_{S'b_1'b_2'b_3'b_4'} v^{[1]}_{a_1b_1 b_1'} A^{[2]}_{Sb_1b_2b_3b_4} v^{[3]}_{a_3b_3 b_3'},\label{eq-M2}\\
M^{[3]}_{a_1b_1b_1',a_3b_3b_3'} = \sum_{S'Sa_2a_4b_2b_2'b_4b_4'} T_{S'Sa_1a_2a_3a_4} A^{[3]\ast}_{S'b_1'b_2'b_3'b_4'} v^{[2]}_{a_2b_2 b_2'} A^{[3]}_{Sb_1b_2b_3b_4} v^{[4]}_{a_4b_4 b_4'},\label{eq-M3}\\
M^{[4]}_{a_2b_2b_2',a_4b_4b_4'} = \sum_{S'Sa_1a_3b_1b_1'b_3b_3'} T_{S'Sa_1a_2a_3a_4} A^{[4]\ast}_{S'b_1'b_2'b_3'b_4'} v^{[1]}_{a_1b_1 b_1'} A^{[4]}_{Sb_1b_2b_3b_4} v^{[3]}_{a_3b_3 b_3'}.\label{eq-M4}
\end{eqnarray}
Eqs. (\ref{eq-effectH}) and (\ref{eq-M2}) are illustrated in Fig. \ref{fig-4SelfConsisEq} as two examples. $A^{[x]}$ is an isometry obtained by the QR\index{QR} decomposition (or SVD\index{SVD}) of the central tensor $\Psi$ referring to the $x$-th virtual bond $b_x$. For example for $x=2$, we have (Fig. \ref{fig-4SelfConsisEq})
\begin{eqnarray}
\Psi_{Sb_1b_2b_3b_4} = \sum_{b} A^{[2]}_{Sb_1bb_3b_4} R^{[2]}_{b b_2}.
\label{eq-4QRPsi}
\end{eqnarray}
$A^{[2]}$ is orthogonal, satisfying
\begin{eqnarray}
\sum_{Sb_1 b_3 b_4} A^{[2] \ast}_{Sb_1bb_3b_4} A^{[2]}_{Sb_1b'b_3b_4} = I_{bb'}.
\label{eq-OrtA3}
\end{eqnarray}

The self-consistent equations can be solved recursively. By solving the leading eigenvector of $\mathcal{H}$ given by Eq. (\ref{eq-effectH}), we update the central tensor $\Psi$. Then according to Eq. (\ref{eq-4QRPsi}), we decompose $\Psi$ to obtain $A^{[x]}$, then update $M^{[x]}$ in Eqs. (\ref{eq-M1})-(\ref{eq-M4}), and update each $v^{[x]}$ by $M^{[x]} v^{[x]}$. Repeat this process until all the five variational tensors converge. The algorithm is the generalized DMRG\index{DMRG} based on infinite tree PEPS\index{PEPS} \cite{LCP00TreeDMRG,NC13TTN}. Each boundary tensor can be understood as the infinite environment of a tree branch, thus the original model is actually approximated at this stage by that defined on an Bethe lattice. Note that when only looking at the tree locally (from one site and its nearest neighbors), it looks the same to the original lattice. Thus, the loss of information is mainly long-range, i.e., from the destruction of loops.

\begin{figure}[tbp]
	\centering
	\includegraphics[angle=0,width=0.7\linewidth]{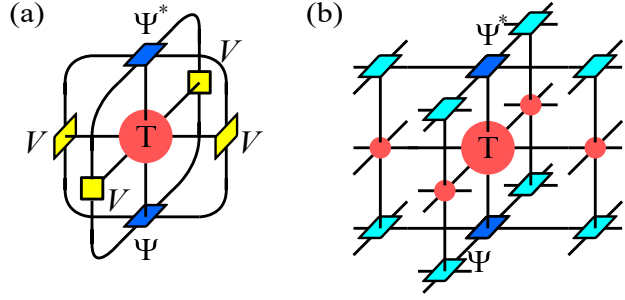}
	\caption{(Color online) The left figure shows the local contraction the encodes the infinite TN\index{TN} for simulating the 2D ground state. By substituting with the self-consistent equations, the TN representing $\tilde{Z} = \langle \tilde{\Phi}| \hat{\rho}_{Bethe}(\tau) | \tilde{\Phi} \rangle$ can be reconstructed, with $\hat{\rho}_{Bethe}(\tau)$ the tree PEPO\index{PEPO} of the Bethe model and $| \tilde{\Phi} \rangle$ a PEPS\index{PEPS}.}
	\label{fig-4ZBethe}
\end{figure}

The Bethe approximation can be understood better from the rank-1 decomposition (see Sec. 4.5). Eqs. (\ref{eq-M1})-(\ref{eq-M4}) encodes a Bethe TN, whose contraction is written as $Z_{Bethe} = \langle \tilde{\Phi}| \hat{\rho}_{Bethe}(\tau) | \tilde{\Phi} \rangle$ with $\hat{\rho}_{Bethe}(\tau)$ the PEPO\index{PEPO} of the Bethe model and $| \tilde{\Phi} \rangle$ a tree iPEPS\index{iPEPS} (Fig.\ref{fig-4ZBethe}). To see this, let us start with the local contraction [Fig.\ref{fig-4ZBethe} (a)] as
\begin{eqnarray}
Z_{Bethe} = \sum \Psi^{\ast}_{S'b_1'b_2'b_3'b_4'} \Psi_{Sb_1b_2b_3b_4} T_{S'Sa_1a_2a_3a_4} v^{[1]}_{a_1b_1 b_1'} v^{[2]}_{a_2b_2 b_2'} v^{[3]}_{a_3b_3 b_3'} v^{[4]}_{a_4b_4 b_4'}.
\label{eq-4Zlocal}
\end{eqnarray}
Then, each $v^{[x]}$ can be replaced by $M^{[x]} v^{[x]}$ because we are at the fixed point of the eigenvalue equations. By repeating this substitution in a similar way as the rank-1 decomposition in Sec. \ref{sec.zeroloop}, we will have the TN\index{TN} for $Z_{Bethe} $, which is maximized at the fixed point [Fig.\ref{fig-4ZBethe} (b)]. With the constraint $\langle \tilde{\Phi}| \tilde{\Phi} \rangle =1$ satisfied, $| \tilde{\Phi} \rangle$ is the ground state of $\hat{\rho}_{Bethe}(\tau)$.

Now, we constrain the growth so that the TN\index{TN} covers the infinite square lattice. Inevitably, some $v^{[x]}$'s will gather at the same site. The tensor product of these $v^{[x]}$'s in fact gives the optimal rank-1 approximation of the ``correct'' full-rank tensor here (Sec. \ref{sec.zeroloop}). Suppose that one uses the full-rank tensor to replace its rank-1 version (the tensor product of four $v^{[x]}$'s), one will have the PEPO\index{PEPO} of $I- \tau \hat{H}$ (with $H$ the Hamiltonian on square lattice), and the tree iPEPS\index{iPEPS} becomes the iPEPS defined on the square lattice. Compared with the NCD\index{NCD} scheme that employs rank-1 decomposition explicitly to solve TN contraction, one difference here for updating iPEPS is that the ``correct'' tensor to be decomposed by rank-1 decomposition contains the variational tensor, thus is in fact unknown before the equations are solved. For this reason, we cannot use rank-1 decomposition directly. Another difference is that the constraint, i.e., the normalization of the tree iPEPS, should be fulfilled. By utilizing the iDMRG\index{iDMRG} algorithm with the tree iPEPS, the rank-1 tensor is obtained without knowing the ``correct'' tensor, and meanwhile the constraints are satisfied. The zero-loop approximation of the ground state is thus given by the tree iPEPS.

\begin{figure}[tbp]
	\centering
	\includegraphics[angle=0,width=1\linewidth]{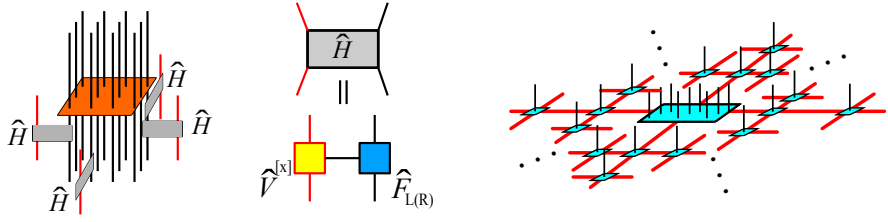}
	\caption{(Color online) The left figure shows the few-body Hamiltonian $\mathcal{\hat{H}}$ in Eq. (\ref{eq-4Heffect2D}). The middle one shows the physical-bath Hamiltonian $\hat{\mathcal{H}}_{\partial}$ that gives the interaction between the corresponding physical and bath site. The right one illustrates the state ansatz for the infinite system. Note that the boundary of the cluster should be surrounded by $\hat{\mathcal{H}}_{\partial}$'s, and each $\hat{\mathcal{H}}_{\partial}$ corresponds to an infinite tree brunch in the state ansatz. For simplicity, we only illustrate four of the $\hat{\mathcal{H}}_{\partial}$'s and the corresponding brunches.}
	\label{fig-4Hbathx}
\end{figure}

The few-body Hamiltonian is constructed in a larger cluster, so that the error brought by zero-loop approximation can be reduced. Similar to the 1D case, we embed a larger cluster in the middle of the entanglement bath. The few-body Hamiltonian (Fig. \ref{fig-4Hbathx}) is written as
\begin{eqnarray}
\mathcal{\hat{H}} = \prod_{\langle n \in cluster, \alpha \in bath \rangle} \hat{\mathcal{H}}_{\partial}(n,\alpha) \prod_{\langle i,j \rangle \in cluster} [I-\tau \hat{H}(s_i,s_j)].
\label{eq-4Heffect2D}
\end{eqnarray}
$\hat{\mathcal{H}}_{\partial}(n,\alpha)$ is defined as the physical-bath Hamiltonian between the $\alpha$-th bath site and the neighboring $n$-th physical site, and it is obtained by the corresponding boundary tensor $v^{[x(\alpha)]}$ and $\hat{F}_{L(R)}(s_n)$ (Fig. \ref{fig-4Hbathx}) as
\begin{eqnarray}
\langle b_{\alpha}'s_n'| \mathcal{\hat{H}}_{\partial}(n,\alpha) |b_{\alpha}s_n\rangle = \sum_{a} v^{[x(\alpha)]}_{ab_{\alpha}' b_{\alpha}} \langle s_n' |\hat{F}_{L(R)}(s_n)_{a} | s_n\rangle.
\label{eq-4Hbathx}
\end{eqnarray}
Here, $\hat{F}_{L(R)}(s_n)_{a}$ is the operator defined in Eq. (\ref{eq-Fsvd}), and $v^{[x(\alpha)]}_{ab_{\alpha}' b_{\alpha}}$ are the solutions of the SEE's\index{SEE} given in Eqs. (\ref{eq-effectH})-(\ref{eq-M4}).

$\hat{\mathcal{H}}$ in Eq. (\ref{eq-4Heffect2D}) can also be rewritten as the shift of a few-body Hamiltonian $\hat{H}_{FB}$, i.e. $\mathcal{\hat{H}} = I-\tau \hat{H}_{FB}$. We have $\hat{H}_{FB}$ possessing the standard summation form as
\begin{eqnarray}
\hat{H}_{FB} = \sum_{\langle i,j \rangle \in cluster} \hat{H}(s_i,s_j) + \sum_{\langle n \in cluster,\alpha \in bath \rangle} \hat{H}_{PB}(n,\alpha),
\label{eq-4Hfb2D}
\end{eqnarray}
with $\mathcal{\hat{H}}_{\partial}(n,\alpha) = I-\tau \hat{H}_{PB}(s_n,b_{\alpha})$. This equations gives a general form of the few-body Hamiltonian: the first term contains all the physical interactions inside the cluster, and the second contains all physical-bath interactions $\hat{H}_{PB}(s_n,b_{\alpha})$. $\hat{\mathcal{H}}$ can be solved by any finite-size algorithms, such as exact diagonalization, QMC\index{QMC}, DMRG\index{DMRG} \cite{W92DMRG, WS98tjDMRG, XLS01DMRG2D} or finite-size PEPS\index{PEPS} \cite{SV09TNQMC, LCB14fPEPS, LDHGH17TNQMC} algorithms. The error from the rank-1 decomposition will be reduced since the loops inside the cluster will be fully considered.

Similar to the 1D cases, the ground-state properties can be extracted by the reduced density matrix $\hat{\rho}(S)$ after tracing over the entanglement-bath degrees of freedom. We have $\hat{\rho}(S) = \text{Tr}_{/(S)} |\Phi \rangle \langle \Phi |$ (with $|\Phi \rangle$ the ground state of the infinite model) that well approximate by
\begin{eqnarray}
\hat{\rho}(S) \simeq \sum_{SS'b_1b_2 \cdots} \Psi^{\ast}_{S'b_1b_2 \cdots} \Psi_{Sb_1b_2 \cdots} |S \rangle \langle S'|.
\label{eq-rhoBath}
\end{eqnarray}
with $\Psi_{Sb_1b_2 \cdots} $ the coefficients of the ground state of $\hat{H}_{FB}$.

Fig. \ref{fig-4Hbathx} illustrates the ground state ansatz behind the few-body model. The cluster in the center is entangled with the surrounding infinite-tree brunches through the entanglement-bath degrees of freedom. Note that solving Eq. (\ref{eq-effectH}) in Stage one is equivalent to solving Eq. (\ref{eq-4Heffect2D}) by choose the cluster as one supercell. 

Some benchmark results of simulating 2D and 3D spin models can be found in Ref. \cite{RPPSL17AOP3D}. For the ground state of Heisenberg model on honeycomb lattice, results of the magnetization and bond energy show that the few-body model of 18 physical and 12 bath sites suffers only a small finite-effect of $O(10^{-3})$. For the ground state of 3D Heisenberg model on cubic lattice, the discrepancy of the energy per site is $O(10^{-3})$ between the few-body model of 8 physical plus 24 bath sites and the model of 1000 sites by QMC\index{QMC}. The quantum phase transition of the quantum Ising model on cubic lattice can also be accurately captured by such a few-body model, including determining the critical field and the critical exponent of the magnetization.

\section{Quantum entanglement simulation by tensor network: summary}

\begin{figure}[tbp]
	\centering
	\includegraphics[angle=0,width=1\linewidth]{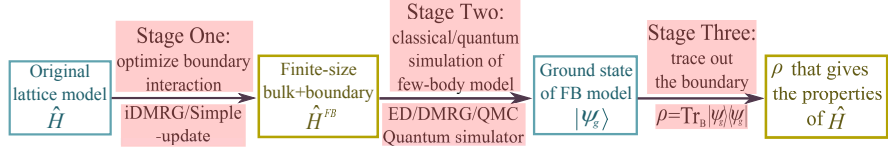}
	\caption{(Color online) The ``\textit{ab-initio} optimization principle'' to simulate quantum many-body systems.}
	\label{fig-4Protocol}
\end{figure}

Below, we summarize the QES\index{QES} approach for quantum many-body systems with few-body models \cite{RPPSL17AOP3D, R16AOP, RXPS+18FTQES}. The QES contains three stages (Fig. \ref{fig-4Protocol}) in general. The first stage is to optimize the physical-bath interactions by classical computations. The algorithm can be iDMRG\index{iDMRG} in one dimension or the zero-loop schemes in higher dimensions. The second stage is to construct the few-body model by embedding a finite-size cluster in the entanglement bath, and simulate the ground state of this few-body model. One can employ any well-established finite-size algorithms by classical computations, or build the quantum simulators according to the few-body Hamiltonian. The third stage is to extract physical information by tracing over all bath degrees of freedom. The QES approach has been generalized to finite-temperature simulations for one-, two-, and three-dimensional quantum lattice models \cite{RXPS+18FTQES}.

As to the classical computations, one will have a high flexibility to balance between the computational complexity and accuracy, according to the required precision and the computational resources at hand. On the one hand, thanks to the zero-loop approximation, one can avoid the conventional finite-size effects faced by the previous exact diagonalization, QMC\index{QMC}, or DMRG\index{DMRG} algorithms with the standard finite-size models. In the QES\index{QES}, the size of the few-body model is finite, but the actual size is infinite as the size of the defective TN\index{TN} (see Sec. \ref{sec.zeroloop}). The approximation is that the loops beyond the supercell are destroyed in the manner of the rank-1 approximation, so that the TN can be computed efficiently by classical computation. On the other hand, the error from the destruction of the loops can be reduced in the second stage by considering a cluster larger than the supercell. It is important that the second stage would introduce no improvement if no larger loops are contained in the enlarged cluster. From this point of view, we have no ``finite-size'' but ``finite-loop'' effects. In addition, this ``loop'' scheme explains why we can flexibly change the size of the cluster in stage two: which is just to restore the rank-1 tensors inside the chosen cluster with the full tensors.

\begin{figure}[tbp]
	\centering
	\includegraphics[angle=0,width=1\linewidth]{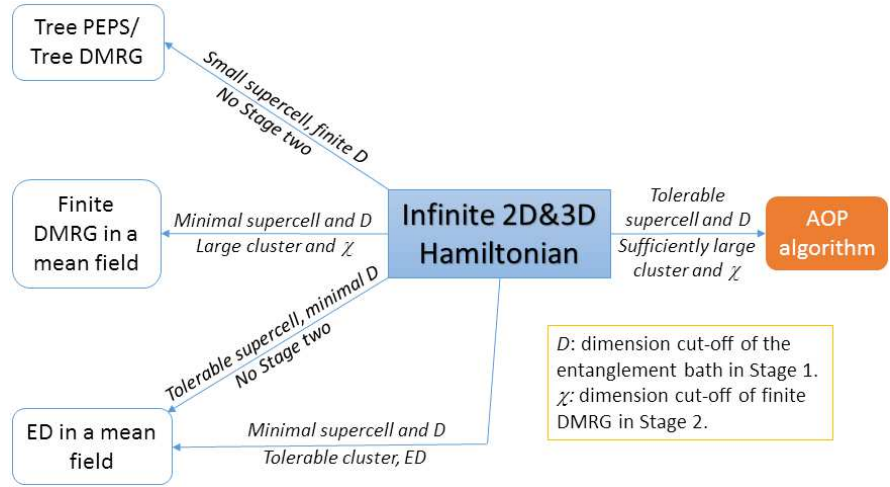}
	\caption{(Color online) Relations to the algorithms (PEPS, DMRG and ED) for the ground-state simulations of 2D and 3D Hamiltonian. The corresponding computational set-ups in the first (bath calculation) and second (solving the few-body Hamiltonian) stages are given above and under the arrows, respectively. Reused from \cite{RPPSL17AOP3D} with permission.}
	\label{fig-4AOPrelation}
\end{figure}

The relations among other algorithms are illustrated in Fig. \ref{fig-4AOPrelation} by taking certain limits of the computational parameters. The simplest situation is to take the dimension of the bath sites $\rm dim (b) =1$, and then $\hat{\mathcal{H}}_{\partial}$ can be written as a linear combination of spin operators (and identity). Thus in this case, $v^{[x]}$ simply plays the role of a classical mean field. If one only uses the bath calculation of the first stage to obtain the ground-state properties, the algorithm will be reduced to the zero-loop schemes such as tree DMRG\index{DMRG} and simple update of iPEPS\index{iPEPS}. By choosing a large cluster and $\rm dim (b) =1$, the DMRG simulation in stage two becomes equivalent to the standard DMRG for solving the cluster in a mean field. By taking proper supercell, cluster, algorithms and other computational parameters, the QES\index{QES} approach can outperform others. 

The QES approach with classical computations can be categorized as a cluster update scheme (see Sec. \ref{sec.updateschemes}) in the sense of classical computations. Compared with the ``traditional'' cluster update schemes \cite{LDX12TTN,  LCB14PEPScontract, WV11PEPSclusterArxiv, LCB14fPEPS}, there exist some essential differences. The``traditional'' cluster update schemes use the super-orthogonal spectra to approximate the environment of the iPEPS. The central idea of QES is different, which is to give an effective finite-size Hamiltonian; the environment is mimicked by the physical-bath Hamiltonians instead of some spectra.

In addition, it is possible to use full update in the first stage to optimize the interactions related to the entanglement bath. For example, one may use TRD\index{TRD} (iDMRG\index{iDMRG}, iTEBD\index{iTEBD} or CTMRG\index{CTMRG}) to compute the environment tensors, instead of the zero-loop schemes. This idea has not been realized yet, but it can be foreseen that the interactions among the bath sites will appear in $\hat{H}_{FB}$. Surely the computation will become much more expensive. It is not clear yet how the performance would be.

The idea of ``bath'' has been utilized in many approaches and gained tremendous successes. The general idea is to mimic the target model of high complexity by a simpler model embedded in a bath. The physics of the target model can be extracted by integrating over the bath degrees of freedom. The approximations are reflected by the underlying effective model. Table \ref{tab-methods} shows the effective models of two recognized methods (DFT\index{DFT} and dynamic mean-field theory (DMFT)\index{DMFT} \cite{AGPTW10DMF}) and the QES\index{QES}. An essential difference is that the effective models of the former two methods are of single-particle or mean-field approximations, and the effective model of the QES is strongly correlated.

\begin{table}[tbp]
	\caption{The effective models under several bath-related methods: density functional theory (DFT\index{DFT}, also known as the \textit{ab-initio} calculations), dynamical mean-field theory (DMFT)\index{DMFT} and QES\index{QES}.}
	\begin{tabular*}{12cm}{@{\extracolsep{\fill}}rccc}
		\\ \hline\hline
		\textbf{Methods} & DFT & DMFT & QES \\ \hline
		\textbf{Effective models} & Tight binding model & Single impurity model & Interacting few-body model
		\\ \hline\hline
		\label{tab-methods}
	\end{tabular*}
\end{table}

The QES\index{QES} allow for quantum simulations of infinite-size many-body systems by realizing the few-body models on the quantum platforms. There are several unique advantages. The first one concerns the size. One of the main challenges to build a quantum simulator is to access a large size. In this scheme, a few-body model of only $O(10)$ sites already shows a high accuracy with the error $\sim O(10^{-3})$ \cite{RPPSL17AOP3D, R16AOP}. Such sizes are accessible by the current platforms. Secondly, the interactions in the few-body model are simple. The bulk just contains the interactions of the original physical model. The physical-bath interactions are only two-body and nearest-neighbor. But there exist several challenges. Firstly, the physical-bath interaction for simulating, e.g., spin-$1/2$ models, is between a spin-$1/2$ and a higher spin. This may require the realization of the interactions between SU(N) spins, which is difficult but possible with current experimental techniques \cite{GHGX+10SUNsimu,BBDRS+13SUNsimu,SHHDB14SUNsimu,ZBBK+14SUNsimu}. The second challenge concerns the non-standard form in the physical-bath interaction, such as the $\hat{S}^x\hat{S}^z$ coupling in $\hat{H}_{FB}$ for simulating quantum Ising chain [see Eq. (\ref{eq-4HLR})] \cite{RPSL17CSW}. With the experimental realization of the few-body models, the numerical simulations of many-body systems will not only be useful to study natural materials. It would become possible to firstly study the many-body phenomena by numerics, and then realize, control, and even utilize these many-body phenomena in the bulk of small quantum devices.

The QES Hamiltonian was shown to also mimics the thermodynamics \cite{RXPS+18FTQES}. The finite-temperature information is extracted from the reduced density matrix
\begin{equation}
\hat{\rho}_R = \text{Tr}_{\text{bath}} \hat{\rho},
\label{eq-6RDMFT}
\end{equation}
with $\hat{\rho}=e^{- \mathcal{\hat{H}_{FB}}/\rm{T}}$ the density matrix of the QES\index{QES} at the temperature $\rm{T}$ and $\text{Tr}_{\text{bath}}$ the trace over the degrees of freedom of the bath sites. $\hat{\rho}_R$ mimics the reduced density matrix of infinite-size system that traces over everything except the bulk. This idea has been used to simulate the quantum models in one, two, and three dimensions. The QES shows good accuracy at all temperatures, where relatively large error appears near the critical/crossover temperature.

One can readily check the consistency with the ground-state QES\index{QES}. When the ground state is unique, the density matrix is defined as $\hat{\rho} = |\Psi \rangle \langle \Psi|$ with $|\Psi \rangle$ the ground state of the QES. In this case, Eqs. (\ref{eq-6RDMFT}) and (\ref{eq-4RDM}) are equivalent. With degenerate ground states, the equivalence should still hold when the spontaneous symmetry breaking occurs. With the symmetry preserved, it is an open question how the ground-state degeneracy affects the QES, where at zero temperature we have $\hat{\rho} = \sum_a^{\mathcal{D}} |\Psi_a \rangle \langle \Psi_a|/ \mathcal{D}$ with $\{|\Psi_a \rangle\}$ the degenerate ground states and $\mathcal{D}$ the degeneracy.

\chapter{Summary}

The explosive progresses of TN\index{TN} that have been made in the recent years opened an interdisciplinary diagram for studying varieties of subjects. What is more, the theories and techniques in the TN algorithms are now evolving into a new numerical field, forming a systematic framework for numerical simulations. Our lecture notes are aimed at presenting this framework from the perspective of the TN contraction algorithms for quantum many-body physics.

The basic steps of the TN contraction algorithms are to contract the tensors and to truncate the bond dimensions to bound the computational cost. For the contraction procedure, the key is the contraction order, which leads to the exponential, linearized, and polynomial contraction algorithms according to how the size of the TN decreases. For the truncation, the key is the environment, which plays the role of the reference for determining the importance of the basis. We have the simple, cluster, and full decimation schemes, where the environment is chosen to be a local tensor, a local but larger cluster, and the whole TN, respectively. When the environment becomes larger, the accuracy increases, but so do the computational costs. Thus, it is important to balance between the efficiency and accuracy. Then, we show that by explicitly writing the truncations in the TN, we are essentially dealing with exactly contractible TN's.

Compared with the existing reviews of TN, a unique perspective that our notes discuss about is the underlying relations between the TN approaches and the multi-linear algebra (MLA)\index{MLA}. Instead of iterating the contraction-and-truncation process, the idea is to build a set of local self-consistent eigenvalue equations that could reconstruct the target TN. These self-consistent equations in fact coincide with or generalize the tensor decompositions in MLA, including Tucker decomposition, rank-1 decomposition and its higher-rank version. The equations are parameterized by both the tensor(s) that define the TN and the variational tensors (the solution of the equations), thus can be solved in a recursive manner. This MLA perspective provides a unified scheme to understand the established TN methods including iDMRG\index{iDMRG}, iTEBD\index{iTEBD}, and CTMRG\index{CTMRG}. In the end, we explain how the eigenvalue equations lead to the quantum entanglement simulation (QES)\index{QES} of the lattice models. The central idea of QES is to construct an effective few-body model surrounded by the entanglement bath, where its bulk mimics the properties of the infinite-size model at both zero and finite temperatures. The interactions between the bulk and the bath are optimized by the TN methods. The QES provides an efficient way for simulating one-, two-, and even three-dimensional infinite-size many-body models by classical computation and/or quantum simulation.

With the lecture notes, we expect that the readers could use the existing TN algorithms to solve their problems. Moreover, we hope that those who are interested in TN itself could get the ideas and the connections behind the algorithms to develop novel TN schemes.


\printindex

%
%

\extrachap{Acronyms}

\begin{description}[CABR]
	\item[AKLT state]{Affleck-Kennedy-Lieb-Tasaki state}
	\item[AOP]{\textit{ab-initio optimization principle}}
	\item[CANDECOMP/PARAFAC]{canonical decomposition / parallel factorization}
	\item[CFT]{conformal field theory}
	\item[CTM]{corner transfer matrix}
	\item[CTMRG]{corner transfer matrix renormalization group}
	\item[DFT]{density functional theory}
	\item[DMFT]{dynamical mean-field theory}
	\item[DMRG]{density matrix renormalization group}
	\item[ECTN]{exactly-contractible tensor network}
	\item[HOOI]{higher-order orthogonal iteration}
	\item[HOSVD]{higher-order singular value decomposition}
	\item[HOTRG]{higher-order tensor renormalization group}
	\item[iDMRG]{infinite density matrix renormalization group}
	\item[iPEPO]{infinite projected entangled pair operator}
	\item[iPEPS]{infinite projected entangled pair state}
	\item[iTEBD]{infinite time-evolving block decimation}
	\item[MERA]{multi-scale entanglement renormalization ansatz}
	\item[MLA]{multi-linear algebra}
	\item[MPO]{matrix product operator}
	\item[MPS]{matrix product state}
	\item[NCD]{network contractor dynamics}
	\item[NP hard]{non-deterministic polynomial hard}
	\item[NRG]{numerical renormalization group}
	\item[NTD]{network Tucker decomposition}
	\item[PEPO]{projected entangled pair operator}
	\item[PEPS]{projected entangled pair state}
	\item[QES]{quantum entanglement simulation/simulator}
	\item[QMC]{quantum Monte Carlo}
	\item[RG]{renormalization group}
	\item[RVB]{resonating valence bond}
	\item[SEE's]{self-consistent eigenvalue equations}
	\item[SRG]{second renormalization group}
	\item[SVD]{singular value decomposition}
	\item[TDVP]{time-dependent variational principle}
	\item[TEBD]{time-evolving block decimation}
	\item[TMRG]{transfer matrix renormalization group}
	\item[TN]{tensor network}
	\item[TNR]{tensor network renormalization}
	\item[TNS]{tensor network state}
	\item[TPO]{tensor product operator}
	\item[TRD]{tensor ring decomposition}
	\item[TRG]{tensor renormalization group}
	\item[TTD]{tensor-train decomposition}
	\item[TTNS]{tree tensor network state}
	\item[VMPS]{variational matrix product state}
	
\end{description}

\bibliographystyle{unsrt} 


\end{document}